\documentclass[twocolumn]{aastex631}

\usepackage{mathtools}
\newcommand{\xx}[1]{$\xi_{#1}$} 

\newcommand{\Mff}{$M_{\rm ff}$}
\newcommand{\MCO}{$M_{\rm CO}$}
\newcommand{\Msib}{$M_{\rm Si, base}$}
\newcommand{\Mob}{$M_{\rm O, base}$}
\newcommand{\Mcb}{$M_{\rm C, base}$}
\newcommand{\Msk}{$M(s_k\!=\!4)$}
\newcommand{\musk}{$\mu(s_k\!=\!4)$}
\newcommand{\Mmusk}{$M\mu(s_k\!=\!4)$}
\newcommand{\gcc}{g cm$^{-3}$}
\newcommand\aastex{AAS\TeX}

\received{\today}
\revised{}
\accepted{}
\submitjournal{ApJ}

\shorttitle{\aastex\ Monotonicity of the cores of massive stars}
\shortauthors{Takahashi, K, et al.}

\begin{document}

\title{Monotonicity of the cores of massive stars}

\correspondingauthor{Koh Takahashi}
\email{koh.takahashi@astr.tohoku.ac.jp}

\author{Koh Takahashi}
\affil{Max-Planck-Institut f\"{u}r Gvravitationsphysik, 14476 Potsdam-Golm, Germany}
\affil{Astronomical Institute, Graduate School of Science, Tohoku University, Sendai, 980-8578, Japan}

\author{Tomoya Takiwaki}
\affil{National Astronomical Observatory of Japan, National Institutes for Natural Science, 2-21-1 Osawa, Mitaka, Tokyo 181-8588, Japan}

\author{Takashi Yoshida}
\affil{Yukawa Institute for Theoretical Physics, Kyoto University, Kitashirakawa Oiwakecho, Sakyo-ku, Kyoto 606-8502, Japan}

\begin{abstract}
Massive stars are linked with diverse astronomical processes and objects including star formation, supernovae and their remnants, cosmic rays, interstellar media, and galaxy evolution. Understanding their properties is of primary importance for modern astronomy, and finding simple rules that characterize them is especially useful.
However, theoretical simulations have not yet realized such relations, instead finding that the late evolutionary phases are significantly affected by a complicated interplay between nuclear reactions, chemical mixing, and neutrino radiation, leading to non-monotonic initial mass dependencies of the iron core mass and the compactness parameter.
We conduct a set of stellar evolution simulations, in which evolutions of He star models are followed until their central densities uniformly reach 10$^{10}$ \gcc, and analyze their final structures as well as their evolutionary properties including the lifetime, surface radius change, and presumable fates after core collapse.
Based on the homogeneous data set, we have found that monotonicity is inherent in the cores of massive stars. We show that not only the density, entropy, and chemical distributions, but also their lifetimes and explosion properties such as the proto-neutron-star mass and the explosion energy can be simultaneously ordered into a monotonic sequence.
This monotonicity can be regarded as an empirical principle that characterizes the cores of massive stars.
\end{abstract}
\keywords{Stellar evolution --- Core Collapse Supernovae}

\section{Introduction}

Massive stars are an important component of the universe.
Throughout their lifetime, they affect ambient environments by emitting intense photon radiation and powerful stellar winds \citep{Langer12}. Massive stars are also important as progenitors of core-collapse supernovae (CCSNe). CCSNe allows us to observe the distant universe as luminous transients \citep[e.g.][]{Modjaz19} and are one of the main drivers of the chemo-dynamical evolution of galaxies \citep[][]{Woosley+07, Nomoto+13}. These explosions produce neutron stars (NSs) and black holes (BHs), and they may also trigger star formation \citep[e.g.][]{Girichidis20}. The remnants left behind after the explosion are observed as supernova remnants, which can be a source of cosmic rays \citep{Vink12, Blasi13}.
Understanding the properties of massive stars is, thus, crucial for entire fields in modern astronomy.

One-dimensional stellar evolution simulations have shown that robust monotonicity is inherent in the structure and the evolution of main-sequence stars \citep[e.g.][]{Kippenhahn&Weigert90}. Namely, strong correlations have been found between the fundamental parameter of the initial mass of the star, and stellar properties such as luminosity, radius, and lifetime. The term, `massive star', already implies that the initial mass is also useful to distinguish stars that eventually experience core collapse from the others which form white dwarves at the end of their lives. However, previous works have also shown that the initial mass is not applicable for characterizing the sub-populations of massive stars. For example, the importance of the mass of the iron core has been recognized for many years \citep[][and references therein]{Woosley&Weaver86, Nomoto&Hashimoto88}, and interestingly, the iron core mass shows a significant non-monotonic dependence on the initial mass. 
In particular, the steep increase of the iron core mass has been attributed to the transition from the convective to the radiative nature of central carbon burning (\citealt{Woosley&Weaver86}, see also, \citealt{Timmes+96}).

The subject above can be rephrased as the question of whether there is a single parameter that has the capability to characterize the evolutionary properties of the cores of massive stars, such as the core mass, the entropy, and the lifetime. 
From a theoretical point of view, the existence of such a parameter is non-trivial. First of all, the hydrostatic structure of a star has infinite degrees of freedom derived from the pressure-density relation, or equation of states (EOS), given the central density. Since EOS is in reality determined by entropy and composition, for a simple relation to holding for the sequence of stellar structures, the composition and entropy distributions must satisfy another simple relation. 
However, it seems difficult to establish such a simple relationship in the cores of massive stars.
This is fundamental because the nuclear reactions and neutrino energy losses that occur inside massive stars are strongly dependent on temperature and density. These reactions not only directly affect the composition and entropy distributions, but also indirectly complicate them by causing core convection.
As a result, the composition and entropy distributions of massive star cores become generally very complex.

Interesting implications have been obtained from recent investigations about the progenitor-explosion connection. A significant increase in the number of observed CCSNe in recent decades has resulted in numerous intriguing correlations. 
For Type II SNe showing lines of hydrogen in the spectrum, correlations have been suggested between $^{56}$Ni ejecta mass, total ejecta mass, plateau luminosity, and expansion velocity \citep{Hamuy03, Anderson+14, Spiro+14, Valenti+15, Mueller+17, Anderson19, Martinez22}.
Similarly, correlations among kinetic energy, ejecta mass, and $^{56}$Ni ejecta mass have been found for the stripped-envelope SNe (SE-SNe) consisting of types IIb, Ib, and Ic (\citealt{Lyman+16, Taddia+18, Anderson19, Barbarino21}, see also \citealt{Meza20, Saito22}).
Given that SN explosions can be defined as the solution to the initial value problem initiated by the collapse of the core of a massive star, the diversity and the correlations in the properties of the explosions should essentially come from the structure of the progenitor star.

Not all massive stars successfully explode and leave NSs behind. Some will explode but still form a BH as a result of accretion after shock revival, while others will create a BH without shock revival and may not explode \citep[the `failed' SNe,][]{Kochanek+08} or may only have a very weak and sub-luminous explosion \citep[the `faint' SNe,][]{Lovegrove&Woosley13}. Indeed, the existence of stellar mass BHs have been indicated by X-ray binaries \citep[cf.][]{Casares14, Corral-Santana16, Tetarenko16} and, more recently, by BH mergers detected by gravitational wave detectors \citep{Abbott21, Abbott21b, Abbott21c}. Moreover, a direct indication of BH formation has been obtained from the disappearance of a red supergiant (RSG) from successive monitoring \citep{Smartt09, Smartt15, Davies18}. Not only the explosion properties but also the explodability, i.e., whether a massive star successfully explodes or not, should also be determined by the progenitor structure. Then, the important question is what is the property that controls the explodability and the properties of successful explosions?

Several studies have investigated the progenitor-explosion connection by conducting systematic simulations of hydrodynamical evolution after core collapse \citep{OConnor&Ott11, Ugliano+12, Pejcha&Thompson15, Perego+15, Sukhbold+16, Ertl+16, Mueller+16, Ebinger18, Ebinger20, Ertl+20} and suggested the possibility of characterizing the explosion properties based on one or two simple parameters that characterize the progenitor structure. One example is the so-called compactness parameter, which was defined by \citet{OConnor&Ott11} as 
\begin{eqnarray}
	\xi_{M} = \frac{M/M_\odot}{R(M) / 1000 \ \mathrm{km}},
\end{eqnarray} 
at the time of core bounce, where $M$ and $R(M)$ are the enclosed mass and the radius as a function of the mass coordinate. The capability for judging the explodability \citep{OConnor&Ott11}, as well as for characterizing the properties of CCSN explosions \citep[e.g.,][]{OConnor&Ott13, Mueller+16}, has been suggested. Similarly, the efficacy of the set of parameters $M_4$ and $\mu_4$, which are related to the density and entropy distributions, for judging the explodability was proposed by \citet{Ertl+16}.
Since the nature of the CCSN explosion should be determined by the time evolution of hydrodynamical quantities such as accretion history, and thus the functional form of the density distribution throughout the core, it is surprising to conclude that a judgment can be made based on such partial and limited information as the compactness parameter.
It should be noted here that the above works have utilized either a parametric one-dimensional hydrodynamic code or a simplified semi-analytical model to estimate the explodability. In this regard, \citet{Burrows20} concluded that the compactness is not a measure of the explodability based on the results of state-of-the-art three-dimensional simulations. Further research is still needed to settle the true efficacy of compactness.

Nevertheless, the concept of discerning the exploding or non-exploding progenitors from a single parameter may be consistent with observations. So far, dozens of SN progenitors that were accidentally imaged before the SN explosion have been identified. By analyzing these pre-explosion images, the nature of the progenitors of SN explosions can be estimated, and it has been reported that there is a lack of luminous progenitors characterized by $\log L/L_\odot \gtrsim 5.1$ (the missing RSG problem; \citealt{Smartt09, Smartt15}, however, see \citealt{Davies18}).
\citet{Horiuchi+14} have further pointed out another problem, the supernova rate problem, i.e., the deficiency of cosmic supernova rate compared to the cosmic star formation rate. And they have shown that if massive stars with compact structures characterized by \xx{2.5} $\gtrsim 0.2$ fail to explode as canonical SNe, then not only the SN rate problem but also the missing RSG problem can be solved simultaneously.

Similar to the iron core mass, the compactness parameter is also known to have a non-monotonic initial mass dependence. The non-monotonicity is affected especially by convective shell burnings of carbon and oxygen \citep{Sukhbold&Woosley14, Chieffi20}, and thus depends on the treatments of input physics including convective boundary mixing, semi-convection, the $^{12}$C($\alpha$,$\gamma$)$^{16}$O reaction rate, as well as the mass-loss rate and the metallicity \citep{Sukhbold&Woosley14, Sukhbold+18, Sukhbold&Adams20, Chieffi20}. As a particularly interesting finding, a strong correlation between the compactness parameter and the iron core mass has been recognized in many works (e.g., \citealt{OConnor&Ott11, Ertl+16, Schneider21}, see also \citealt{Sukhbold&Woosley14} for the binding energy outside the iron core, \citealt{Chieffi20} for compactness parameters defined at different locations, and \citealt{Schneider21} for the core entropy and masses of carbon- and neon-free regions). 

In this work, we aim to find a simple relation, which can be used to characterize the core properties of massive stars, by conducting the simulation of the massive star evolution. Even if it exists, such a relation could be subtle, and hence, there is concern that some influence, whether physical or numerical, may obscure the relation. The wind mass-loss and H shell burning could be such an effect that impacts the initial-mass to core mass relations as well as the core structure. In order to avoid such complications, we follow the evolution of the helium star model, which is not affected by the aforementioned effects and can be regarded as an idealized helium core of a massive star. Although the models cannot be directly compared to observations, we assume that the qualitative tendencies are still common to more realistic models and, hopefully, to real stars.

In the next section, we describe the two theoretical frameworks that we utilize in this work; the stellar evolution code and the semi-analytic code that is developed following \citet{Mueller+16} and used to estimate properties of the post-collapse evolutions. In section 3, first, an alternative indicator of the density structure, \Mff, is introduced, which has a more intuitive definition as well as a better convergency than the compactness parameter during the late time evolution. Then we show that the global (but still inner) density structure of the CCSN progenitors can be well sorted according to the \Mff{} order. In section 4, correlations between \Mff{} and two other evolutionary properties, the remaining time till collapse and the stellar radius, as well as explosion properties including the explodability are investigated. We discuss the robustness of the correlations by comparing stellar models obtained from different codes and settings and discuss the observational relevance of the newly found lifetime-core structure correlation to the pre-collapse mass ejection in section 5. The summary and conclusion are given in section 6.

\section{Method}

\subsection{The stellar evolution code}
The evolution of single He stars is calculated using the {\it HOSHI} code \citep{Takahashi+18, Takahashi&Langer21}. The initial chemical composition is set to pure helium, and the initial metallicity is zero. The input physics used in the code is almost the same as that used in \citet{Takahashi&Langer21}, so we omit writing the details here and describe only the differences. First, while the code is capable of treating the time evolution of the stellar rotation and the magnetic fields, these are neglected in the present models. Also, wind mass loss is not taken into account in the current modeling. The nuclear reaction network includes 300 isotopes ranging from n and p to $^{80}$Br. The complete list is given in \citet{Takahashi+18}. The $^{12}$C($\alpha$,$\gamma$)$^{16}$O rate of \citet{deBoer+17} is applied for our fiducial models. The effect of convective boundary mixing is treated as a diffusive process as described in \citet{Takahashi&Langer21}, however, the effect is set to be minimal as a very small control parameter of $f_\mathrm{ov} = 0.001$ is applied for the current models.

We calculate stellar evolution for a total of 128 models with different initial masses. The initial mass interval ($d M_{\rm ini}$) is changed for light, intermediate-mass, and heavy stars. For the lightest stars in the range $M_{\rm ini}/M_\odot \in [2.0, 9.1]$, we calculate 72 models using $d M_{\rm ini}/M_\odot = 0.1$, and for the intermediate mass range $M_{\rm ini}/M_\odot \in [9.2, 14.0]$, 25 models are calculated with $d M_{\rm ini}/M_\odot = 0.2$. For the heavier side $M_{\rm ini}/M_\odot \in [14.5, 29.5]$, an increment of $d M_{\rm ini}/M_\odot = 0.5$ is used (31 models).

For each model, the evolution from the He-zero-age-main-sequence (HeZAMS) phase until the central density, $\rho_c$, reaches $10^{10}$\,g\,cm$^{-1}$ is calculated unless the simulation stops due to convergence problems. We have confirmed that the star has already lost hydrostatic stability with this final central density. Convergence problems happen for the lowest mass models with $M_{\rm ini} \leq 2.7 M_\odot$, where two models stop during the shell carbon burning phase ($M_{\rm ini}/M_\odot = 2.0, 2.1$), five models stop after the formation of the ONe core ($M_{\rm ini}/M_\odot \in [2.2, 2.6]$), and one model of $M_{\rm ini}/M_\odot = 2.7$ stops during the shell O+Ne burning phase. After removing the 8 non-convergent models, we obtained in total of 120 progenitor models of CCSNe.

\begin{figure}[t]
 \centering
 \includegraphics[width=\hsize]{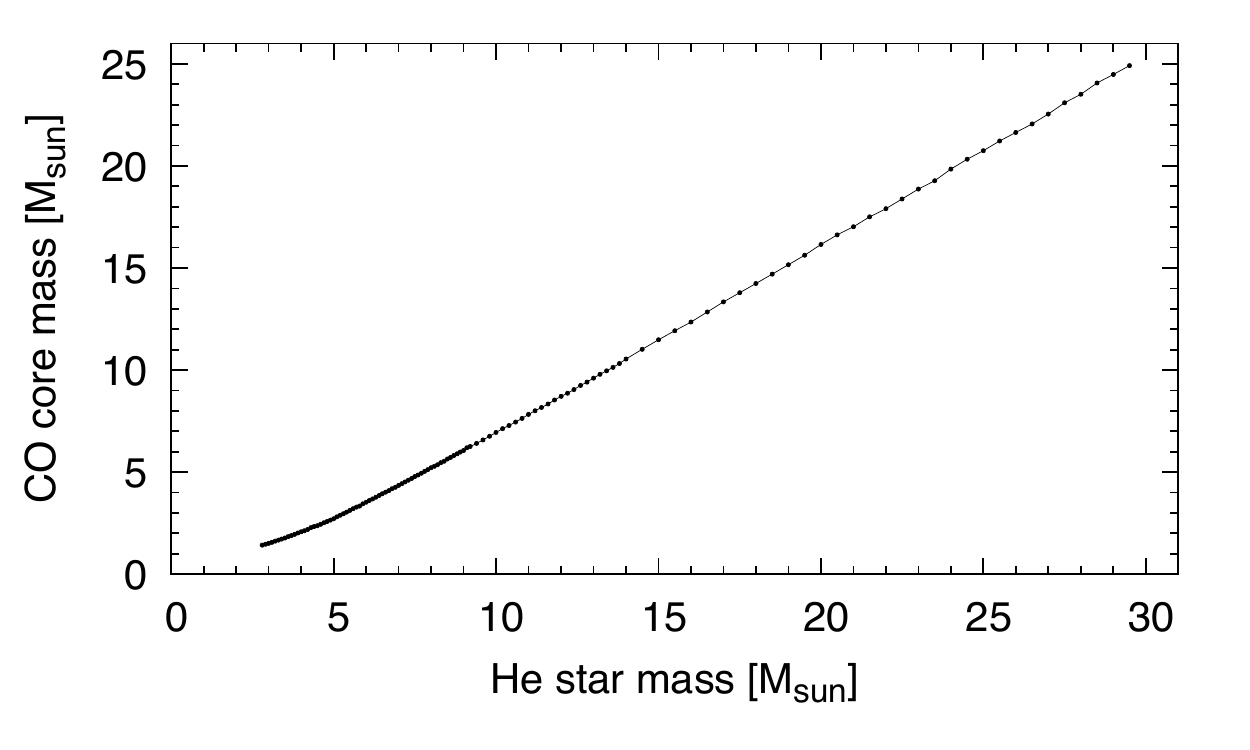}
 \caption{Relation between the He star mass and the CO core mass. As a result of applying a minimal convective boundary mixing for the core He burning, this relation is almost identical to models with completely neglect the convective boundary mixing.
 }
 \label{plot-MHe-MCO}
\end{figure}

As a result of neglecting the wind mass loss, the CO core mass (\MCO{}) distribution obeys a highly monotonic relation with the He star mass (Fig.~\ref{plot-MHe-MCO}). In this work, \MCO{} is defined as the lower mass limit of the region where the helium mass fraction exceeds 0.01. The usage of other definitions (such as based on the heating rate) is possible but the qualitative results would be unchanged. Hereafter, we will use the CO core mass, instead of the He star mass, as the model indicator.

\subsection{M\"{u}ller's semi-analytic model}

In order to estimate the fate after core collapse, a semi-analytic code has been developed following the description in \citet{Mueller+16}. By integrating a few ordinary differential equations, M\"{u}ller's semi-analytic model provides the result of the post-collapse evolution including the fate and the explosion properties such as the explosion energy and the remnant (NS) mass if the progenitor is estimated to successfully explode. 

This model relies on the delayed neutrino-heating mechanism for CCSNe, in which a fraction of the gravitational energy released by the accretion is converted into thermal energy as a result of the neutrino energy transport. Hence, the shock revival is assumed to happen if the neutrino heating timescale becomes shorter than the advection timescale, 
$
	\tau_{\rm heat} < \tau_{\rm adv}.
$
Each of the timescales is estimated based on scaling relations and fitting formulae obtained from realistic simulations. Shock propagation after the revival is treated differently depending on whether the shock is strong enough to blow off the post-shock material. In the earlier phase, the post-shock material is still bound and a fraction of the shocked material is assumed to accrete onto the central remnant, leading to the growth of the remnant mass as well as the additional energy injection. On the other hand, all the shocked material is ejected in the later phase, and accordingly, the remnant mass becomes constant and the explosion energy is changed only by the explosive nucleosynthesis. The transition is assumed to take place when the post-shock velocity exceeds the local escape velocity, 
$
	v_{\rm post} > v_{\rm esc}.
$

An outline of the flow that determines fate is as follows. First, it is assumed that stars that do not experience shock revival fail to explode and eventually form BHs. Of those that experience shock revival, stars that are affected by significant matter accretion after shock revival are also assumed not to explode and form BHs. The judgment on this is based on the evolution of either the proto-neutron star (PNS) mass, diagnostic explosion energy, or redshift correction at the surface of the PNS. Eventually, stars that experience shock revival and are affected by minimal matter accretion are assumed to successfully explode and form NSs. BH formation can be accompanied by matter ejection if an accretion disk surrounding the central BH is formed \citep[e.g.,][]{Just22, Fujibayashi22} or if a large energy loss due to neutrinos occurs \citep{Lovegrove&Woosley13}.
However, the possibility of a successful explosion from a BH forming progenitor is not considered in our model as in M\"{u}ller's original work, given the huge uncertainties in the current theory.

We have carefully constructed the code and have confirmed that it yields largely consistent results with the original work. Hence, we believe that the conclusions regarding the fate presented in this work will be unaffected by the different implementations of this model. Nevertheless, some disagreements have remained. For the traceability, we provide our implementation of the model in the Appendix. 

The advantage of utilizing a semi-analytic model is the low cost of the computation, which enables us to compare the explosion properties for hundreds of progenitors \citep[e.g., ][]{Schneider21, Aguilera-Dena22}. 
On the other hand, care must be taken since the model relies on many approximated relations. Comparison with the most realistic, and thus the most computationally expensive, simulations \citep[e.g.,][]{Takiwaki16, Takiwaki21, Mueller17, Nagakura18, Vartanyan19, Bollig21} and with statistical properties derived from a number of multi-dimensional simulations \citep{Nakamura15, Summa16, Suwa16a, Pan16, Vartanyan18, Ott18, OConnor18, Burrows20} will be needed for verification. 
For this purpose, a discussion of whether the trends obtained in this model are consistent with previous studies is given in the Appendix.

\section{Characterizing the structure of core-collapse supernova progenitors}

In this section, we focus on quantities that characterize the structure of massive stars at the pre-collapse stage. In general, these quantities are divided into three categories; quantities related to the density structure, quantities related to the chemical structure, and quantities related to the thermal structure. The first category includes the compactness parameter $\xi_M$, and a quantity newly introduced in this work, $M_{\rm ff}$. Although we do not discuss the compactness parameter defined at different locations in detail in this work, correlations between them have been noted in the literature, e.g., \citet{Pejcha&Thompson15, Chieffi20}. The masses of bases of the chemically defined layers are involved in the second category. In particular, the mass of the iron core, which should be identical to the base mass of the silicon layer used in this work, is known to correlate with the compactness parameter. 
Furthermore, \citet{Schneider21} have reported that the masses of the C-free and Ne-free regions have a CO core mass dependence similar to that of the compactness and the iron core mass. \citet{Schneider21} have also reported qualitatively similar trends of the core entropy as a function of the CO core mass, which is included in the third category. The Ertl's parameters, $M_4$ and $\mu_4$, may also be included in this category since they utilize the entropy distribution. We will confirm that these correlations indeed arise in our models, and moreover, will show that they follow an identical monotonic sequence.

\subsection{The compactness parameter}

\begin{figure}[t]
 \centering
 \includegraphics[width=\hsize]{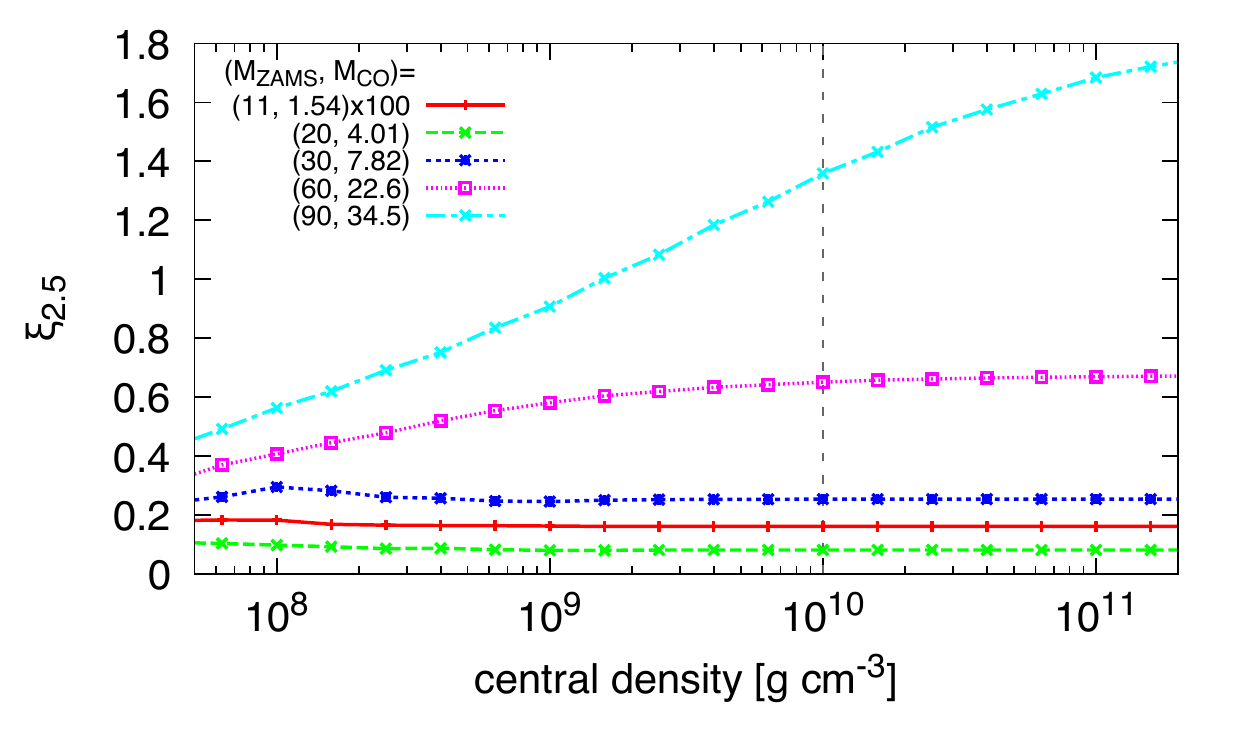}
 \caption{Evolution of \xx{2.5} as a function of central density. Plot is made for zero-metallicity models of $M_{\rm ZAMS} =$ 11, 20, 30, 60, and 90 $M_\odot$, and the ZAMS and the CO core masses are indicated by the legends. For the $M_{\rm ZAMS} =$ 11 $M_\odot$ model, \xx{2.5} multiplied by 100 is shown. 
The reason for multiplying by the large factor of 100 is that this model has a CO core mass smaller than 2.5 $M_\odot$ and the reference mass of \xx{2.5} is located in the inflated He layer.
The vertical dotted line is set at $\rho_\mathrm{c} = 10^{10}$ g cm$^{-3}$ to make the comparison with our fiducial He star models easier. }
 \label{plot-cevo1}
\end{figure}

As noted by \citet{OConnor&Ott11}, care must be taken for the timing of evaluation of the compactness parameter because the value can change as the star collapses. To avoid this uncertainty, \citet{OConnor&Ott11} uniformly evaluate the compactness parameter at the core bounce time. However, calculations until core bounce may be expensive for a stellar evolution simulation. Instead, it is beneficial if one has a criterion about the timing, after which the constancy of the compactness parameter is guaranteed.

Throughout this work, we set 2.5 $M_\odot$ as the reference mass coordinate to assign the compactness parameter for a given progenitor structure (\xx{2.5}). The evolution of \xx{2.5} as a function of central density is shown in Fig.~\ref{plot-cevo1}. For this purpose, a sequence of zero-metallicity stellar models is additionally calculated, and results of models with initial masses 11, 20, 30, 60, and 90 $M_\odot$ are plotted. Their ZAMS and the CO core masses are indicated in the figure as well. 
It shows that more compact models require a larger central density for the time evolution of \xx{2.5} to converge. Less compact models that have \xx{2.5} $\lesssim$ 1.0 when the central density reaches $10^{10}$\,g\,cm$^{-3}$ do not change their compactness values thereafter. Therefore, the \xx{2.5} measured when the central density reaches $10^{10}$\,g\,cm$^{-3}$ for models with \xx{2.5} $\lesssim$ 1.0 is expected to be maintained until core bounce. On the other hand, to measure converged values of compactness for more compact models characterized by \xx{2.5 }$\gtrsim$ 1.0, a central density greater than $10^{11}$\,g\,cm$^{-3}$ is needed\footnote{Here, we have implicitly assumed that the convergence of compactness with respect to time evolution depends monotonically on compactness. This assumption can be confirmed from Fig.~\ref{plot-cevo1} and can be inferred from the monotonicity of the cores found in this study.}.

\begin{figure}[t]
 \centering
 \includegraphics[width=\hsize]{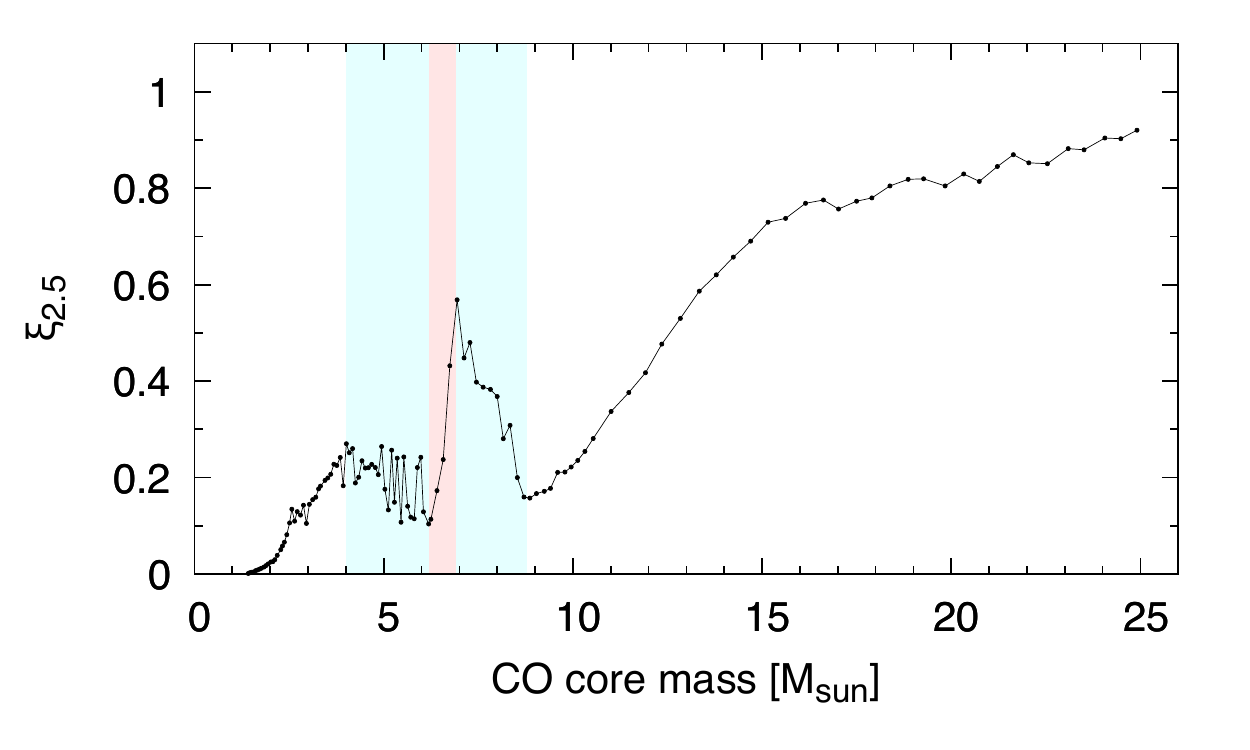}
 \caption{Distribution of \xx{2.5} evaluated at $\rho_c = 10^{10}$ \gcc{}. To emphasize the non-monotonicity, the decreasing trends in $M_\mathrm{CO} \in [4.0, 6.2] \ M_\odot$ and $M_\mathrm{CO} \in [6.9, 8.8] \ M_\odot$ and the increasing trend at $M_\mathrm{CO} \in [6.2, 6.9] \ M_\odot$ are overlaid with cyan and red bands.}
 \label{plot-MCO-xi25}
\end{figure}

For our fiducial He star models, Fig.~\ref{plot-MCO-xi25} shows the distribution of \xx{2.5} as a function of the CO core mass, which is evaluated at $\rho_c = 10^{10}$ \gcc{}. Note that the results for models with $M_{\rm CO} \leq 1.4 M_\odot$ ($M_{\rm He} \leq 2.7 M_\odot$) are not included here since evolution simulations for these models are halted before their central densities reach $10^{10}$ \gcc{}. 
Given the results of the examination of the compactness convergence presented above, the central density $\rho_c = 10^{10}$ \gcc{} is large enough to converge \xx{2.5} for these He star models since all of these models have \xx{2.5} $<$ 1.0, especially \xx{2.5} $\lesssim$ 0.7 for $M_{\rm CO} \leq 15 M_\odot$.

The feature of the \xx{2.5} distribution will be summarized as follows: \xx{2.5} monotonically increases with the mass in the less massive end with $M_{\rm CO} < 2.6 M_\odot$. Except for some offsets, the monotonic behavior is kept until $M_{\rm CO} < 4.0 M_\odot$. The compactness follows an interesting decreasing trend with significant scatter and reaches a local minimum at $M_{\rm CO} = 6.2 M_\odot$. After showing a steep increase, it shows a prominent peak at $M_{\rm CO} = 6.9 M_\odot$. Then a second decreasing trend follows, after which the second local minimum exists at $M_{\rm CO} = 8.8 M_\odot$. At last, the compactness follows a monotonically increasing trend. These are, in particular, very consistent with the KU series in \citet{Sukhbold&Woosley14}, which is a CO core model series with a very metal-poor $10^{-4}$ $Z_\odot$ initial composition\footnote{More smooth \xx{2.5} distributions as a function of the CO core mass can be found in Fig.~21 and 22 of \citet{Limongi18}. This may be because they have applied wide initial mass spacing between models and have mixed models with different metallicities in the figures. Also, the higher carbon mass fraction in their models \citep{Chieffi20} could account for the difference.}.

The non-monotonic behavior has been described in detail by \citet{Sukhbold&Woosley14}. 
In accordance with their analysis, we have confirmed that the first decreasing trend in $M_{\rm CO} \in [4.0, 6.2] \ M_\odot$ is due to the decreasing widths of C burning convective regions (both the core and the shell convective regions), which is related to the transition of the convective to radiative nature of central C burning.
We have confirmed that the sudden increase in the range of $M_{\rm CO} \in [6.2, 6.9] \ M_\odot$ as well as the more gentle increase in $M_{\rm CO} \ge 8.8 M_\odot$ result from outward migration of the third or second C burning shells in these mass ranges, and the transitional inward migration of the second C burning explains the decreasing trend between them.

There are changes of inclination at $M_{\rm CO} = 2.6 M_\odot$ and $15.6 M_\odot$. We note that these changes are artificially introduced through the definition of the compactness parameter. Namely, the inclination in the compactness distribution is affected by the chemical composition of the location where it is estimated because layers of two different chemical compositions can have different density gradients. \xx{2.5} is evaluated at the enclosed mass of 2.5 $M_\odot$. While this location is inside the oxygen-carbon layer in the majority of cases, it is included in the helium layer for the less massive models with $M_{\rm CO} < 2.6 M_\odot$, and is included in the inner carbon-free layer for more massive models with $M_{\rm CO} > 15.6 M_\odot$. Therefore, the two changes of inclination should not mean qualitative changes in the density structure.

\subsection{The enclosed mass inside an iso-free-fall-time surface}

The compactness parameter may not be the unique indicator for characterizing the density structure of a progenitor model. To find another indicator, we utilize the free-fall time, which can be defined as a function of the mass coordinate as 
\begin{eqnarray}
	\tau_{\rm ff}(M) = \frac{\pi}{2\sqrt{2}} \sqrt{ \frac{R(M)^3}{GM} }.
\end{eqnarray}
The free-fall time is in proportion to the inverse of the average density, hence, it becomes a monotonically increasing function as long as the density distribution is monotonically decreasing. We define the mass coordinate at which the free-fall timescale exceeds a provided time reference as \Mff{}. In this work, a reference time of 1 s is used\footnote{We do not have a strong motivation to decide the reference time of 1 s, though it is perhaps closer to the timescale of CCSN explosion than 0.1 or 10 s. In fact, the results shown later are insensitive to the choice, and the fact that there is no specific way to determine the reference time at least up to the zeroth order is an important property that supports the monotonicity of the core discussed later.}. Owing to the monotonicity of the free-fall time, a unique solution is found for \Mff{}.

\begin{figure}[t]
 \centering
 \includegraphics[width=\hsize]{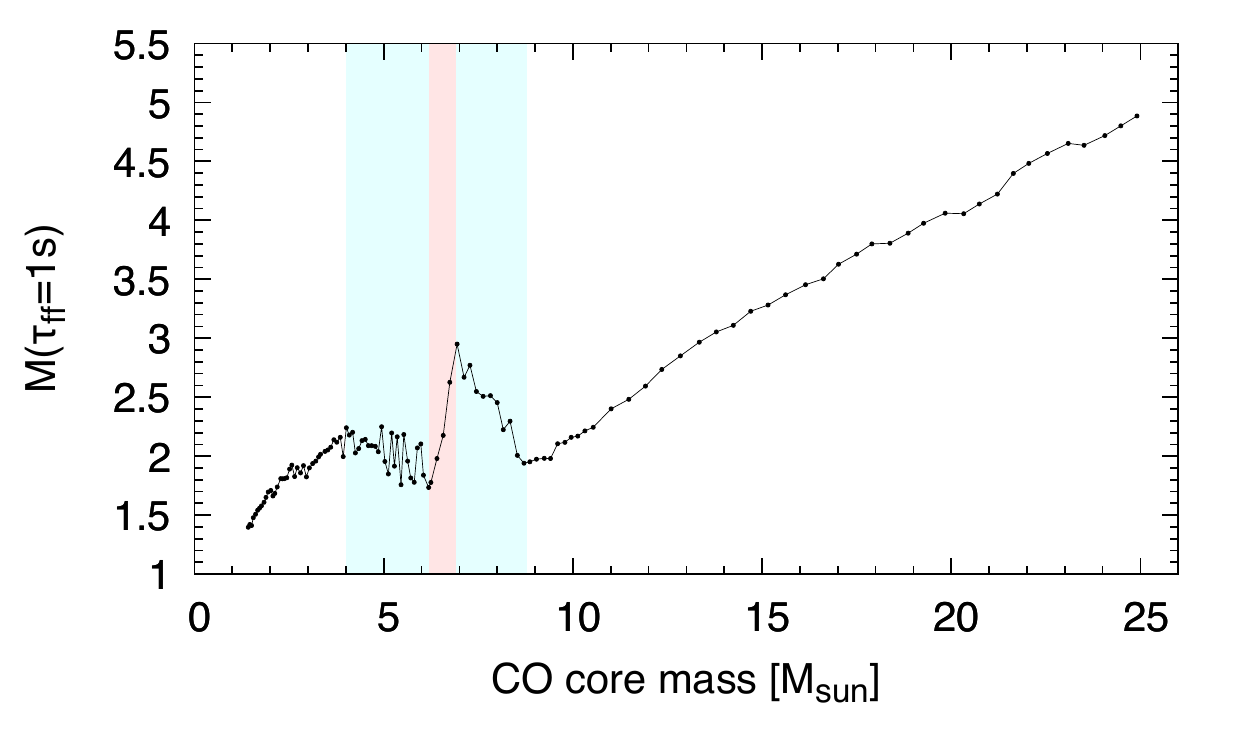}
 \caption{Distribution of \Mff{} evaluated at $\rho_c = 10^{10}$ \gcc{}. The cyan and red bands are the same as those in Fig.~\ref{plot-MCO-xi25}.
 }
 \label{plot-MCO-Mtau}
\end{figure}

Figure \ref{plot-MCO-Mtau} shows the distribution of \Mff{} evaluated when $\rho_c = 10^{10}$ \gcc{}. The similarity to the compactness distribution is apparent. For instance, both \xx{2.5} and \Mff{} follow decreasing trends for $M_\mathrm{CO} \in [4.0, 6.2] \ M_\odot$ and $M_\mathrm{CO} \in [6.9, 8.8] \ M_\odot$, which are overlaid with the cyan bands as well as the sudden increasing trends between them overlaid with the red band. The accurate correspondence may not be so surprising because both quantities are merely determined by ratios between (powers of) the mass coordinate and the radius.

\begin{figure}[t]
 \centering
 \includegraphics[width=\hsize]{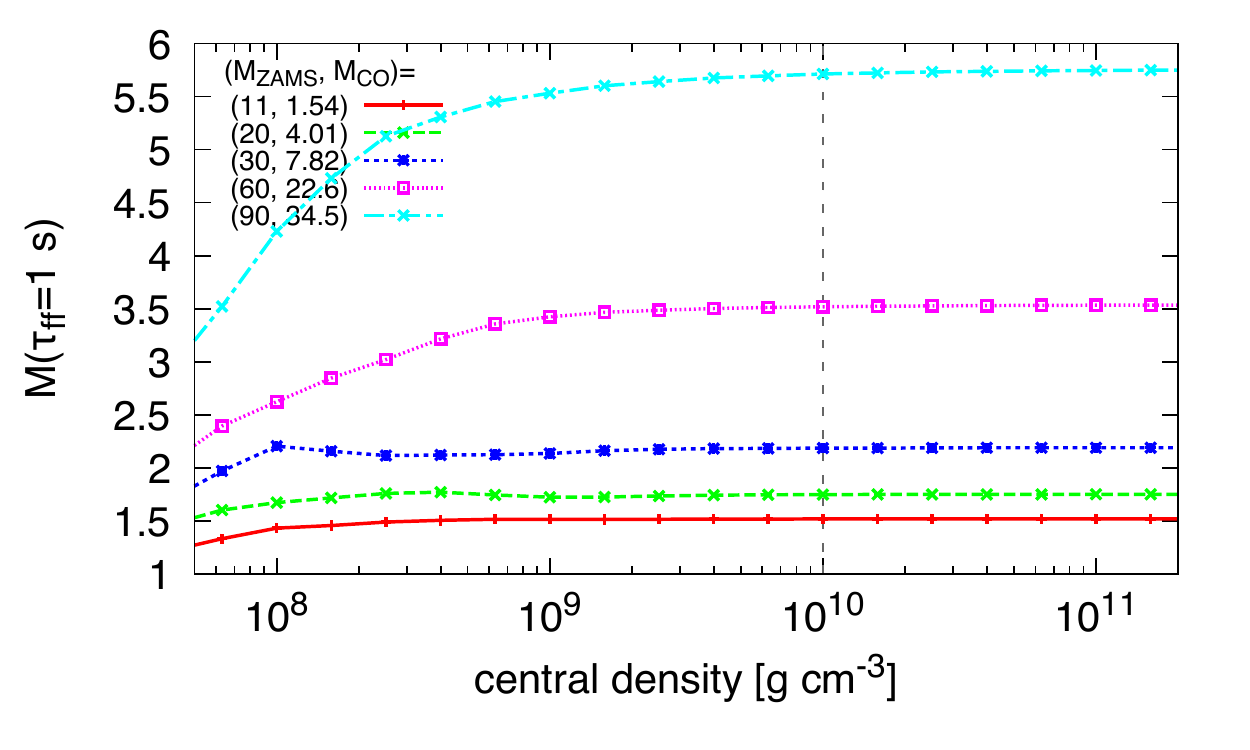}
 \caption{Same as Fig.\ref{plot-cevo1}, but for \Mff{}.}
 \label{plot-cevo2}
\end{figure}

Nevertheless, there are several benefits of utilizing \Mff{} instead of \xx{2.5}. Firstly, the value of \Mff{} is more intuitive since it provides a rough estimate of the averaged mass accretion rate during the formation of the PNS. At the same time, this value can be regarded as an estimate of the remnant mass, assuming that shock revival occurs in about 1 s after core collapse and that subsequent mass accretion is negligible. Secondly, \Mff{} is applicable to a less massive progenitor model in which the reference mass coordinate ($M$ of $\xi_M$) is outside the CO core. In such a case, the compactness parameter is evaluated to be extremely small, while \Mff{} is basically set to be the CO core mass. Similarly, the monotonically increasing trend in more massive models of $M_{\rm CO} > 15.6 M_\odot$ is now linearly followed. Hence, fewer artifacts are included in the \Mff{} distribution. Thirdly, \Mff{} has a better convergence on the evolutionary phases than \xx{2.5}. The evolution of \Mff{} as a function of the central density is shown in Fig.~\ref{plot-cevo2} for the same zero-metallicity models as in Fig.~\ref{plot-cevo1}, which shows better convergence of \Mff{} for the most massive $M_{\rm CO} = 34.5 M_\odot$ model than \xx{2.5}.

\subsection{The mass coordinates of bases of chemically defined layers}

\begin{figure}[t]
 \centering
 \includegraphics[width=\hsize]{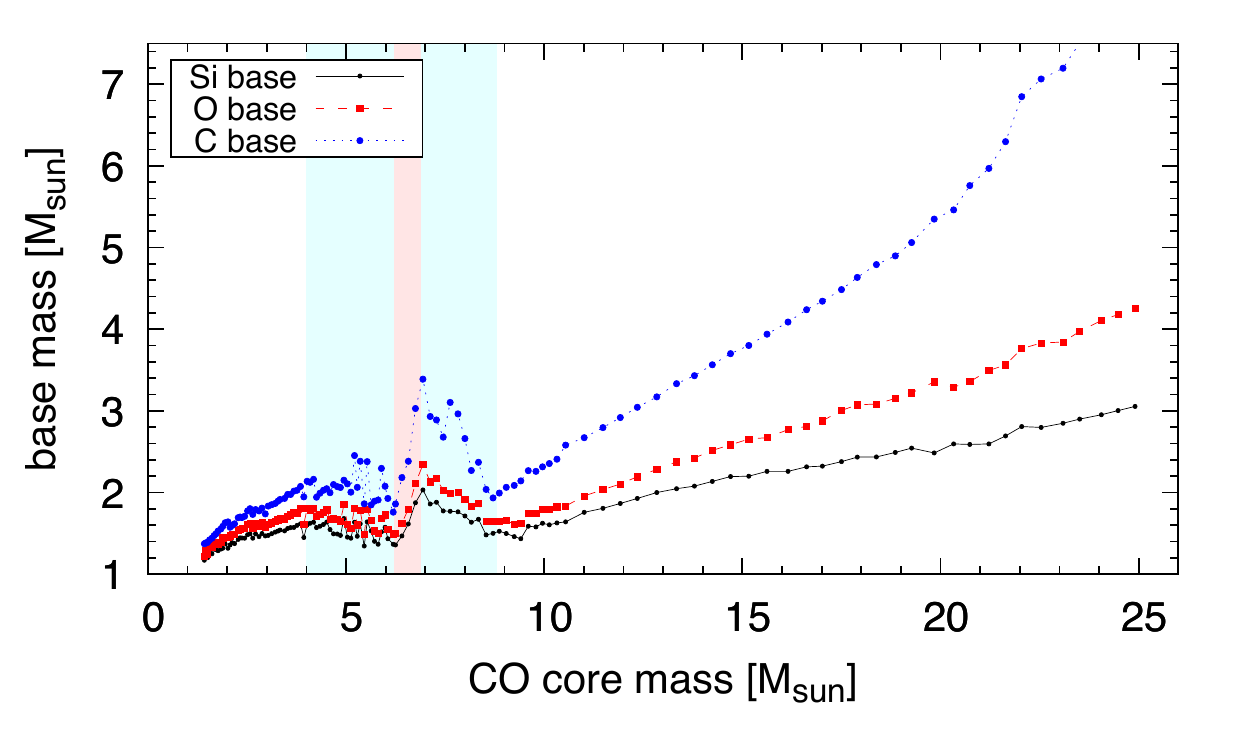}
 \caption{Distribution of mass coordinates at the base of the silicon layer (\Msib{}, black solid), the oxygen layer (\Mob{}, red dashed), and the carbon-rich layer (\Mcb{}, blue dotted) evaluated at $\rho_c = 10^{10}$\,g\,cm$^{-3}$. The cyan and red bands are the same as those in Fig.~\ref{plot-MCO-xi25}.
 }
 \label{plot-MCO-Mbase}
\end{figure}

A massive star is considered to form an onion-like chemical structure, in which layers composed of heavier elements are located closer to the stellar center. In addition to the density distribution, the chemical distribution is also fundamental to the progenitor structure. In order to characterize the chemical distribution, we define base masses, that is, the mass coordinates of bases of chemically defined layers. For example, the silicon base mass, \Msib{}, is defined as the innermost mass coordinate where the mass fraction of $^{28}$Si firstly exceeds $X(^{28}\rm{Si})$ = $10^{-2}$. Similarly, the oxygen base mass (\Mob{}) and the carbon base mass (\Mcb{}) are defined by the conditions of $X(^{16}\rm{O})$ = $10^{-3}$ and $X(^{12}\rm{C})$ = $10^{-3}$. Although the reference mass fractions are set arbitrarily, the base masses defined here will correspond to the traditional core masses such as the Fe and Si core masses.

The relation between the base masses and $M_\mathrm{CO}$ is shown in Fig.~\ref{plot-MCO-Mbase}. Similar to the \Mff{} distribution, the distributions of the three base masses clearly share the basic features of non-monotonicity obtained for the \xx{2.5} distribution. The correlation of \Msib{} is comparable to the correlation between the compactness parameter and the iron core mass shown in \citet{OConnor&Ott11}. In addition, we find that \Mob{} and \Mcb{}, which are defined at outer layers than \Msib{}, also show strong correlations with \xx{2.5} and \Mff{}.

\begin{figure}[t]
 \centering
 \includegraphics[width=\hsize]{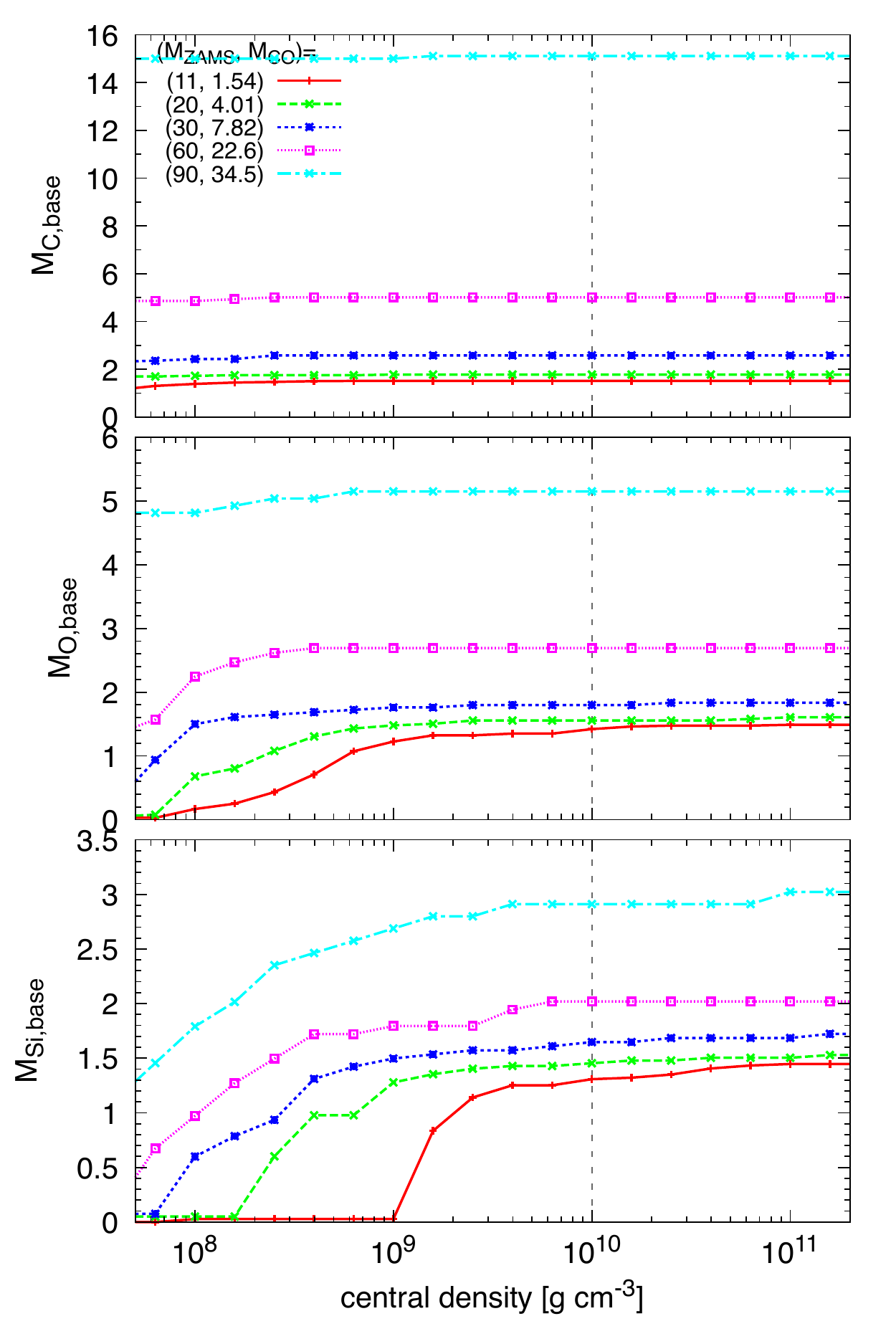}
 \caption{Same as Fig.\ref{plot-cevo1}, but for \Mcb{} (top), \Mob{} (middle) and \Msib{} (bottom).}
 \label{plot-cevo-Mbase}
\end{figure}

The time evolution of \Mob{} and \Msib{} as a function of central density is shown in Fig.~\ref{plot-cevo-Mbase}.
\Mob{} is basically kept constant if $\rho_\mathrm{c} > 10^{9}$ \gcc. An exception is the model with \MCO{} $= 1.54 \ M_\odot$, but the change is $\sim 15$\% and it approaches convergence if $\rho_\mathrm{c} \gtrsim 10^{10}$ \gcc.
On the other hand, \Msib{} is farther from convergence since it is defined more inside than \Mob{}, where the temperature is higher and the nuclear reaction timescale is shorter. Conversely, we have confirmed that \Mcb{} becomes nearly constant at least for $\rho_\mathrm{c} > 10^{8}$ \gcc.

\subsection{Ertl's parameters}

\citet{Ertl+16} have analyzed the interaction between matter accretion and neutrino heating and have proposed a criterion to judge the explodability of CCSN progenitors, which can discriminate between explosion and non-explosion with high accuracy of only a few percent exceptions. They have defined two parameters, $M_4$ and $\mu_4$, which are related to both the density and entropy distributions, and have used $M_4$ and the product of the two, $M_4\mu_4$, for the criterion.
$M_4$ is defined as the mass coordinate at which the entropy per baryon firstly exceeds the reference value of 4 $\mathrm{k_B}$. For clarity, we express the parameter by \Msk{} hereafter in this work. The other parameter, $\mu_4$, is the normalized mass derivative, (dM/dr)/($M_\odot$/1000 km), evaluated at \Msk{}. Again, we express the parameter by \musk{}. In practice, \musk{} has been evaluated by numerical differentiation as \musk{} $= \Delta M/[r(M(s_k\!=\!4) + \Delta M)-r(M(s_k\!=\!4))]$ with the mass interval of $\Delta M = 0.3 M_\odot$ in the original work, and has not been directly related to the density at $M(s_k=4)$ that is implied by the formal definition.

\begin{figure}[t]
 \centering
 \includegraphics[width=\hsize]{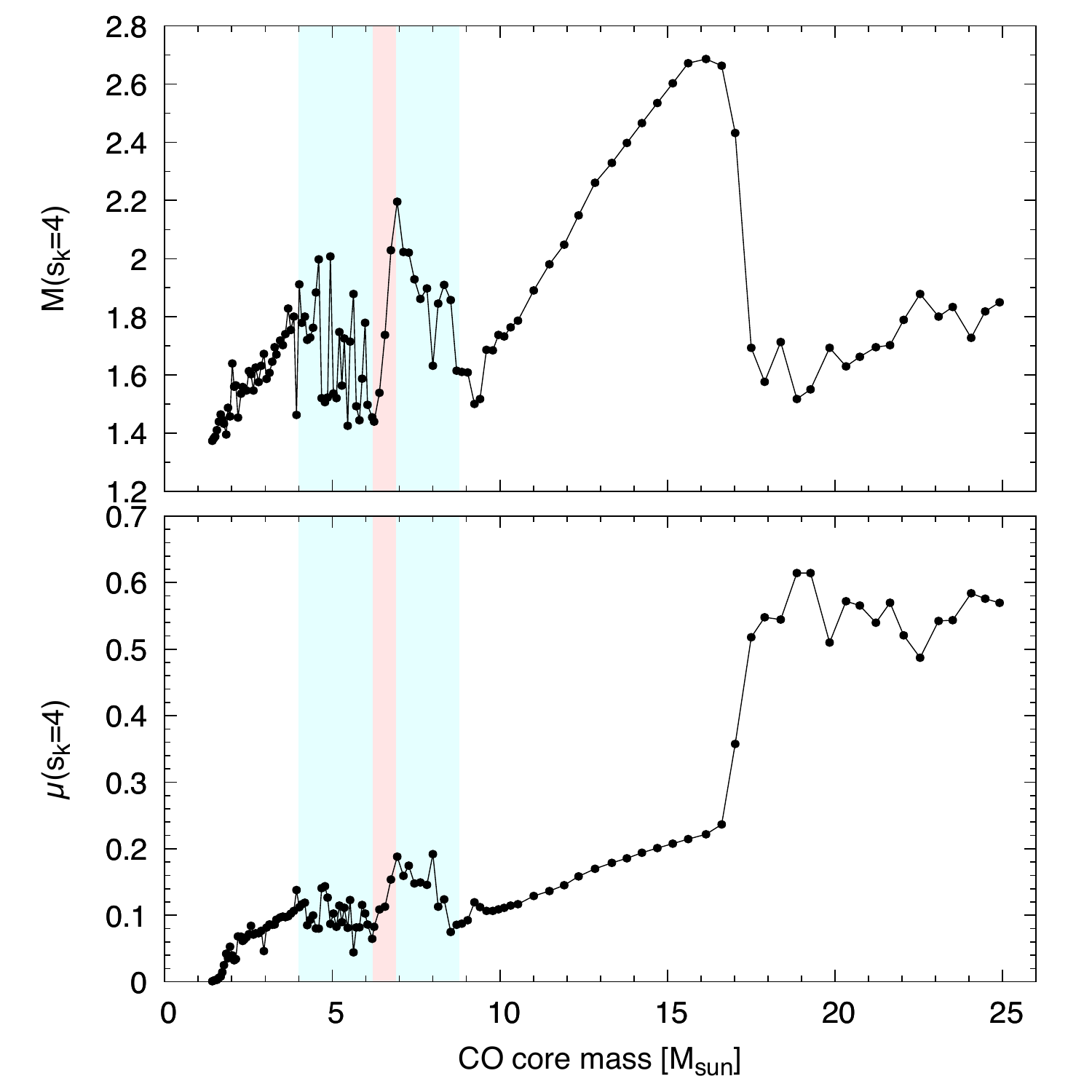}
 \caption{Distribution of \Msk{} (top) and \musk{} (bottom) evaluated at $\rho_c = 10^{10}$\,g\,cm$^{-3}$. The cyan and red bands are the same as those in Fig.~\ref{plot-MCO-xi25}.
 }
 \label{plot-MCO-Ertl}
\end{figure}

\begin{figure}[t]
 \centering
 \includegraphics[width=\hsize]{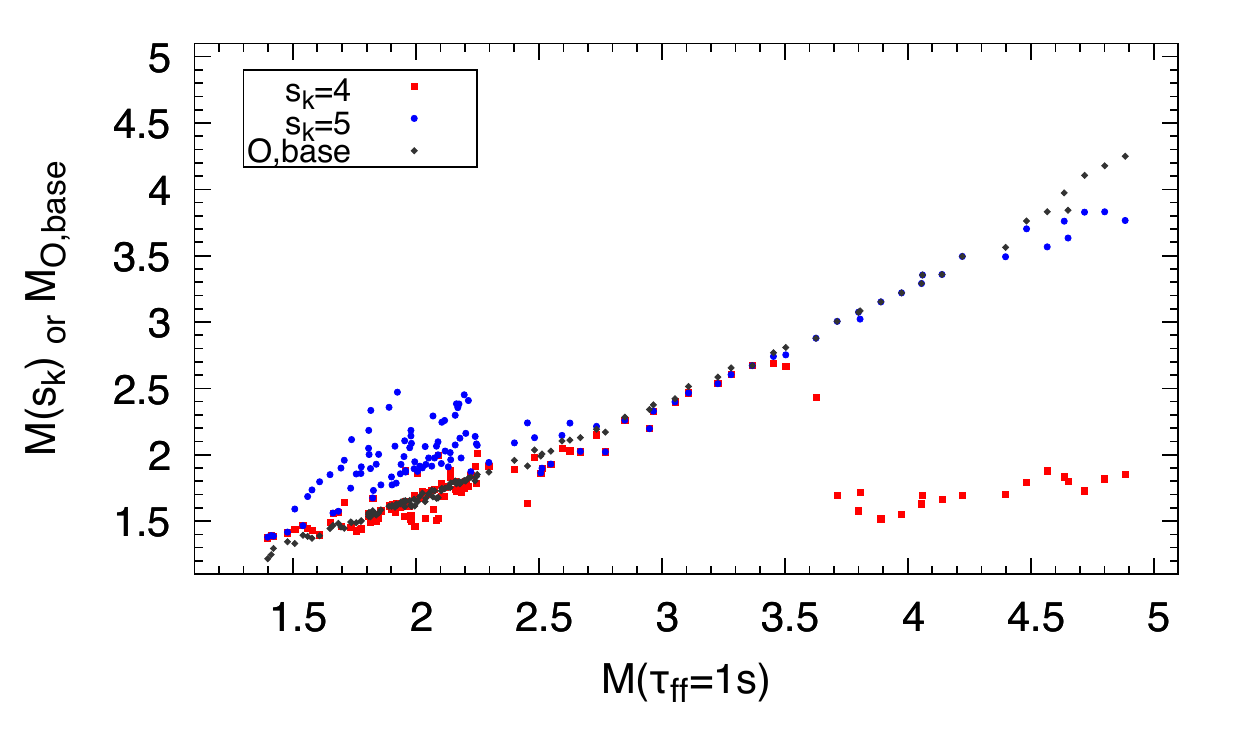}
 \caption{Distribution of \Msk{} (red, square), $M(s_k\!=\!5)$ (blue, circle), and \Mob{} (black, diamond) as a function of \Mff. Values are evaluated at $\rho_c = 10^{10}$\,g\,cm$^{-3}$. }
 \label{plot-Mff-M4}
\end{figure}

The distributions of \Msk{} and \musk{} as a function of \MCO{} are shown in Fig.~\ref{plot-MCO-Ertl}. Both parameters exhibit non-monotonic CO core mass dependencies, which are very similar to the \xx{2.5} and \Mff{} distributions in the less massive models characterized by \MCO{} $\lesssim 17 \ M_\odot$. This property originates from the fact that \Msk{} is almost identical to \Mob{} on the less compact side \Mff{} $\lesssim 3.5 \ M_\odot$ corresponding to \MCO{} $\lesssim 17 \ M_\odot$, as Fig.~\ref{plot-Mff-M4} plotting the distribution of \Msk{} and \Mob{} as a function of \Mff{} shows. This is because oxygen burning forms a strong entropy jump; in fact, \Msk{} has been used as an indicator to specify the O-burning shell \citep[cf.][]{Heger&Woosley10}. Hence, the correlation between \Msk{} and \Mff{} shown here is essentially identical to the one between \Mob{} and \Mff{}, which has been discussed earlier. 
Meanwhile, the \Msk{} and \musk{} distributions in models characterized by \MCO{} $\gtrsim 17 \ M_\odot$ show completely different trends. The distribution of $M(s_k\!=\!5)$ as a function of \Mff{} in Fig.~\ref{plot-Mff-M4} shows that, in models with \Mff{} $\gtrsim 3.5 \ M_\odot$ corresponding to the heavier models, the indicator tracing \Mob{} shifts from \Msk{} to $M(s_k\!=\!5)$ explaining the changes in the trends. The shift is due to the effect that the entropy of the entire core is larger for more compact models (see Section 3.5), and the entropy after the jump exceeds $s_k = 5$.

\begin{figure}[t]
 \centering
 \includegraphics[width=\hsize]{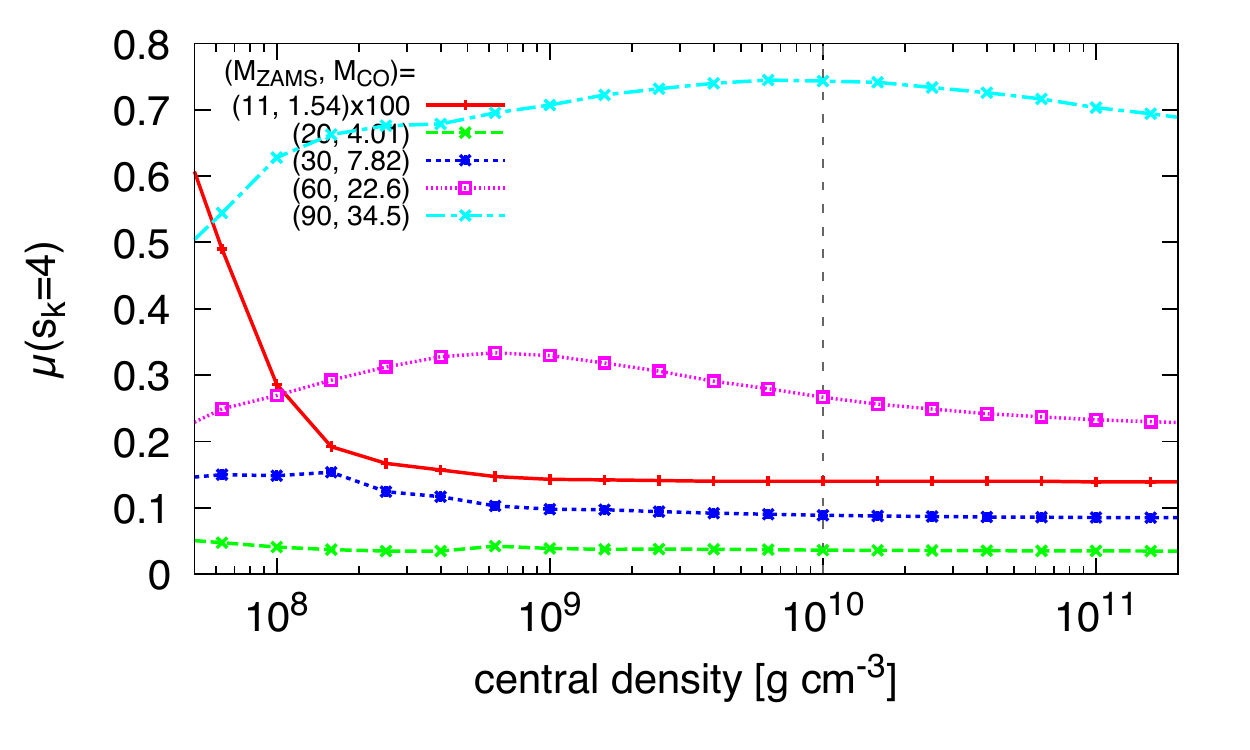}
 \caption{Same as Fig.\ref{plot-cevo1} but for \musk. For the $M_{\rm ZAMS} =$ 11 $M_\odot$ model, \musk{} multiplied by 100 is shown. The reason for multiplying by the large factor is that the finite $\Delta M = 0.3 M_\odot$ is used and its outer boundary was outside the CO core.}
 \label{plot-cevo8}
\end{figure}

Similar to Fig.~\ref{plot-cevo2}, the evolution of \musk{} in the later evolutionary phase is shown in Fig.~\ref{plot-cevo8}. It shows that the less massive models with \MCO{} $\lesssim 20 M_\odot$ and \musk{} $\lesssim 0.2$ keep \musk{} nearly constant in the later phase of $\rho_c > 10^9$ g cm$^{-3}$. Besides, we have confirmed that $M(s_k=4)$ stays constant independent of the CO core mass, which is consistent with the identity between \Msk{} and \Mob{}. Hence, as far as the identity between \Msk{} and \Mob{} is established, both \musk{} and \Msk{} stay constant during the collapsing phase. The figure also shows that \musk{} changes by $\sim$ 30\% for the more massive models. However, this result may not be relevant to us, because \musk{} of such massive models will have incompatible characteristics with that of the less massive models because it lacks the identity to \Mob{}. 

\subsection{The core entropy}

\begin{figure}[t]
 \centering
 \includegraphics[width=\hsize]{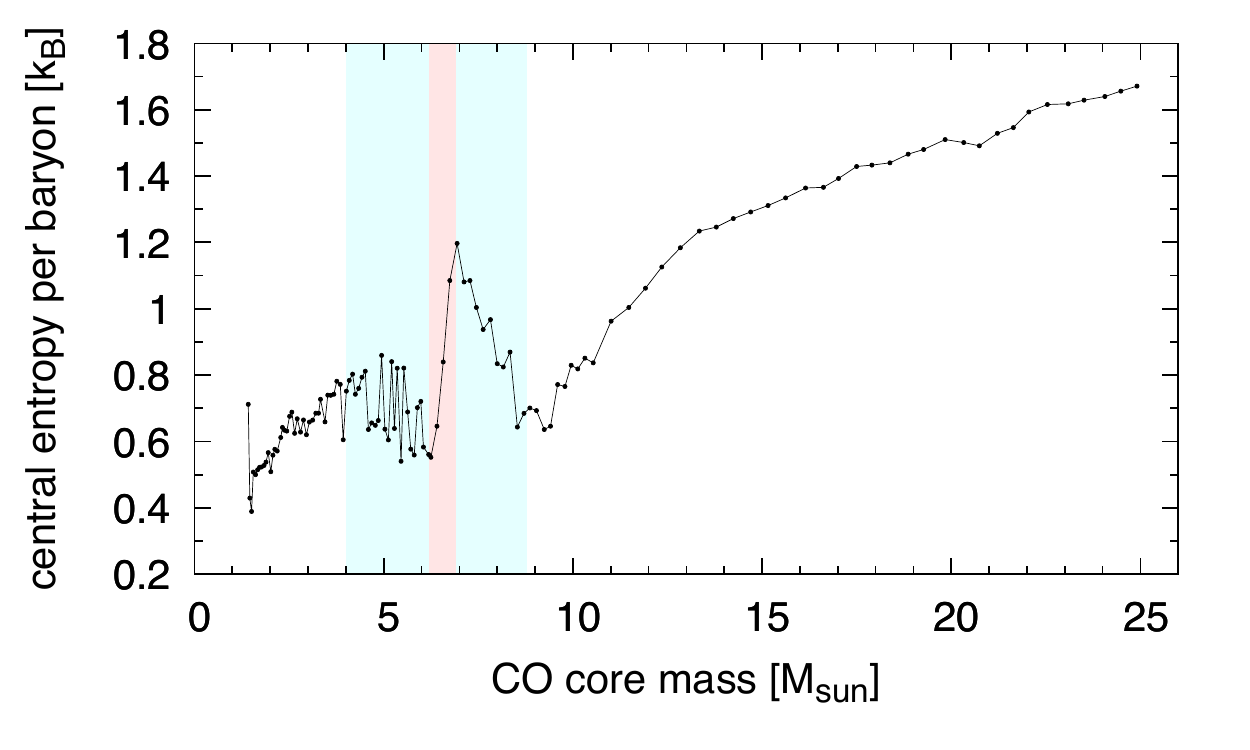}
 \caption{Distribution of the entropy per baryon at the center evaluated at $\rho_\mathrm{c} = 10^{10}$ \gcc. The cyan and red bands are the same as those in Fig.~\ref{plot-MCO-xi25}.}
 \label{plot-MCO-skcore}
\end{figure}

In Fig.~\ref{plot-MCO-skcore}, the entropy of the stellar center is shown as a function of CO core mass. Given the similar shapes of the distributions, this figure shows that the central entropy is strongly correlated with the base masses of the Si, O, and C layers as well as \Mff{}. It is noteworthy that \citet{Schneider21} attributes the correlation between the iron core mass and the central entropy to the property of quasi-isentropic iron cores. Our results are consistent with this understanding, but more so, imply that this correlation can be traced back to earlier evolutionary phases as the correlation extends not only to the base masses of the inner Si and O layers but also to the base mass of the C layer.

\begin{figure}[t]
 \centering
 \includegraphics[width=\hsize]{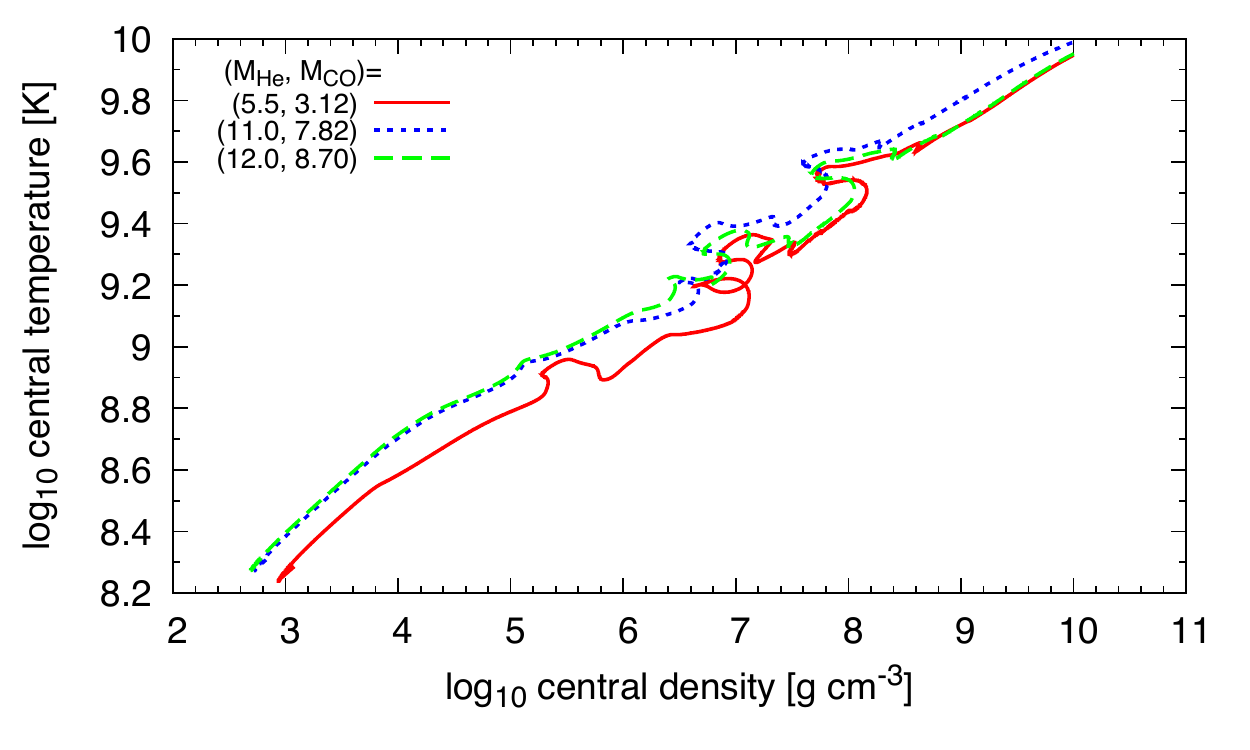}
 \caption{Central density and temperature evolution of the models of \MCO = 3.12 (red, solid), 7.82 (blue, dotted),  and 8.70 M$_\odot$ (green, dashed). In this density-temperature diagram, the entropy is smaller in the lower right region and larger in the upper left region.}
 \label{plot-comp-rhot}
\end{figure}

However, the correlation is not so trivial from the point of view of stellar evolution because the core entropy should initially correlate with the total mass of the star, and the total mass is not correlated well with \Mff{}. In order to develop the monotonic dependence on \Mff{}, the entropy order must reverse in some models during the stellar evolution. 
The example of the reversal is illustrated by the evolution in the central density and temperature plane shown in Fig.~\ref{plot-comp-rhot}, in which results of models with $M_{\rm CO} =$ 3.12, 7.82,  and 8.70 M$_\odot$ are compared. They have \Mff{} = 1.94, 2.51, and 1.94 $M_\odot$ and the mass coordinates of the last shell C burning, \Mcb, of 1.85, 2.96, and 1.93 $M_\odot$ for models with \MCO = 3.12, 7.82, and 8.70 $M_\odot$, respectively.
In the beginning, the entropy order coincides with the mass order as expected. The lowest mass model follows the track of the lowest entropy, in which several bumpy features (e.g., a bump at $T_c \sim 10^{8.8}$ K, a loop at $T_c \sim 10^{9.2}$ K, bumps at $T_c \sim 10^{9.3}$ and $\sim 10^{9.6}$ K, respectively due to the C, Ne, O, and Si burnings) appear due to the relatively high electron degeneracy. 
The higher mass models initially follow higher entropy tracks, which have fewer bumpy features. However, the entropy order reverses after the central Ne burning ($T_c \sim 10^{9.2}$ K). After this phase, the most massive model with \MCO = 8.70 M$_\odot$ eventually follows a converging evolution on the track of the lowest mass model with \MCO = 3.12 M$_\odot$. 

Based on the rough concurrence of the start of the last C burning with the start of the reversal of the entropy order, we speculate that the late evolution after the central Ne burning can be described as an evolution of a core that has an effective core mass given by the base mass of the last C burning shell. The reversal taking place in the model with \MCO = 8.70 M$_\odot$ would be understood as a relaxation process, in which a core having a high initial entropy eventually cools to adopt a lower entropy that is required to contract with a given small effective mass.

\begin{figure}[t]
 \centering
 \includegraphics[width=\hsize]{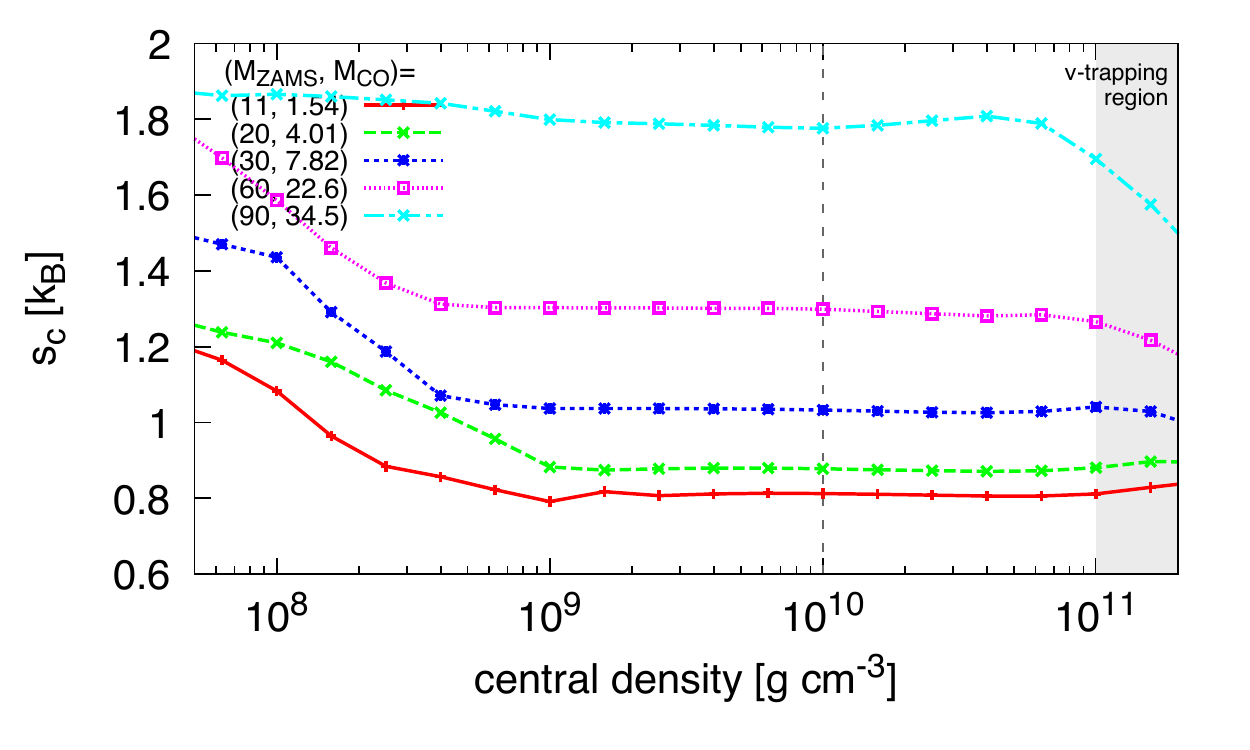}
 \caption{Same as Fig.\ref{plot-cevo1} but for the central entropy per baryon. Note that the later result of $\rho_\mathrm{c} \gtrsim 10^{11}$ \gcc{} is unreliable because the neutrino trapping is not taken into account in this work.}
 \label{plot-cevo6}
\end{figure}

The late-time evolution of the central entropy is plotted in Fig.~\ref{plot-cevo6}. It shows that the central entropy becomes nearly constant for $\rho_\mathrm{c} \gtrsim 10^{9}$ \gcc{} irrespective of \MCO. This result indicates that the rate of change of $Y_e$ and the accompanied neutrino emission do not significantly affect the entropy in the central NSE region during the early collapsing phase. 
We note that the entropy change after $\rho_\mathrm{c} \gtrsim 10^{11}$ \gcc{} shown in the figure is unreliable. This is because our stellar evolution code does not treat neutrino trapping, which will take place in such a high-density region. Both the $Y_e$ evolution and neutrino emission will be overestimated in the high-density regions. The entropy evolution will be less substantial in the later collapsing phase if the neutrino processes are properly treated.

\subsection{Convergent internal structures of progenitors with similar \Mff{}}

\begin{figure}[t]
 \centering
 \includegraphics[width=\hsize]{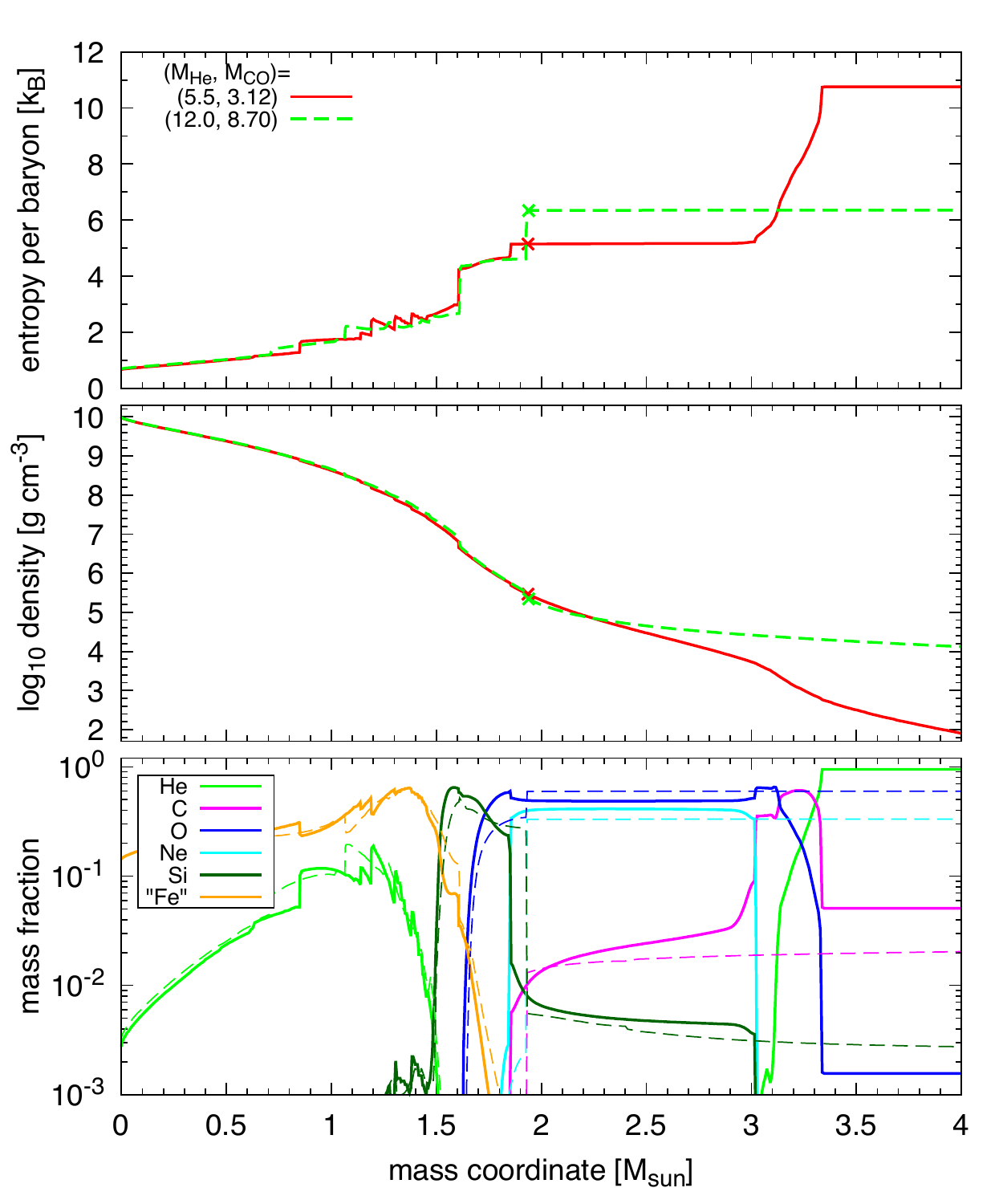}
 \caption{Distributions of the entropy per baryon (top), the density (middle), and the mass fractions of chemical species (bottom) as a function of the mass coordinate are compared for models of \MCO = 3.12 (red, solid) and 8.70 M$_\odot$ (green, dashed). Crosses in the entropy and density plots indicate the locations of \Mff{} for each model (\Mff = 1.93 and 1.94 $M_\odot$ for \MCO = 3.12 and 8.70 M$_\odot$ models, respectively).}
 \label{plot-comp-str}
\end{figure}

So far, we have investigated the behavior of parameters that characterize the progenitor structure such as \xx{2.5} and \Mff{}. Although interesting correlations among these parameters have been found, relevance to the global structure is not clear. In order to link these parameters to the global progenitor structure, distributions of the entropy, density, and chemical elements are plotted in Fig.~\ref{plot-comp-str}, in which the two models with \MCO = 3.12 and 8.70 M$_\odot$, the converging evolutions of which have been discussed in the previous section, are compared.

In spite of the difference in the CO core masses, these models show a striking resemblance of the entropy distributions, in particular for the inner regions of $\sim 1.9 \ M_\odot$, which is shown in the top panel. The most important feature is the coincidence of the significant jumps at $\sim 1.6 \ M_\odot$, which trace the strong heating of the shell O burning. Not only the locations but also the level differences are similar. The entropy structures inside the jump are also similar, though less significant saw-shaped bumps, which are remnants from previous shell Si burning phases, are involved. Mass coordinates of the last shell C burnings that are indicated by more or less significant jumps outside the O burning bases are close as well. Meanwhile, the outer distributions of shell C burning are totally different. The entropy of the convective C burning layers are $\sim 5.1$ and $\sim 6.2$ respectively for the less and more massive models. This order originates from the entropy order of the CO cores. Furthermore, while the plot includes the high entropy helium envelope surrounding the CO core of $3.12 \ M_\odot$ for the less massive model, the corresponding structure is out of the plot for the more massive model as it has a more extended CO core of $8.70 \ M_\odot$.

The density distributions shown in the middle panel also validate the close resemblance of the progenitor's internal structures. As we have selected the progenitor models having the same central density of $10^{10}$ g cm$^{-3}$, they show nearly perfect agreement up to $\sim 1.9 M_\odot$, where carbon layers begin. Inside the carbon layer, density jumps locate at $\sim 1.6 M_\odot$ in both models, which, of course, coincide with the entropy jump due to the shell O burning. On the other hand, density distributions outside the carbon layers are totally different, as they must connect with the helium layers at different locations.

Crosses in the top and middle panels indicate the locations of \Mff{} for both models. By chance, the locations roughly correspond to the bases of the shell C layers at $\sim 1.9 M_\odot$. The \Mff{} depends on the mass and radius distributions, hence only on the density distribution. Therefore, by having the nearly same density structures of inner $\sim 1.9 M_\odot$ regions, these two models give the same \Mff{} for reference times shorter than 1 s. In turn, these models are also expected to have nearly the same mass accretion histories up to $\sim 1$ s after core collapse.

Consistent with the similarities confirmed for the entropy and density distributions, the chemical distributions shown in the bottom panel are also similar for these two models. Accordingly, the models have similar base masses of (\Msib/$M_\odot$, \Mob/$M_\odot$, \Mcb/$M_\odot$) = (1.49, 1.63, 1.85) and (1.50, 1.64, 1.93), respectively. 

\subsection{One parameter characterization}

\begin{figure*}[t]
 \centering
 \includegraphics[width=\hsize]{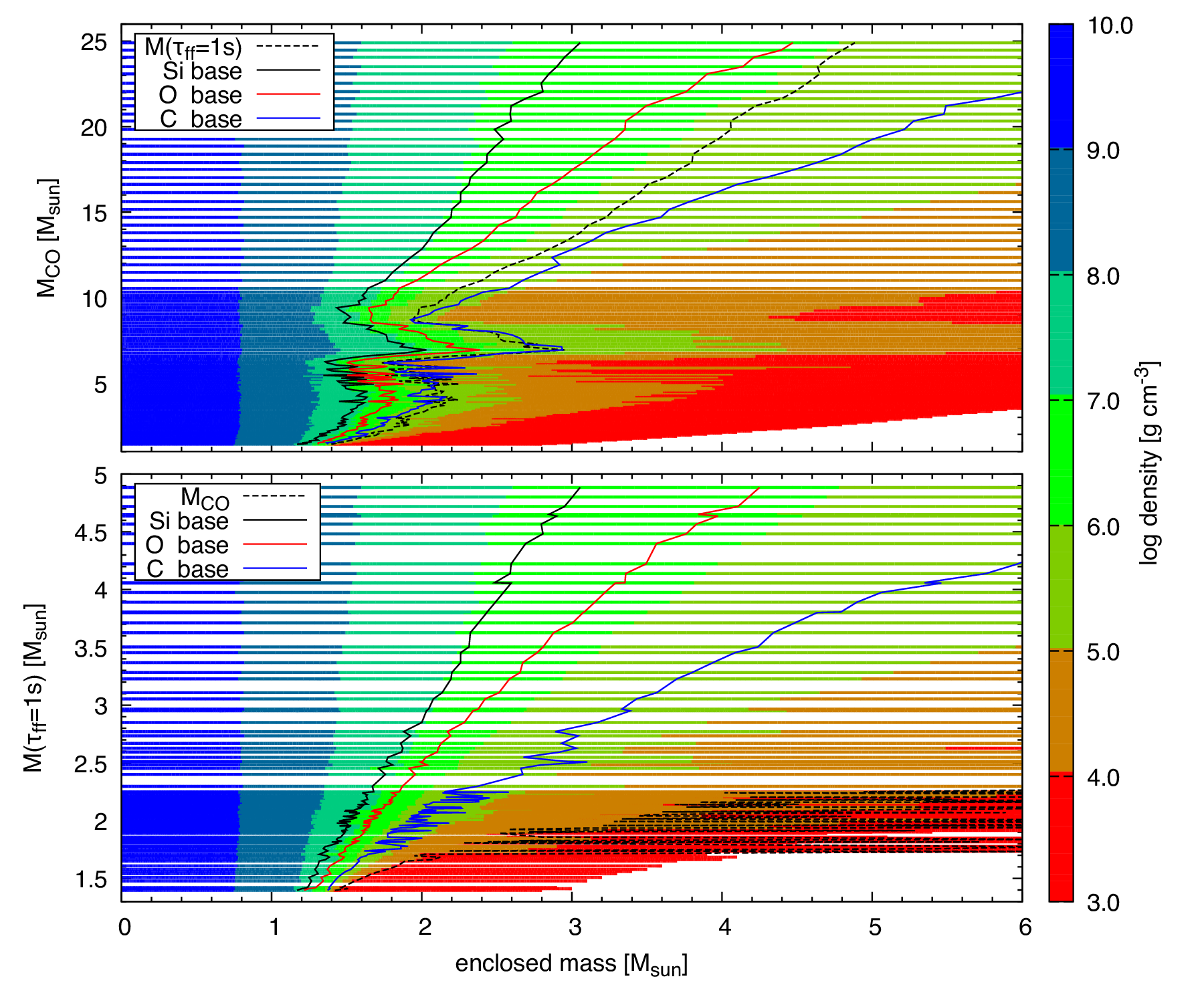}
 \caption{Density distributions of all models with a different order, with the \MCO{} order (top panel) and with the \Mff{} order (bottom panel). Each line shows the density distribution of the inner $M \leq 6 \ M_\odot$ region of one model using the color coordinate, the definition of which is indicated by the right color box. In the top panel, mass coordinates of \Mff{}, \Msib{}, \Mob{}, and \Mcb{} are additionally plotted by black dashed, black solid, red solid, and blue solid lines, respectively. Instead of \Mff{}, \MCO{} is plotted by the black-dashed line in the bottom panel.
 }
 \label{plot-sort-dens}
\end{figure*}

The similarities discussed in the above section imply that it is possible to represent the global core structure based on the parameter \Mff{}, which is calculated from only partial information about the core structure. To further investigate this implication, density distributions of all models are projected by using a color coordinate in Fig.~\ref{plot-sort-dens}. 
Models are sorted according to \MCO{} in the top panel, and each horizontal line shows the density distribution of one model. The location of \Mff{} is shown by the black dashed line, and locations of base masses of \Msib{}, \Mob{}, and \Mcb{} are overplotted by the black, red, and blue solid lines. 
Because the central densities of our models are adjusted to be $\rho_\mathrm{c} = 10^{10}$ \gcc, inner density structures of $\rho \gtrsim 10^{9}$ \gcc{} are nearly identical. On the other hand, density structures marked by the color boundaries of $\rho =$ $10^{8}$, $10^{7}$, and $10^{6}$ \gcc{} show similar \MCO{} dependencies to \xx{2.5} and \Mff.
The models are sorted according to the \Mff{} order in the bottom panel. This plot demonstrates that, by sorting the models with \Mff{}, the density structure approximately inside \Mcb{} can be sorted into a highly monotonic sequence for the wide range of \MCO{}. 

\begin{figure}[t]
 \centering
 \includegraphics[width=\hsize]{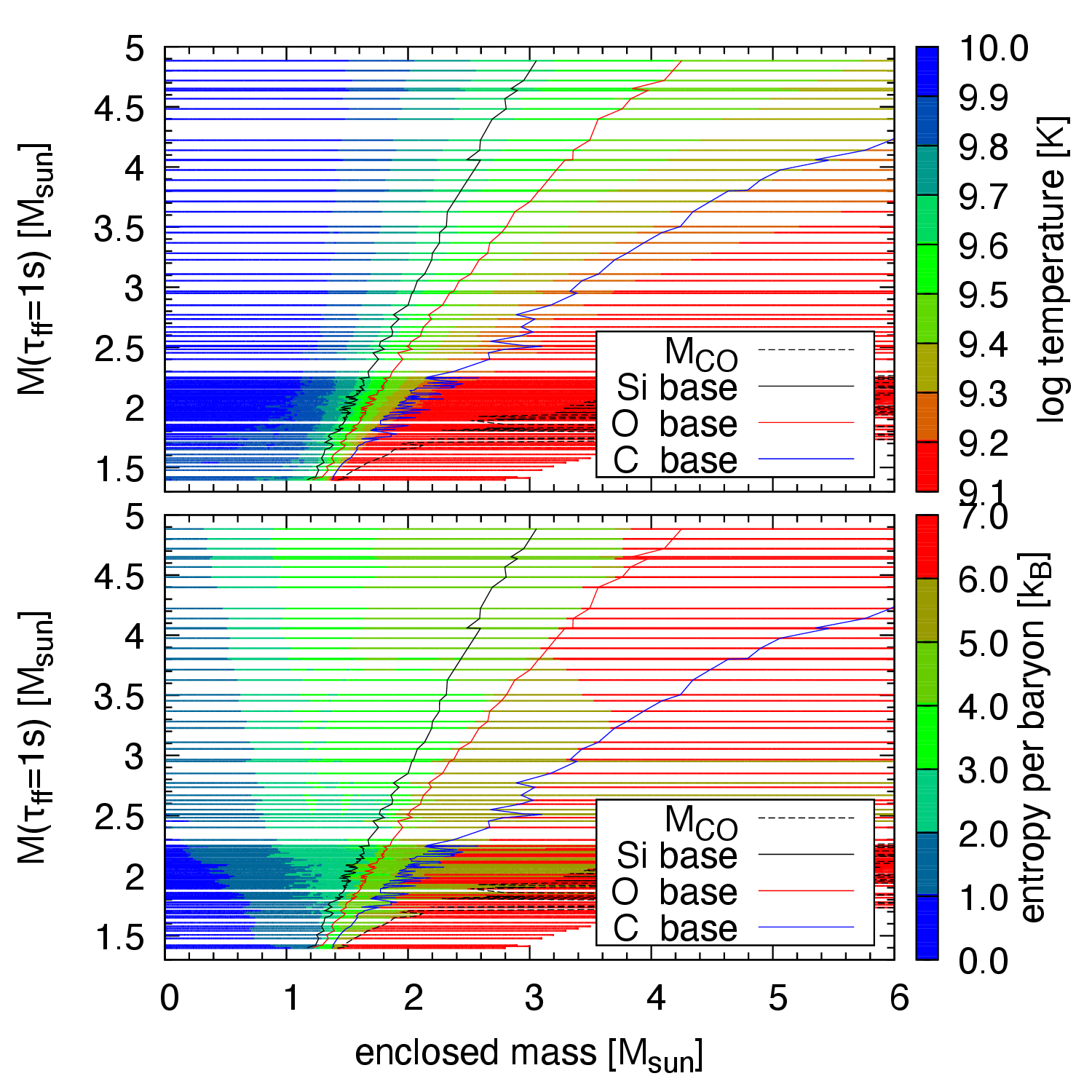}
 \caption{The same as the bottom panel of Fig.~\ref{plot-sort-dens}, but for the temperature distributions (top panel) and the entropy distributions (bottom panel).}
 \label{plot-sort-therm}
 \end{figure}

In addition to the density structure, the thermal structure follows a monotonic sequence once the models are sorted with \Mff{}. This is shown by Fig.~\ref{plot-sort-therm}, in which the temperature distributions ordered by \Mff{} are shown in the top panel, and the entropy distributions are in the bottom.
The high monotonicity of the temperature distributions will explain the strong correlations between \Mff{} and the chemically defined base masses. This is because the temperature is the chief determinant of nuclear reaction rates so the locations of the elemental bases are well traced by the contour of constant temperature, such as \Msib{} by $T \sim 10^{9.6}$ K, \Mob{} by $T \sim 10^{9.5}$ K, and \Mcb{} by $T \sim 10^{9.3}$ K. 
Although the monotonicity in the entropy distributions is much more complicated than in the density and temperature distributions, the figure shows that the central entropy follows the \Mff{} order as discussed in the previous section. Also, note that the coincidence of \Mob{} and the color boundary of the entropy of $s_k = 4.0$ for models with \Mff{} $\lesssim 3.5 M_\odot$ or $s_k = 5.0$ for models with \Mff{} $\gtrsim 2.5 M_\odot$ indicates that the location of the most significant entropy jump in a collapsing star also correlates well with \Mff{} as discussed above.

From this result, we further deduce that the inner structure of a collapsing star can be identified as first order by specifying only a single parameter, even though the late-time stellar evolution, especially the convective evolution, of CCSN progenitors is quite complicated. The density-dependent parameter \Mff{}, or equivalently \xx{2.5}, is applicable for the sorting. Besides, provided the strong correlations, other parameters such as base masses of \Msib{}, \Mob{}, and \Mcb{} or the central entropy are also plausible.

\section{Correlations with observables}
So far, we have discussed correlations between quantities that characterize the core structure at the onset of core collapse. Although these correlations are fundamental for improving our understanding of massive star structure, they are difficult to confirm directly from observations because the core is hidden deep inside the star. Therefore, correlations involving observable quantities are equally interesting. In this section, we intend to find such correlations from quantities that could be observed, or at least constrained, from current and future observations.

Firstly, we analyze the remaining lifetimes from particular evolutionary phases till core collapse. Secondly, the surface quantities of the radii and the luminosities of the models, which would be comparable to those of envelope-stripped stars in the real universe, are discussed. Finally, properties of the CCSN including the PNS mass, the explosion energy, and the explodability are highlighted because of their particularly high accessibility through transient surveys.

\subsection{Remaining time till core-collapse}

Once an evolutionary phase is properly defined, the remaining time from that phase to core collapse can be evaluated based on a stellar evolution calculation. In this work, we define evolutionary phases mainly based on chemical composition. The location of the maximum temperature is first taken as a reference position for one time-snapshot (most of the time, it is at the stellar center). The evolutionary phase, $iphase$, is set according to the chemical composition at the reference position as
\begin{eqnarray}
iphase = \begin{cases*}
        2, & if        $X_{^{4}\mathrm{He}} > 2\%$,\\
        3, & else if $X_{^{12}\mathrm{C}} > 2\%$,\\
        4, & else if $X_{^{20}\mathrm{Ne}} > 2\%$,\\
        5, & else if $X_{^{16}\mathrm{O}} > 2\%$,\\
        6, & else if $X_{^{28}\mathrm{Si}} > 2\%$,\\
        7, & otherwise,
        \end{cases*}
\end{eqnarray}
and the initial value of $iphase = 1$ is set for the start of the simulation, the He ZAMS phase.
For each $iphase$, the reference element is defined as
\begin{eqnarray}
elem(iphase) = \begin{cases*}
        ^{4}\mathrm{He}, & ($iphase = 2$),\\
        ^{12}\mathrm{C}, & ($iphase = 3$),\\
        ^{20}\mathrm{Ne},&($iphase = 4$),\\
        ^{16}\mathrm{O}, & ($iphase = 5$),\\
        ^{28}\mathrm{Si}, & ($iphase = 6$),
        \end{cases*}
\end{eqnarray}
and furthermore, the mass fraction of the reference element,
$X_\mathrm{ref} = $(mass fraction of $elem(iphase)$ at the reference position),
is defined.

\begin{figure*}[t]
 \centering
 \includegraphics[width=\hsize]{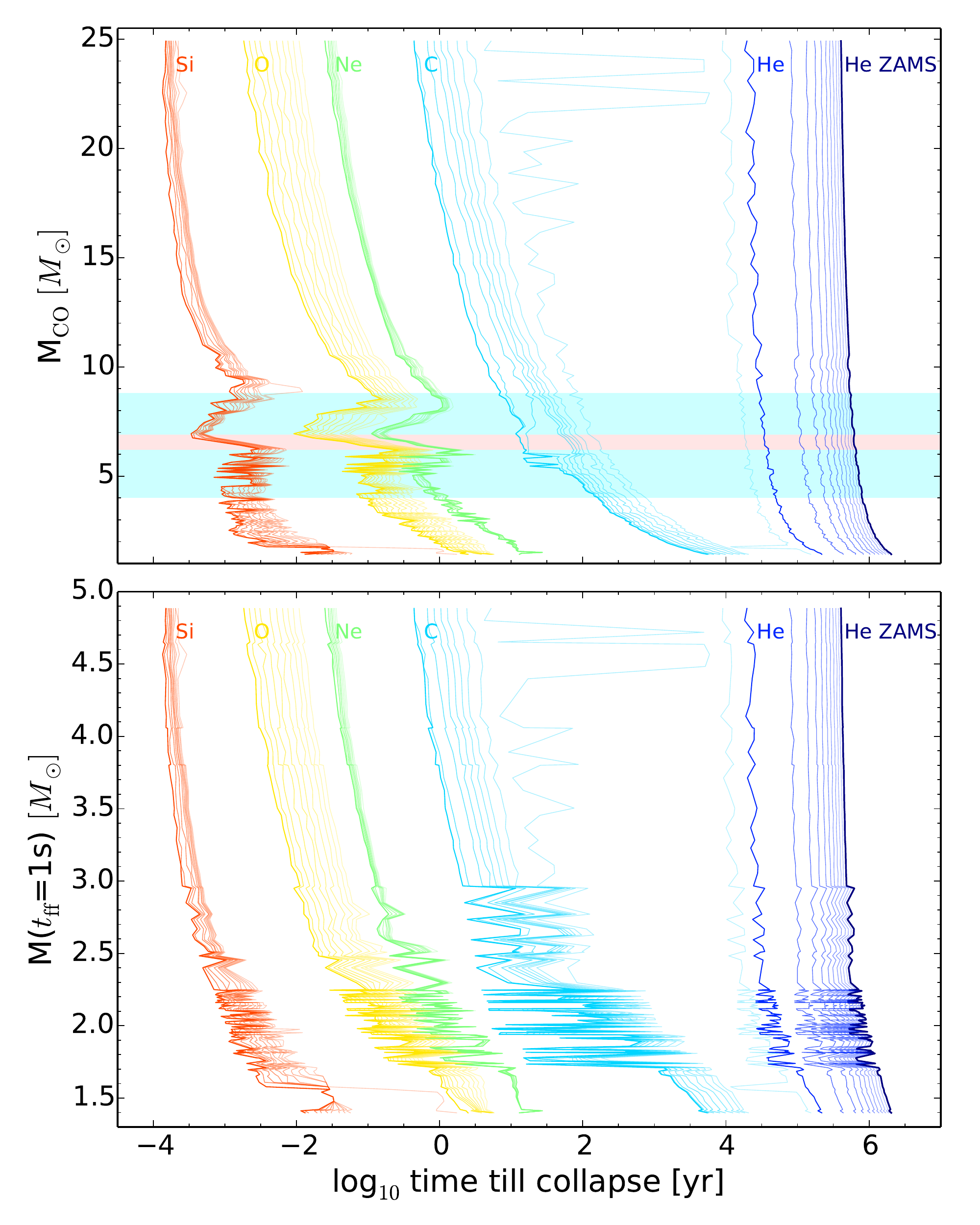}
 \caption{The remaining time until core-collapse from several evolutionary phases, which are defined based on the chemical composition. The beginning of the simulation is traced by the black line (He ZAMS), and phases of the core He, C, Ne, O, and Si burnings are indicated by the blue, cyan, green, yellow, and red lines, respectively. For the definitions of evolutionary phases, see the text. In the top panel, models are sorted according to the \MCO{} order, and in the bottom, they are sorted according to the \Mff{} order. The cyan and red bands in the top panel are the same as those in Fig.~\ref{plot-MCO-xi25}.
 }
 \label{plot-lifetime}
\end{figure*}

The remaining time till core-collapse estimated for our model set is shown in Fig~\ref{plot-lifetime}. Similar to Fig.~\ref{plot-sort-dens}, the remaining times are shown according to the \MCO{} order in the top panel and shown according to the \Mff{} order in the bottom panel.
The rightmost black line indicates the start of the simulation of the He ZAMS phase.
The leftmost lines of each color, corresponding to the boundaries between different $iphase$, are set to indicate the depletion times of the corresponding elements. Other thinner lines indicate the depleting processes; in the beginning of each $iphase$, the initial reference mass fraction $X_\mathrm{ref,0}$ is recorded, and lines are drawn when $X_\mathrm{ref}/X_\mathrm{ref,0} =$ 0.9, 0.8, ..., and 0.1.
Hence, the thinner lines of $X_\mathrm{ref}/X_\mathrm{ref,0} = 0.9$ roughly indicate the beginning of each nuclear-burning phase but carbon. Since the mass fraction of $^{12}$C has already begun to decrease due to the $^{12}$C($\alpha$,$\gamma$)$^{16}$O reaction in the late core He burning phase, $X_\mathrm{ref}/X_\mathrm{ref,0} \sim 0.7$ may provide a better proxy for the initiation of the core C burning.
Also note that a depletion line indicates the time when the reference element is depleted at the reference location for the first time, but it does not necessarily mean the complete depletion of the reference element from the whole core. On the contrary, the depletion is usually followed by successive shell burnings. In particular, the shell C burning phase starts after central C depletion. 

The remaining times for the He burning phase show clear correlations to the CO core mass, thus, to the He core mass and presumably to the ZAMS mass. Similarly, the remaining times of the C burning phase also show strong correlations to $M_\mathrm{CO}$, though the mass dependency is stronger than the He burning phase. The least massive models take $\sim 10^4$ yr from the initiation of core C burning till collapse, while it takes only $\sim 1$ yr for the most massive models in our sample. This huge difference is due to the significant temperature dependency of the neutrino cooling rate. Besides, a jump at $M_\mathrm{CO} \sim 5 M_\odot$ indicates a transition from the convective to the radiative nature of the central C burning. Above this transition, the duration of the central C burning is reduced because convective transport no longer supplies nuclear fuel to the center.

The later Ne, O, and Si burning phases show peak structures; the durations are longest locally around $M_\mathrm{CO} \sim 6 M_\odot$, decrease towards the local shortest peak at $M_\mathrm{CO} \sim 7 M_\odot$, increase until $M_\mathrm{CO} \sim 9 M_\odot$, then decrease constantly. These features are quite consistent with the trends obtained for the \xx{2.5} and \Mff{} distributions, which are indicated in the top panel as cyan and red bands. Therefore, clear monotonic correlations can be manifested when the remaining times are expressed as a function of \Mff{}, as shown in the bottom panel. 

The plot still involves a huge scatter, especially in the less massive range with \Mff{} $\in [1.6, 2.3] \ M_\odot$. This scatter is not due to the shuffling of models having smaller and larger \MCO{}, but rather originates from the scatter seen in the less massive models with \MCO{} $\in [3, 6] \ M_\odot$. At present, it is unclear whether this kind of scatter is realistic or not. The large scatter is likely induced by the highly non-linear interplay of nuclear burning and convective mixing, which can significantly affect the lifetimes of nuclear burning phases. Hence, real stars would show the same scatter. However, from a numerical point of view, such behavior might be enhanced due to coarse resolutions both in space and time. To answer the question, further investigation is needed.

It is noteworthy that the least massive models with \Mff{} $\leq 1.56 \ M_\odot$ (\MCO $\leq 1.72 \ M_\odot$) start the `core Si burning' $\sim 1$ year before core-collapse. This Si burning is induced by the off-center O+Ne flash, which takes place in a low mass oxygen core having a high electron degeneracy \citep{Umeda+12, Woosley&Heger15}. The relevance of observations is discussed later.

\subsection{Radius, luminosity, and effective temperature}

\begin{figure}[t]
 \centering
 \includegraphics[width=\hsize]{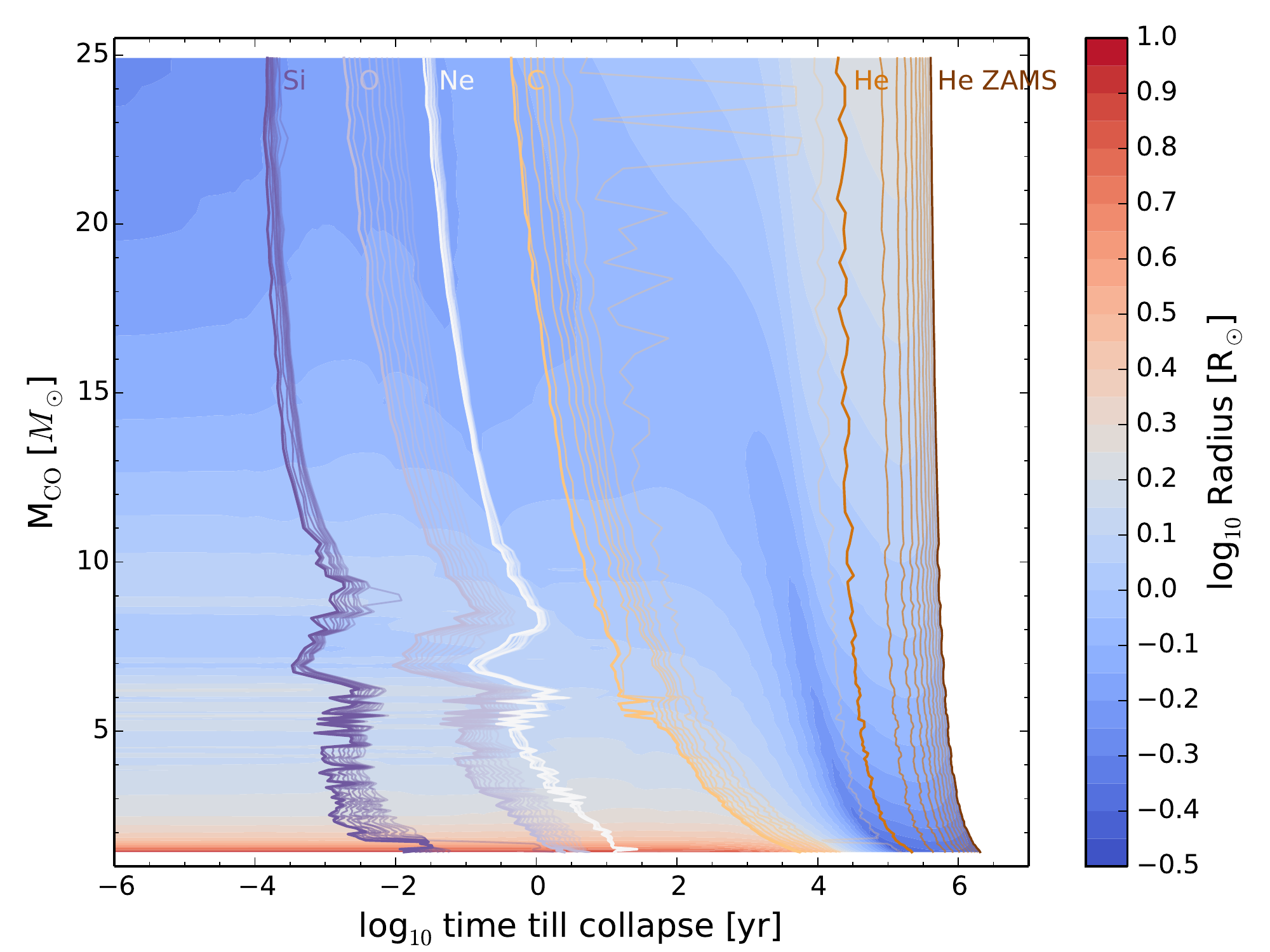}
 \caption{Radius evolution of He star models shown by the color map overplotted with the distributions of remaining time till collapse from evolutionary phases. The y-axis is \MCO of the model and the remaining time till collapse is shown by the x-axis: the radius evolution of one model is indicated by the color change on a horizontal line from the right side to the left. The color coordinate for the radius change is shown in the right column.
 }
 \label{plot-radevol}
\end{figure}

The radius evolution of all models is plotted in Fig.~\ref{plot-radevol}, in which the evolutionary phases are also shown.
The mass dependence of the radius evolution up to core C depletion is rather simple. As a common feature, a He star model first expands and then contracts during the core He burning phase. In this phase, the smaller the initial mass is, the smaller the stellar radius is. Later, this relation is reversed by the shell He burning; the helium envelope expands more for less massive models, while it keeps contracting for more massive models after core He depletion. The expansion/contraction bifurcation takes place at $M_{\mathrm{CO}} \sim 15$ $M_\odot$. 

On the other hand, the evolution after core C depletion has a more complicated mass dependency.
The less massive models with $M_{\mathrm{CO}} \lesssim 4.3$ $M_\odot$ except for the least massive model hardly change the radii for the later phases, thus $|R_{\mathrm{collapse}}/R_{\mathrm{Cdep}}| <$ 0.05 dex, where $R_{\mathrm{collapse}}$ and $R_{\mathrm{Cdep}}$ are the stellar radii at the core-collapse and core C depletion phases. Models with $M_{\mathrm{CO}} \sim$ 4.5--6.8 $M_\odot$ expand their radii during Ne and O burning phases. This is particularly true for models with $M_{\mathrm{CO}} \sim$ 5.6--6.4 $M_\odot$, in which $R_{\mathrm{collapse}}/R_{\mathrm{Cdep}}$ can be as large as +0.1 dex. Conversely, models with $M_{\mathrm{CO}} \sim$ 7.1--8.4 $M_\odot$ contract during the later phases, resulting in $R_{\mathrm{collapse}}/R_{\mathrm{Cdep}} \sim -0.1$ dex. More massive models with $M_{\mathrm{CO}} \gtrsim$ 8.7 $M_\odot$ expand after the core Ne burning phase. Among them, less massive models with $M_{\mathrm{CO}} \sim$ 8.7--15.0 $M_\odot$ expand also after the O burning phase similar to the models with $M_{\mathrm{CO}} \sim$ 4.5--6.8 $M_\odot$, while the radii decrease during the later phases for more massive models with $M_{\mathrm{CO}} \gtrsim$ 15.0 $M_\odot$.

\begin{figure}[t]
 \centering
 \includegraphics[width=\hsize]{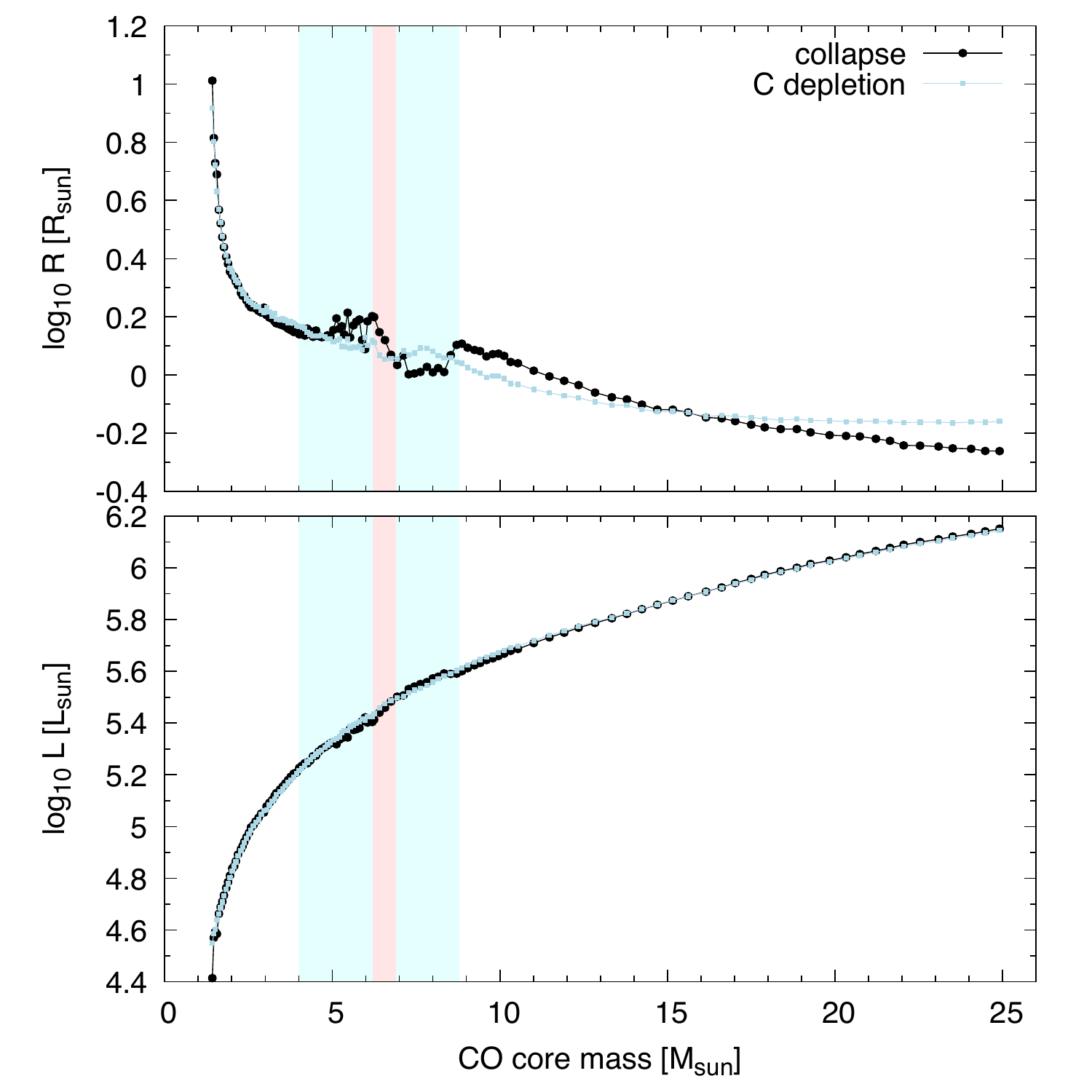}
 \caption{Radius (top) and luminosity (bottom) distributions as a function of \MCO. In both panels, distributions recorded at the central C depletion are shown by the cyan lines, and at the core-collapse by the black lines. The cyan and red bands are the same as those in Fig.~\ref{plot-MCO-xi25}. 
 }
 \label{plot-MCO-RLT}
\end{figure}

As a result, the stellar radius shows a smooth relation with the CO core mass up to C depletion, but eventually forms distinctive peaks ($M_{\mathrm{CO}} \sim$5--7 $M_\odot$ and $\sim$ 9--15 $M_\odot$) and a valley between them by core-collapse. In the top panel of Fig.~\ref{plot-MCO-RLT}, the distribution of the stellar radius at the core-collapse phase and the C depletion is shown. The coincidence between the first peak (valley) and small (large) \Mff{} at $M_{\mathrm{CO}} \sim$ 5--7 (7--9) $M_\odot$ strongly indicates that this structure originates from different inner core evolutions. Meanwhile, such a structure does not develop significantly for the luminosity distribution shown in the bottom panel. Consequently, the peak-valley structure in the radius distribution appears as a valley-peak structure in the effective temperature distribution.

\subsection{Properties of supernova explosions}

\begin{figure}[t]
 \centering
 \includegraphics[width=\hsize]{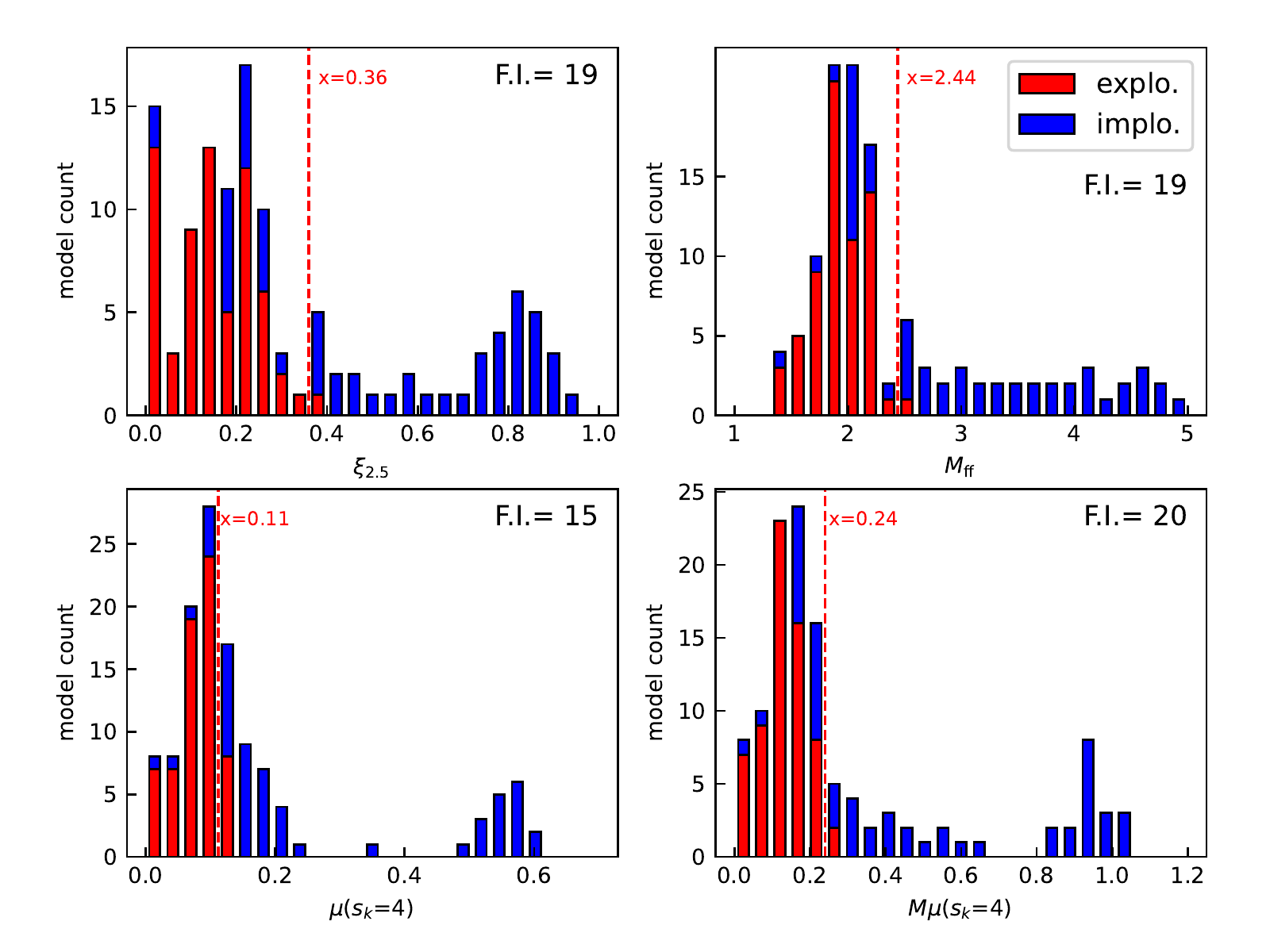}
 \caption{Histograms of models labeled with successful or unsuccessful CCSN explosions, which are assessed by Muller's semi-analytic model, are shown. Exploding models are shown in the red bins and non-exploding models are in the blue bins. The x-axes are set from \xx{2.5} (top-left), \Mff (top-right), \musk \ (bottom-left), and \Mmusk \ (bottom-right). The threshold lines, below which the model is supposed to explode, are shown by red dashed lines with indications of the values, and the false identification numbers are indicated in the upper right.}
 \label{plot-explodability2}
\end{figure}

Figure~\ref{plot-explodability2} shows the relation between the explodability and density indicators (\xx{2.5} and \Mff) and the Ertl's parameters. The explodability is estimated based on the semi-analytic model developed by \citet{Mueller+16}, with which we assign `explosion' for models experiencing the shock revival and forming NSs and `implosion' for models never experiencing the shock revival or forming BHs due to the late time accretion.
The critical value of each indicator is determined as the value that would minimize the false identification number of
\begin{eqnarray}
	\nonumber (number \ of \ imploded \ models \ with \ x < x_{\mathrm{crit}} \\
	 \nonumber + number \ of \ exploded \ models \ with \ x \ge x_{\mathrm{crit}}),
\end{eqnarray}
where $x \in \{ \xi_{2.5}, M_{\mathrm{ff}}, \mu(s_k=4), M\mu(s_k=4) \}$. The false identification rate is the ratio of the false identification number to the total model number.

As for \xx{2.5}, the figure clearly shows that the absolute value can be used to determine the explodability, which is consistent with previous works. Besides, we also find that \Mff{} is as useful as the compactness for the identification, showing the same false identification number. Originally, the Ertl's parameters of \musk{} and \Mmusk{} (product of \Msk{} and \musk{}) were used in combination for the fate identification in \citet{Ertl+16}. However, \citet{Mueller+16} has reported that a single-parameter classification only depending on \musk{} can yield a better false identification with their semi-analytic model because the possibility of the late time BH formation after the shock revival has been neglected in \citet{Ertl+16}. This is why we compare the results of the fate classification utilizing one of \musk{} and \Mmusk{} in this work. These parameters are also capable of identifying the fate, resulting in similar false identification rates. In summary, we have found that, based on any of these indicators, explodability can be judged with roughly equal accuracy.

Although we compute the false identification number and rate, these are only for determining the optimal value to identify different fates, and it is not our purpose to determine the precise values. This is because we do not expect that a complete identification is possible from approximate methods such as those performed in this work. False identification rates of about 10\% are obtained for all the indicators in this work, and it should be interpreted as a typical accuracy when using such an approximate method.
Furthermore, the critical values derived in this work are not accurate enough for quantitative comparisons. This is because we have found that the critical values are sensitive to the method applied for the fate estimate. For example, implosion more likely takes place if \xx{2.5} $>$ 0.36 for our model set, which is larger than 0.278 obtained in \citet{Mueller+16}. However, we speculate this is not due to the different model set but chiefly due to the different implementations of M\"{u}ller's semi-analytic prescription since a critical value of 0.33, which is closer to ours than that of \citet{Mueller+16}, is obtained even if we apply our implementation to the same progenitor models used in \citet{Mueller+16}. If another method based on, for example, 1D hydrodynamical simulations were used, even different values could be obtained.
Therefore, we conclude that the qualitative feature of being able to identify fate is more robust and reliable than the quantitative features including the false identification rate and the critical values.

\begin{figure}[t]
 \centering
 \includegraphics[width=\hsize]{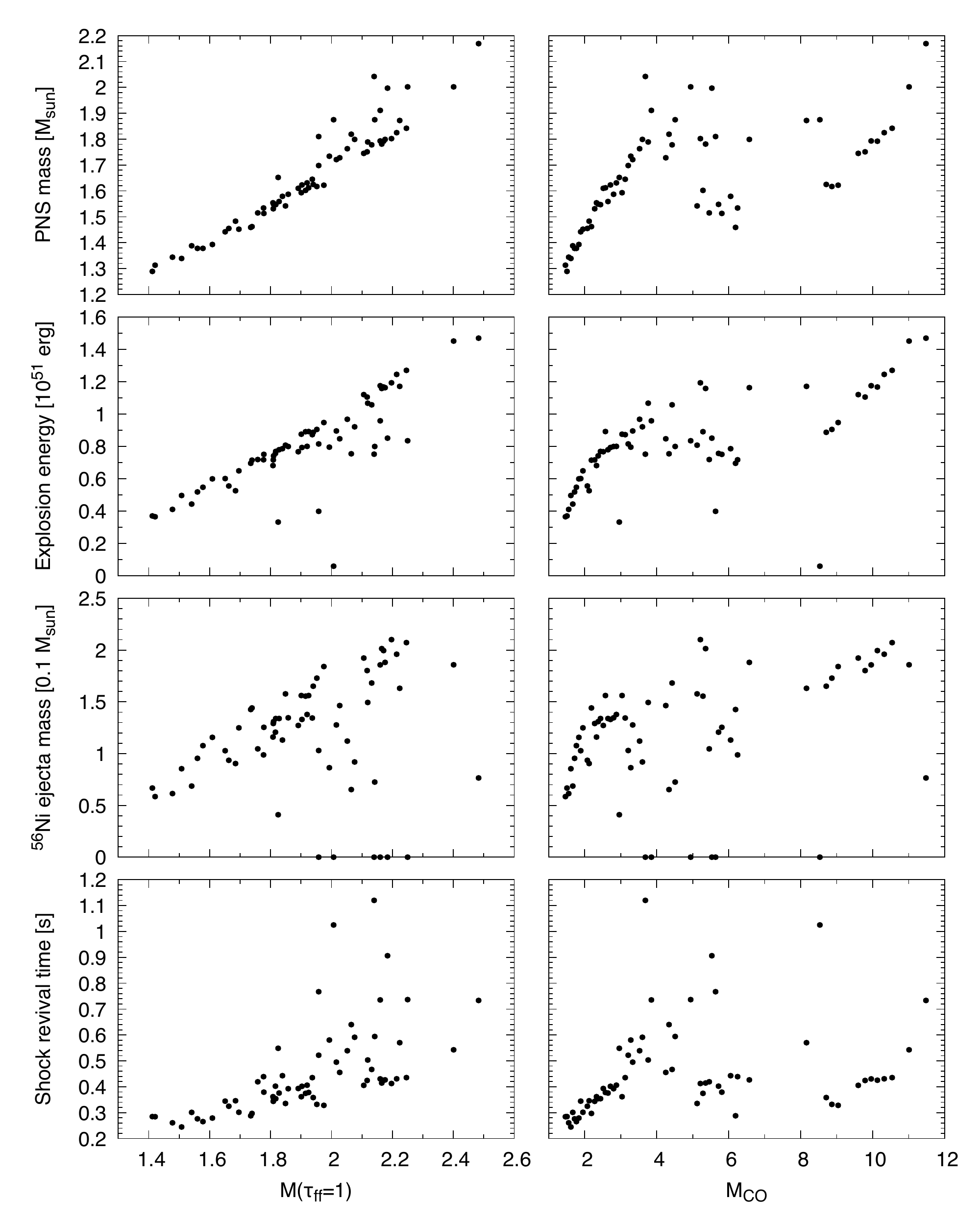}
 \caption{Relation between explosion properties estimated by the semi-analytic model of \citet{Mueller+16} and model indicators.
 The explosion properties of the PNS mass (top panels), the explosion energy (second top), the $^{56}$Ni ejecta mass (third top), and the shock revival time (bottom) are shown by the vertical axis, and the model indicators of \Mff{} (left panels) and the CO core mass (right) are shown by the horizontal axis.
 }
 \label{plot-Mff-exp}
\end{figure}

Explosion properties of (baryonic) PNS mass, explosion energy, nickel ejecta mass, and shock revival time are presented as a function of \Mff{} and $M_\mathrm{CO}$ in Fig.~\ref{plot-Mff-exp}\footnote{The PNS mass, explosion energy, and nickel ejecta mass are $M_{\rm PNS}$, $E_{\rm diag}$, and $M_{\rm Ni}$ calculated at the simulation end, respectively, and the shock revival time is the time when the condition $t_{\rm heat} < t_{\rm adv}$ is met. For detailed definitions, see the Appendix.}. 
It shows that these explosion properties, especially the PNS mass, have strong positive correlations with \Mff. These correlations suggest that the progenitor density structure would determine not only the explodability but also the detailed properties of the supernova explosions. 
Taking the fact that the supernova explosion is a genuine non-linear phenomenon into consideration, the existence of this kind of correlation is non-trivial and thus interesting. Since the present analysis is based on the approximate model, further investigations with more realistic simulations are required.
Nevertheless, it is noteworthy that the correlations shown here are consistent with an interesting correlation between the mass and the entropy of PNSs that is found from more realistic and systematic 1D explosion simulations \citep{daSilvaSchneider20}. 
This is because, provided a likely correlation between the entropies of the nascent NS and the progenitor's iron core, the aforementioned correlation results from the correlation between the PNS mass and the entropy of the progenitor core.

$M_\mathrm{CO}$ is probably a more accessible parameter by observations than \Mff{} as it could correlate with the total ejecta mass both in cases of type II SNe and SE-SNe. The right column of Fig.~\ref{plot-Mff-exp} indicates that explosion properties show different tendencies depending on $M_\mathrm{CO}$.
In the lower end of $M_\mathrm{CO} \lesssim 4 M_\odot$, explosion properties, in particular the PNS mass and the explosion energy, obey linear correlations with $M_\mathrm{CO}$. This is due to the linear correlation between $M_\mathrm{CO}$ and \Mff{} for these less massive progenitors. Because the CO core mass range roughly corresponds to the ZAMS mass of $M_\mathrm{ZAMS} \lesssim 20 M_\odot$, and because most SNe may emerge from the less massive range considering the nature of the initial mass function, this coincides with the correlations observed for type II SNe \citep[e.g.,][]{Mueller+17}.
An island of explosion exists for $M_\mathrm{CO} \in [8.1, 11.6] \ M_\odot$, which is consistent with earlier theoretical studies \citep[e.g.,][]{Ugliano+12}. These massive exploding models are estimated to yield explosions with relatively larger NS masses, explosion energies, and $^{56}$Ni ejecta masses, and this could be consistent with observations suggesting the positive correlation between the total ejecta mass, the $^{56}$Ni ejecta mass, and the kinetic energy of SE-SNe \citep[e.g.,][]{Taddia+18}.

\section{Discussion}

\subsection{Monotonicity in other model sets}

In this subsection, we aim to check the degree to which the monotonic relation between the indicator, \Mff{}, and the global density and temperature distributions is robust. For this purpose, a similar analysis has been performed for four additional sets of models, in addition to the set we have described so far (hereafter referred to as H1).
Three of them, H2, H3, and H4 are calculated using the same stellar evolution code but with different initial compositions; they have pure helium (H2), solar- (H3), or zero-metallicity (H4) compositions initially (full stellar evolution with hydrogen envelopes are treated in H3 and H4). Another difference is that a reaction rate of $^{12}\mathrm{C}(\alpha,\gamma)^{16}\mathrm{O}$ of \citet{Caughlan&Fowler88} multiplied by a factor of 1.2 is applied for models in these sets.
The fourth set is the one provided by \citet{Mueller+16}, which consists of models with solar-metallicity calculated by the stellar evolution code KEPLER applying $^{12}\mathrm{C}(\alpha,\gamma)^{16}\mathrm{O}$ of \citet{Buchmann96} multiplied by a factor of 1.2. This set is referred to as M16 hereafter.
In addition, these model sets use different termination conditions for stellar evolution simulations; H2 uses the same condition as H1, stopping the simulation at $\rho_c = 10^{10}$ \gcc{}, and H3 and H4 use the condition $T_c = 10^{9.9}$ K. On the other hand, simulations in M16 terminate when the collapse velocity anywhere in the core exceeds 1,000 km s$^{-1}$ (Heger, private communication).

\begin{figure}[t]
 \centering
 \includegraphics[width=\hsize]{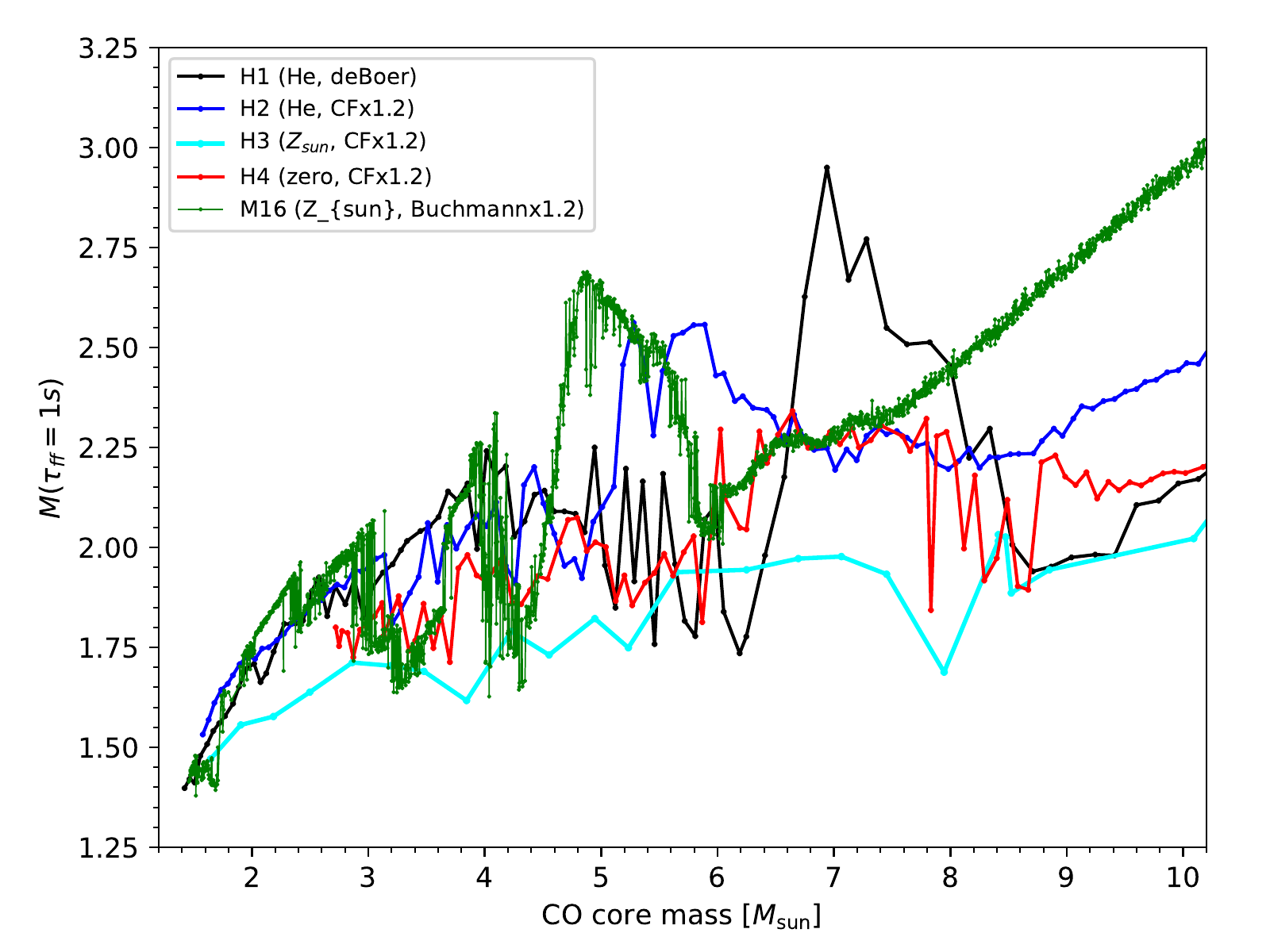}
 \caption{The \MCO-\Mff{} relation for different sets. H1 (shown by the black line) is the set we have so far discussed. Using the same code, but applying different initial compositions and $^{12}$C($\alpha$,$\gamma$)$^{16}$O reaction rate, sets H2 (He star, blue), H3 (solar composition with H envelopes, cyan), and H4 (zero-metallicity with H envelopes, red) are calculated. M16 (green) consists of models analyzed in \citet{Mueller+16}.}
 \label{plot-MCO-Mtau-comp}
\end{figure}

The $M_\mathrm{CO}$--\Mff{} relations compared in Fig.~\ref{plot-MCO-Mtau-comp} show different properties among the model sets. In particular, the locations, widths, and heights of the peaks seen at \MCO$\sim$ 5--9 $M_\odot$ are different for all sets (The major peaks are around 6--9 $M_\odot$ for H1, 5--7 $M_\odot$ for H2, 5--8 $M_\odot$ for H3, 6--9 $M_\odot$ for H4, and 5--6 $M_\odot$ for M16). The diversity seen in models H1 to H4 indicates that the $M_\mathrm{CO}$--\Mff{} relation is sensitive to the different computational settings of hydrogen envelopes, metallicity, and $^{12}\mathrm{C}(\alpha,\gamma)^{16}\mathrm{O}$ rate. This result is understandable since any of those differences result in different C/O ratios that the CO cores have at their birth \citep[e.g.,][]{Sukhbold&Woosley14, Patton20, Sukhbold&Adams20}. Moreover, more significant offsets between H models and model M16 may indicate that the difference in the stellar evolution code including the reaction network, EOS, opacity, convective boundary mixing, etc, is as influential as the other settings. We do not perform a comprehensive analysis in this work as it is beyond its scope, however, performing such an analysis is clearly important for future realistic predictions.

\begin{figure*}[t]
	\centering
	\begin{tabular}{cc}
	\begin{minipage}[t]{0.5\hsize}
		\centering
		\includegraphics[width=\hsize]{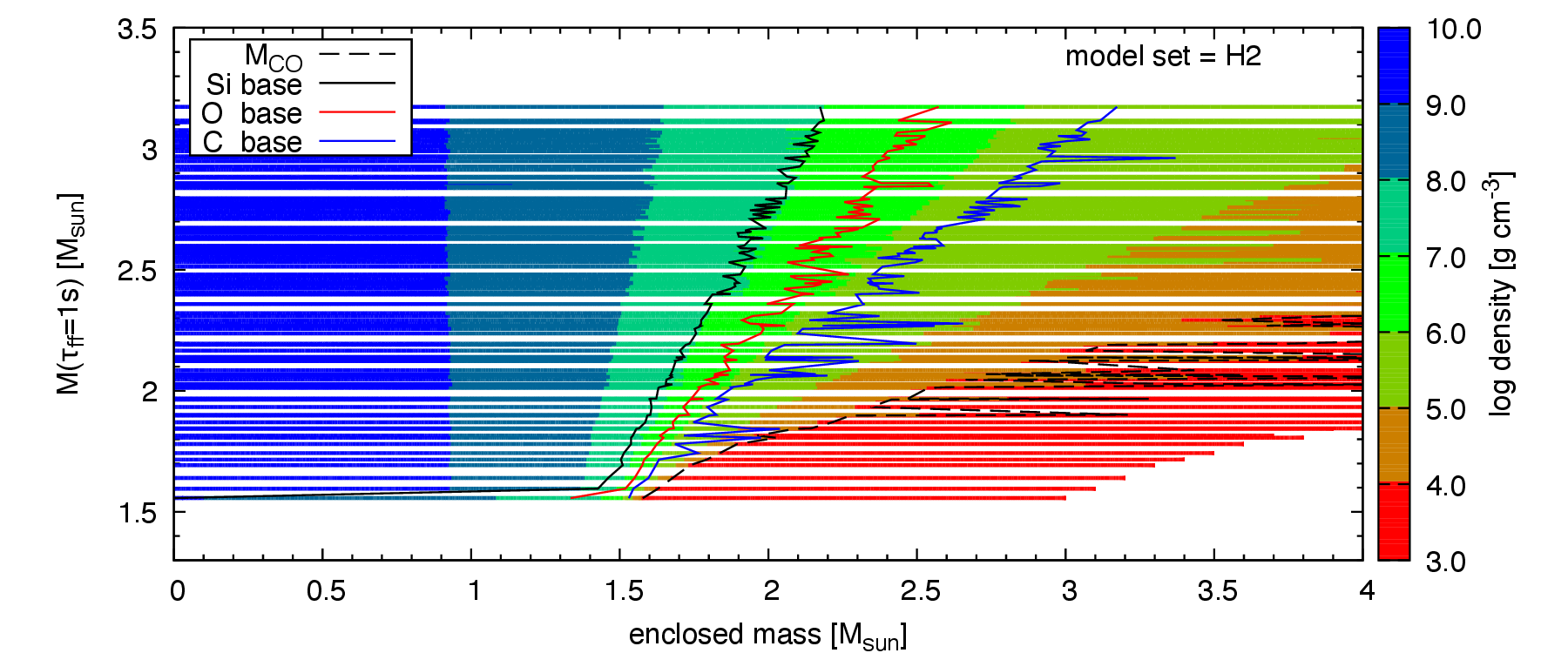}
 	\end{minipage} &
	\begin{minipage}[t]{0.5\hsize}
		\centering
		\includegraphics[width=\hsize]{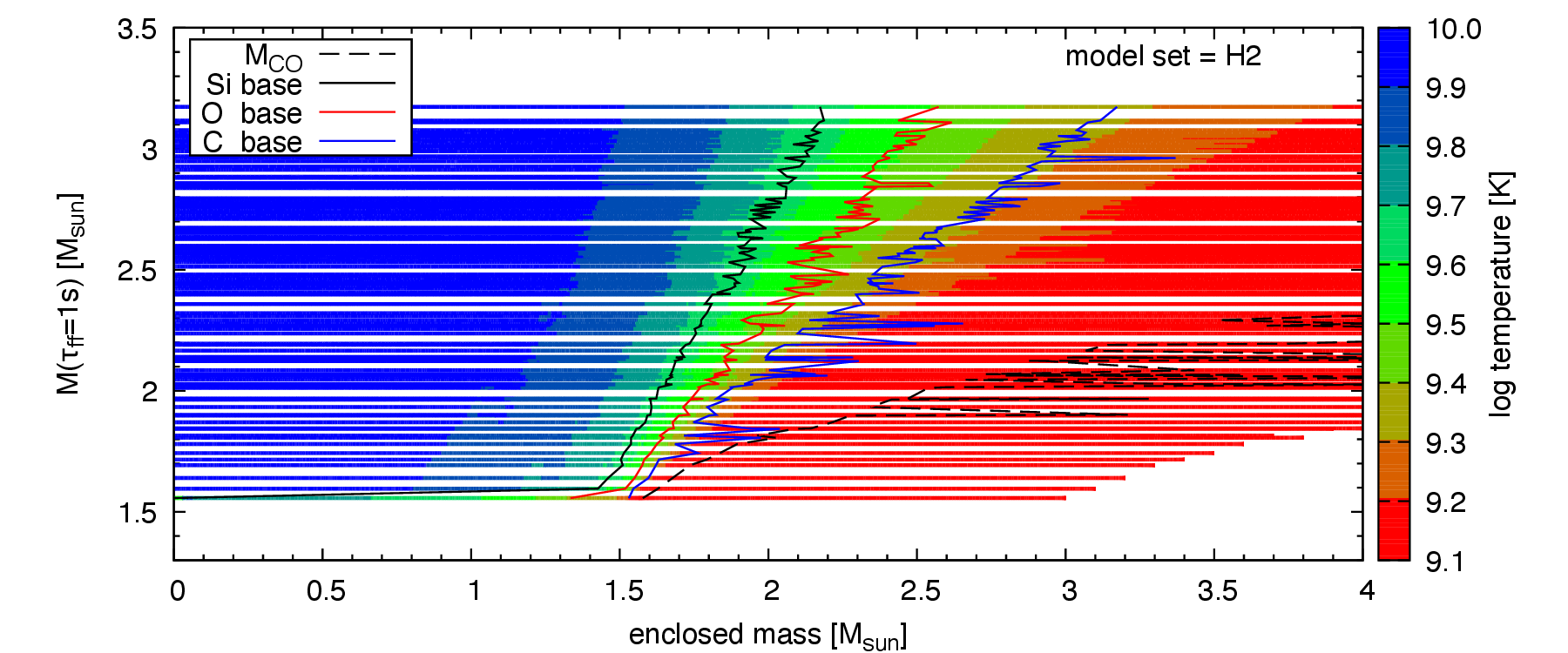}
 	\end{minipage}
	\end{tabular}

	\begin{tabular}{cc}
	\begin{minipage}[t]{0.5\hsize}
		\centering
		\includegraphics[width=\hsize]{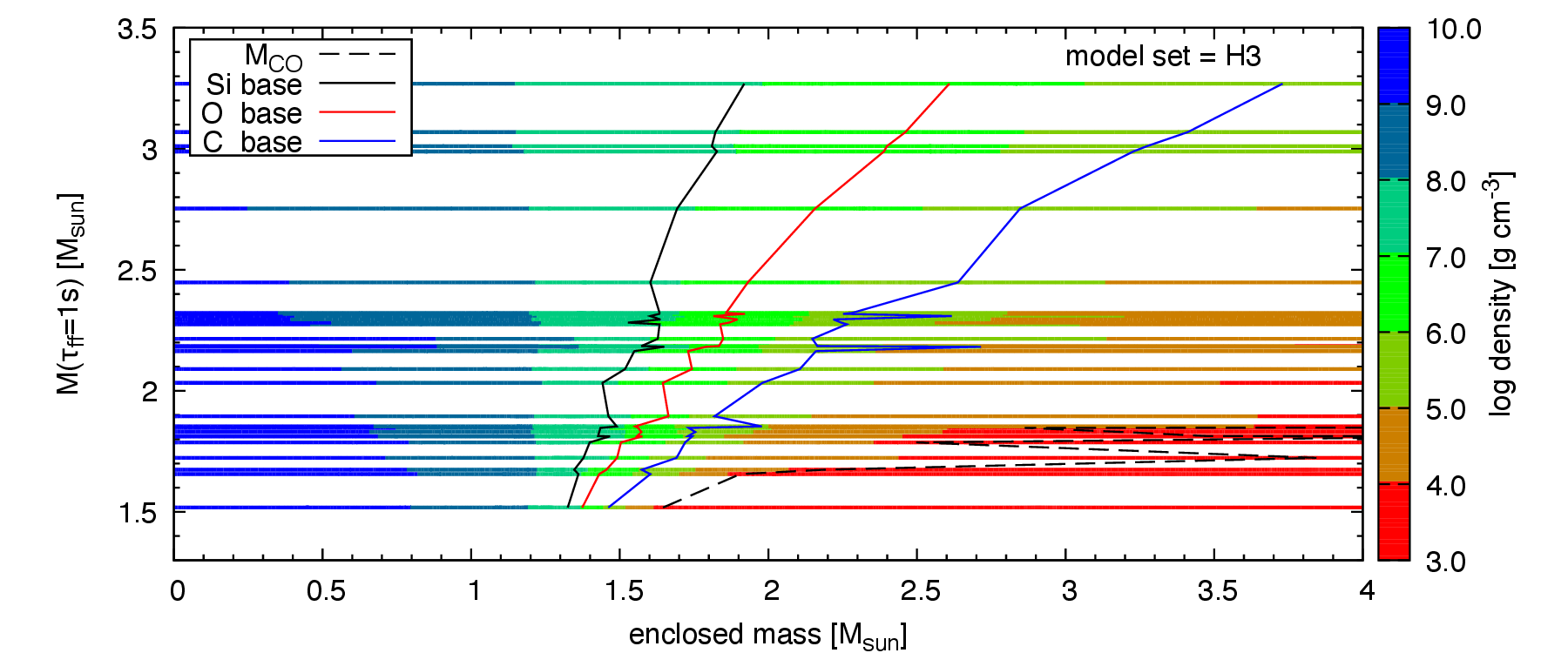}
 	\end{minipage} &
	\begin{minipage}[t]{0.5\hsize}
		\centering
		\includegraphics[width=\hsize]{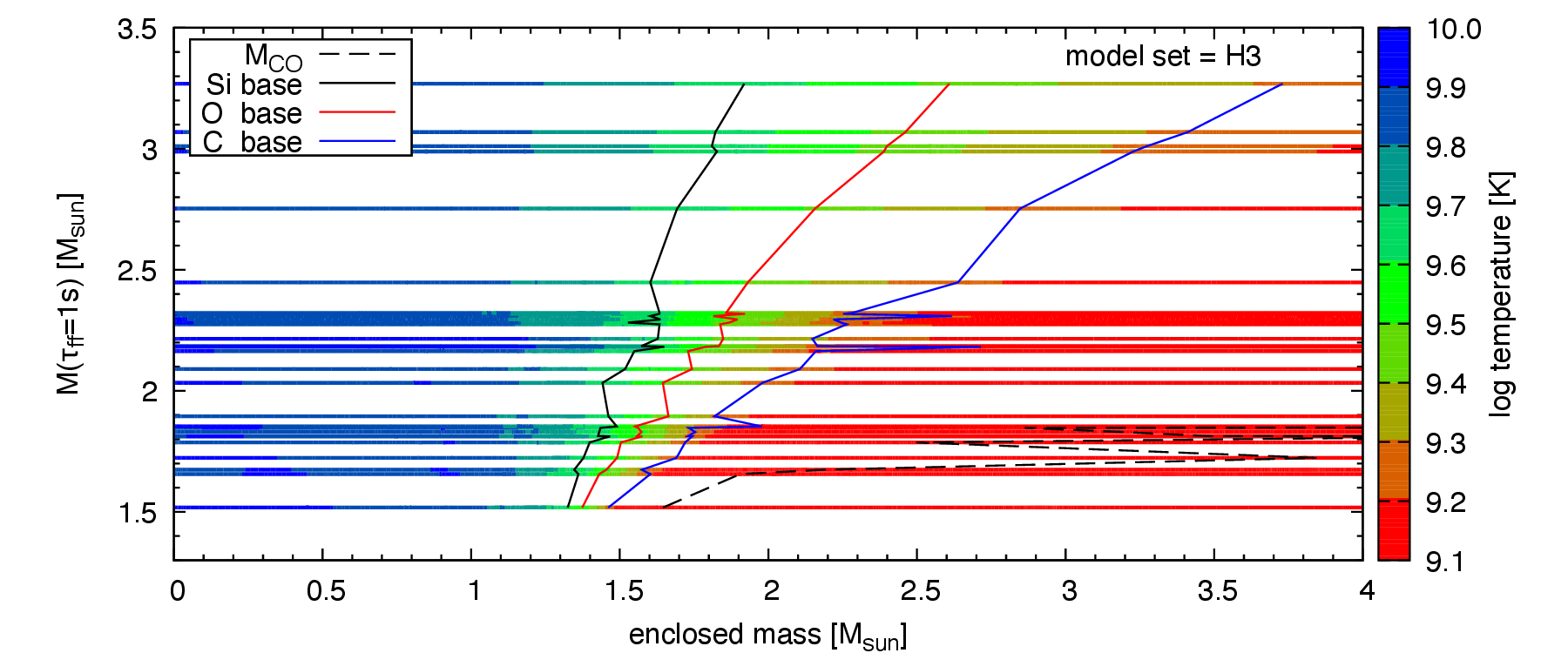}
 	\end{minipage}
	\end{tabular}

	\begin{tabular}{cc}
	\begin{minipage}[t]{0.5\hsize}
		\centering
		\includegraphics[width=\hsize]{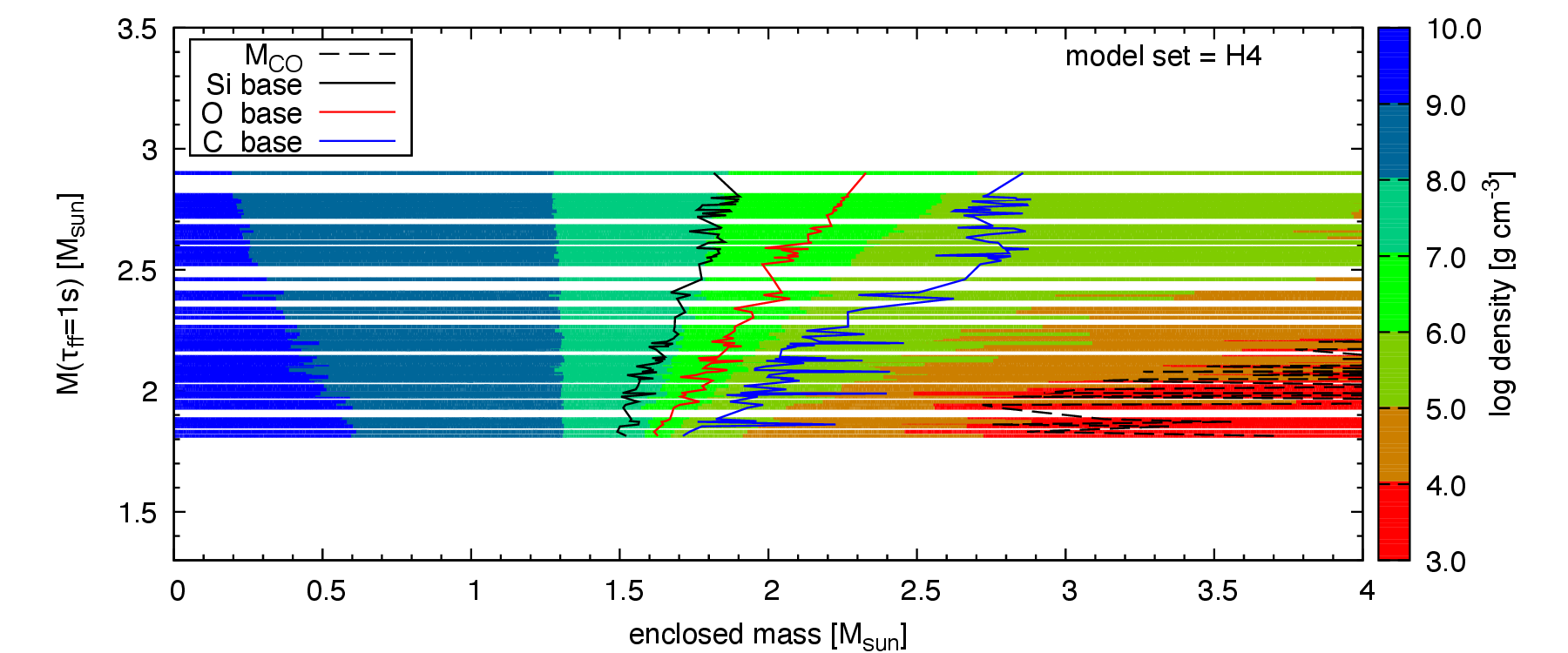}
 	\end{minipage} &
	\begin{minipage}[t]{0.5\hsize}
		\centering
		\includegraphics[width=\hsize]{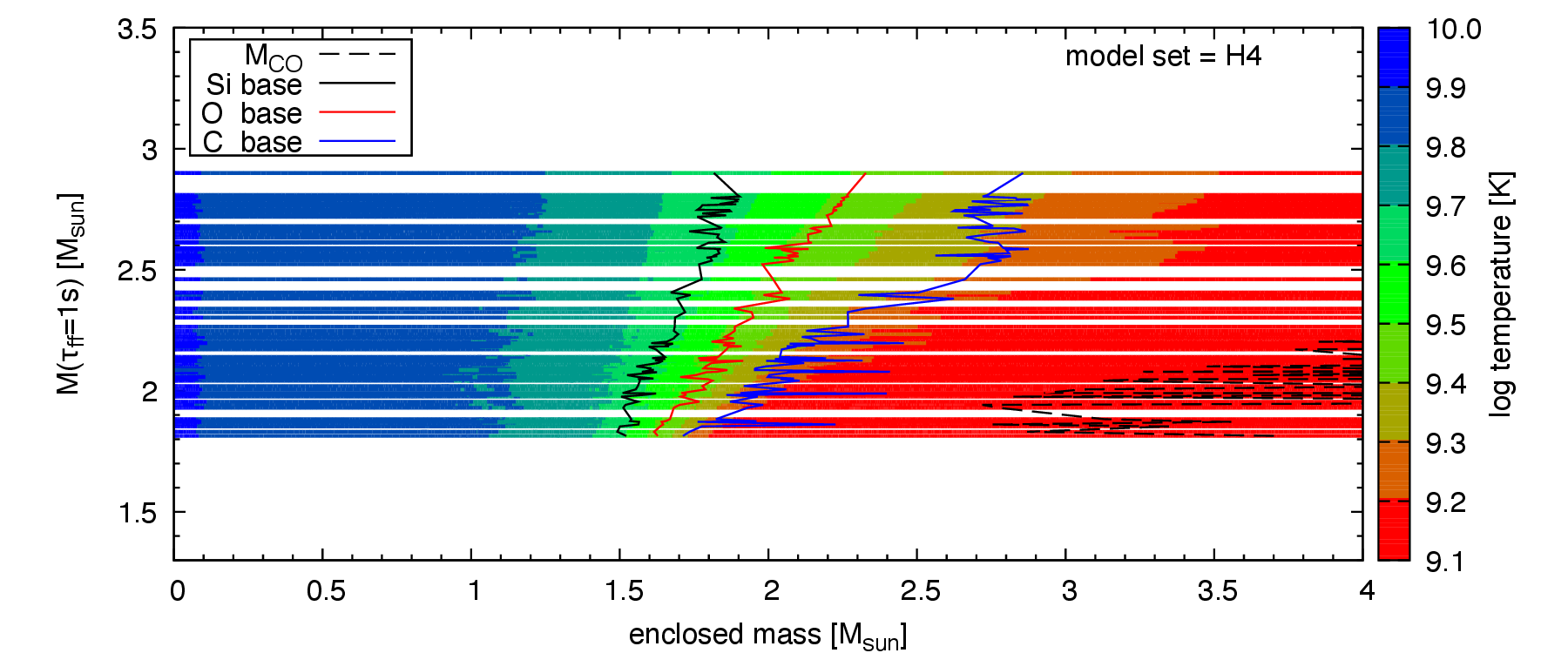}
 	\end{minipage}
	\end{tabular}

	\begin{tabular}{cc}
	\begin{minipage}[t]{0.5\hsize}
		\centering
		\includegraphics[width=\hsize]{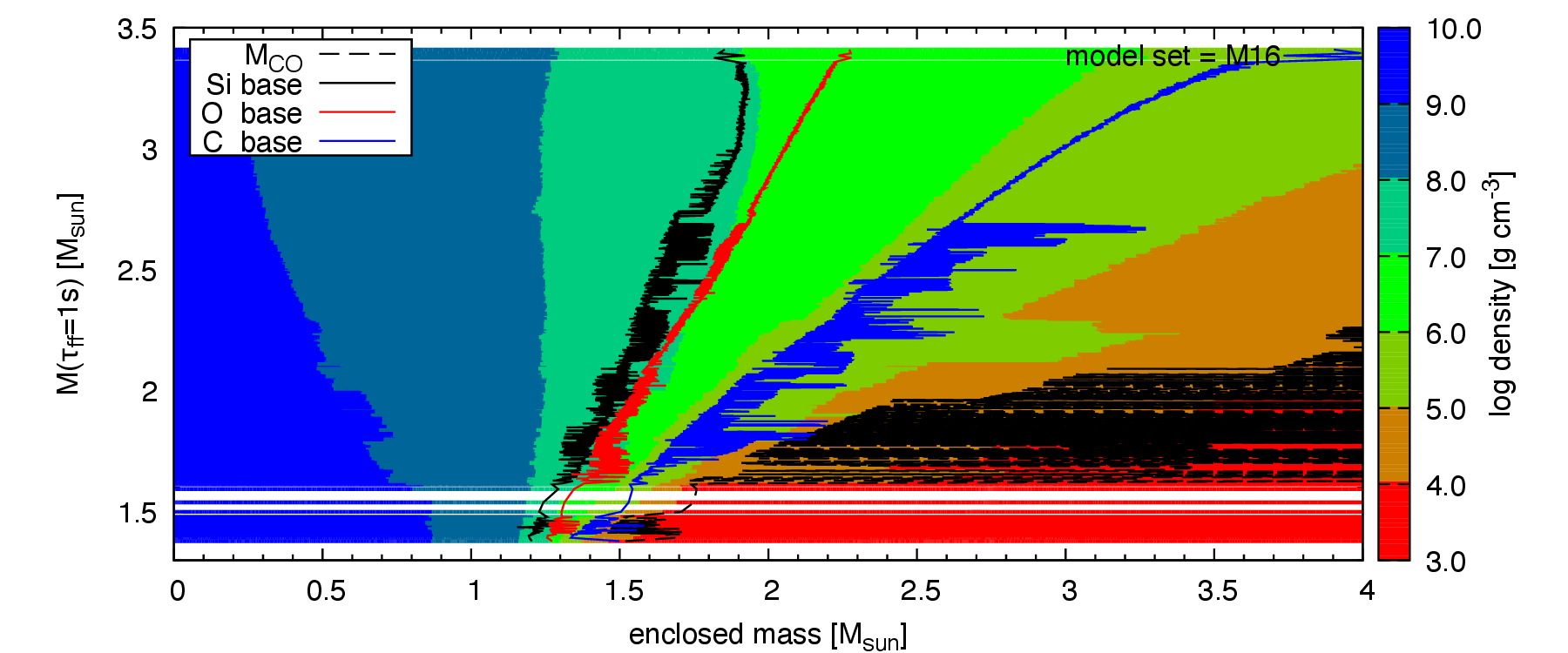}
 	\end{minipage} &
	\begin{minipage}[t]{0.5\hsize}
		\centering
		\includegraphics[width=\hsize]{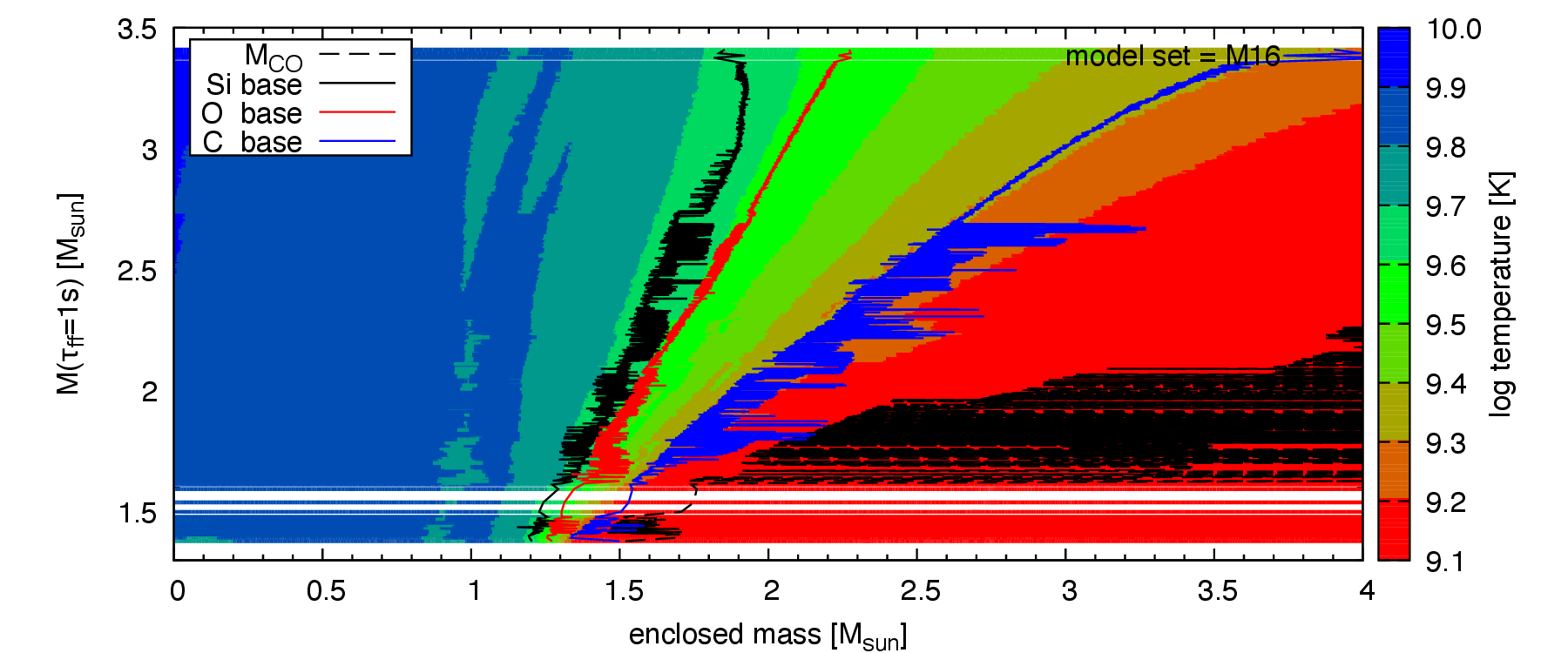}
 	\end{minipage}
	\end{tabular}
 \caption{The same as the bottom panel of Fig.~\ref{plot-sort-dens} but for density (left column) and temperature (right column) distributions of different sets H2 (top), H3 (second), H4 (third), and M16 (bottom).}
 \label{plot-rename-mff-comp}
\end{figure*}

Figure \ref{plot-rename-mff-comp} shows the projections of density and temperature distributions similar to Fig.~\ref{plot-sort-dens} and \ref{plot-sort-therm} but for other model sets. Differences exist in the details. For example, in the M16 model set, the \Mob{} (shown by the red line) traces a constant temperature $T \sim 10^{9.6}$ K, whereas, in other model sets calculated with the HOSHI code, it traces a lower constant temperature $T \sim 10^{9.5}$ K. This would indicate the strong impact of applying different reaction rates, reaction network, or QSE/NSE treatments on determining the innermost stellar structures. Also, due to different termination conditions, the density and temperature structures below $\sim$1 $M_\odot$ show different trends for different model sets, i.e., when compared at the same \Mff{}, the H1 and H2 models show higher densities and temperatures in the inner regions than the H3, H4, and M16 models. Nevertheless, the \Mff{}-based sorting clearly reveals a significant correlation in density, temperature, and compositional structure for all model sets. Hence, this correlation is likely to be universal and independent of the prescriptions for stellar evolution simulations.

\subsection{Mass ejection prior to core-collapse}

Recent high-cadence SN surveys have revealed that many CCSN progenitors experience enhanced mass loss in the final years before core-collapse, which leads to the formation of dense circumstellar medium (CSM) \citep[e.g.,][]{Bruch+21}. The CSM-ejecta interaction is believed to be the origin of the narrow lines of Type IIn SNe \citep[e.g.,][]{Chevalier&Fransson94} as well as Type Ibn SNe, and the existence of a dense CSM can also be inferred from the so-called `flash-spectroscopy' \citep{Khazov+16, Yaron+17}. More direct evidence can be obtained from pre-explosion images. \citet{Ofek+14} has conducted a systematic search for the precursor eruptions for progenitors of Type IIn SNe and concluded that most Type IIn progenitors undergo the precursor eruptions prior to the SN explosion \citep[see also][]{Strotjohann+21}. Furthermore, it will be possible to estimate the onset time of the enhanced mass loss by monitoring the evolution of the spectroscopic features changing from optically thick to optically thin.

Several theoretical explanations have been proposed for the mechanism of enhanced mass loss. The high mass loss rate may be achievable by line-driven winds \citep{Vink&deKoter02} or super-Eddington continuum-driven winds \citep{Shaviv01, VanMarle+08}. However, considering the peculiar proximity to the core collapse, other mechanisms such as wave-driven mass loss \citep{Quataert&Shiode12, Shiode&Quataert13} or mass ejection powered by off-center nuclear flashes \citep{Dessart+10} might be more plausible because these mechanisms will operate for only later evolutionary phases.

The convective motion inside the star will excite waves when it hits the convective boundary layers. After the waves are transported to the surface evanescent region, some of the energy will be dissipated leading to heating in the stellar envelope. The convective motion is more energetic for the later evolutionary phases, so at some point, this energy transfer may result in mass ejection from the surface. Theoretical studies have estimated that this wave-driven mass loss can operate during and after the Ne burning phase \citep{Quataert&Shiode12, Shiode&Quataert13, Fuller17, Fuller+18}. 
As we have shown, the remaining lifetime till collapse for the later burning phases of Ne, O, and Si burnings have rough anti-correlations to \Mff{} (Fig.~\ref{plot-lifetime}). Hence, we expect that the onset time of the wave-driven mass loss will also show anti-correlations to \Mff{}\footnote{This expectation should be consistent with the anti-correlation between the onset time and the He core mass indicated by \citet{Shiode&Quataert13}.}. 

In addition, Figure~\ref{plot-lifetime} illustrates that the least massive models of $M_\mathrm{CO} \in [1.42, 1.72] \ M_\odot$ ($M(\tau_\mathrm{ff}=1s) \in [1.40, 1.56] \ M_\odot$) experience off-center O+Ne flashes due to the high electron degeneracy, which takes place $\sim$ 1--10 yr before core-collapse. Depending on the injected energy, such flashes may result in mass ejection \citep{Woosley&Heger15}, which itself could be observed as SN-like transients or SN impostors \citep{Dessart+10}. From the small \Mff s, it is expected that the additional transients triggered by the off-center flashes will be associated only with the least energetic CCSNe that finally form the least massive NSs \citep{Suwa18}.
In the coming decades, the number of SNe, in which both the explosion properties and the onset time of the final enhanced mass loss are estimated, will significantly increase thanks to the large surveys such as the Rubin Observatory LSST \citep{Ivezic19}. We expect that the further correlations linked via the fundamental correlations with \Mff{} will be verified with future statistics.

\section{Summary and Conclusion}

\begin{figure*}
    \centering
	\includegraphics[width=0.7\hsize]{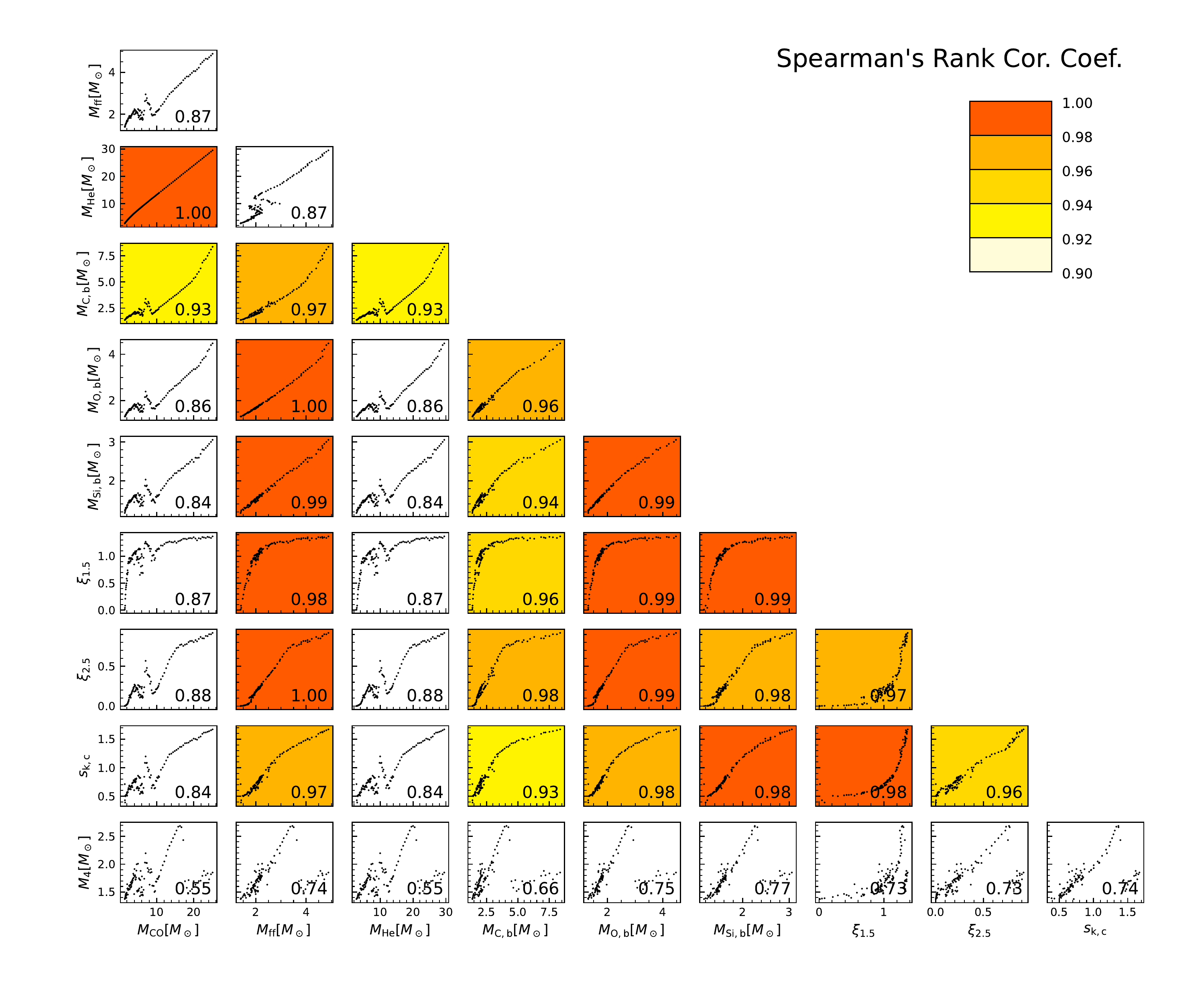}
	\includegraphics[width=0.7\hsize]{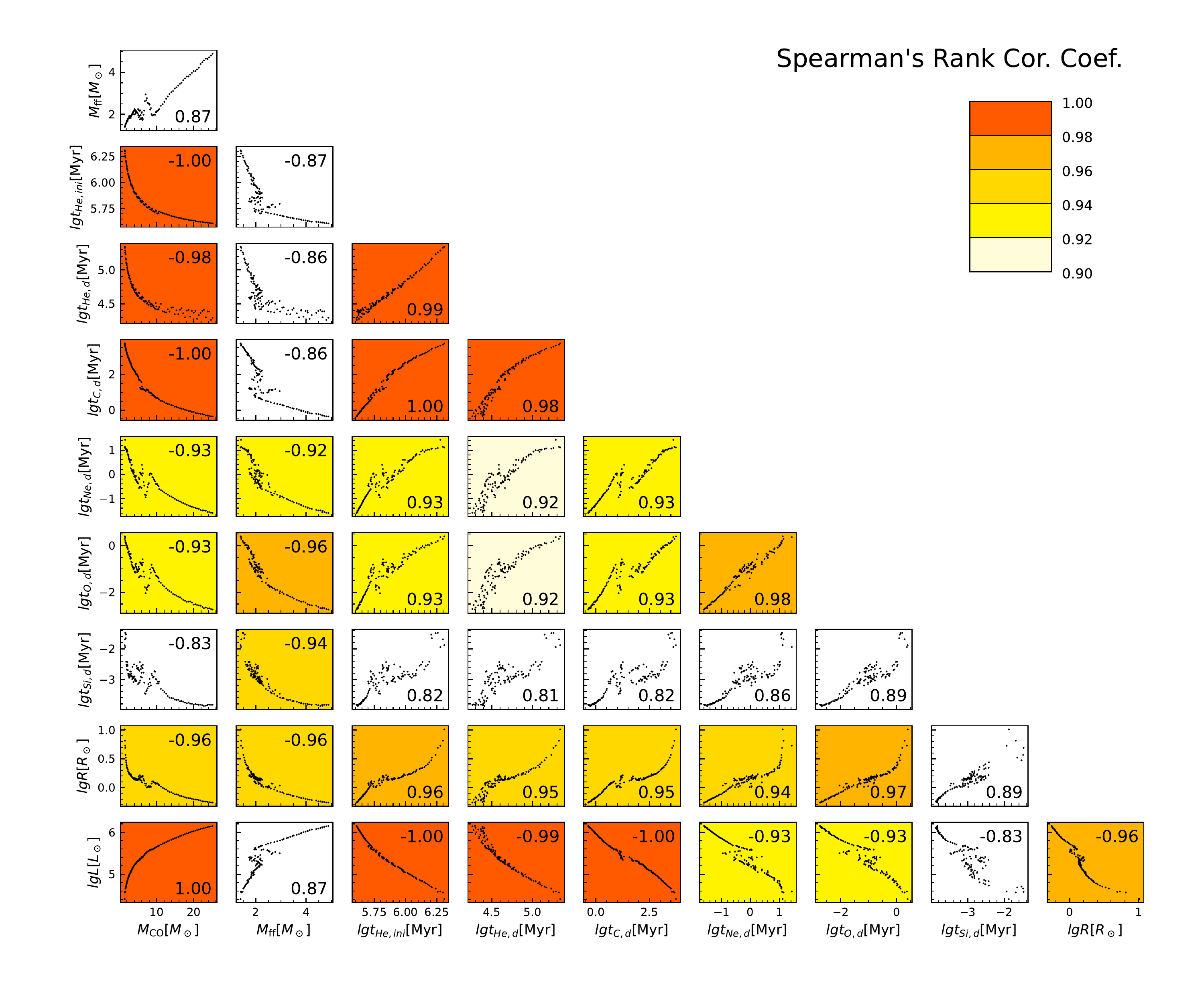}
 \caption{Matrices showing correlations between structure properties (top) and evolutionary properties (bottom). In the top panel, the characterizing parameters of \MCO{}, \Mff{}, $M_\mathrm{He}$, \Mcb{}, \Mob{}, \Msib{} (in units of $M_\odot$), \xx{1.5}, \xx{2.5}, $s_{\mathrm{k, c}}$, and $M_4$ are compared, and in the bottom panel, the logarithm of the remaining lifetimes from the beginning of the He burning phase and from the depletion of He, C, Ne, O, and Si at the reference points (in units of Myr), as well as the logarithm of radius ($R_\odot$) and luminosity ($L_\odot$) at the surface at core-collapse, are shown together with \MCO{} and \Mff{}. The face colors indicate Spearman's rank correlation coefficients, which are also indicated by the numbers included in each sub-panel, with the color scale shown in the top-right color bar.}
 \label{plot-matrix0}
\end{figure*}

We have found that monotonicity is inherent in the cores of massive stars. The density, entropy, and chemical distributions inside the base of the C burning layer can be sorted simultaneously if a characterizing parameter for the sorting is appropriately given. The correlations between the structural properties discussed in this work are summarized in the top panel of Fig.~\ref{plot-matrix0}. Because of the monotonicity, choosing the characterizing parameter is arbitrary. The compactness parameter $\xi$ could be one possibility, but other parameters such as \Mff{}, the chemically defined base masses, and the core entropy have the same qualitative functionality. We have also found that not only the final core structure but also the evolutionary properties of the remaining lifetimes after neon ignition and the final He star radius obey the monotonicity (the bottom panel of Fig.~\ref{plot-matrix0}).

We stress that the existence of such monotonicity is non-trivial. Indeed, it is well known that the core structure has no monotonic correlation to the initial stellar mass. This is because stiff nuclear reaction rates and neutrino energy loss rates, as well as the non-linear interplay between the nuclear reactions and chemical mixing, bring significant complexity to the entropy and chemical distributions, and hence, the hydrostatic density structure inside the core of the massive star.

\begin{figure}
    \centering
	\includegraphics[width=\hsize]{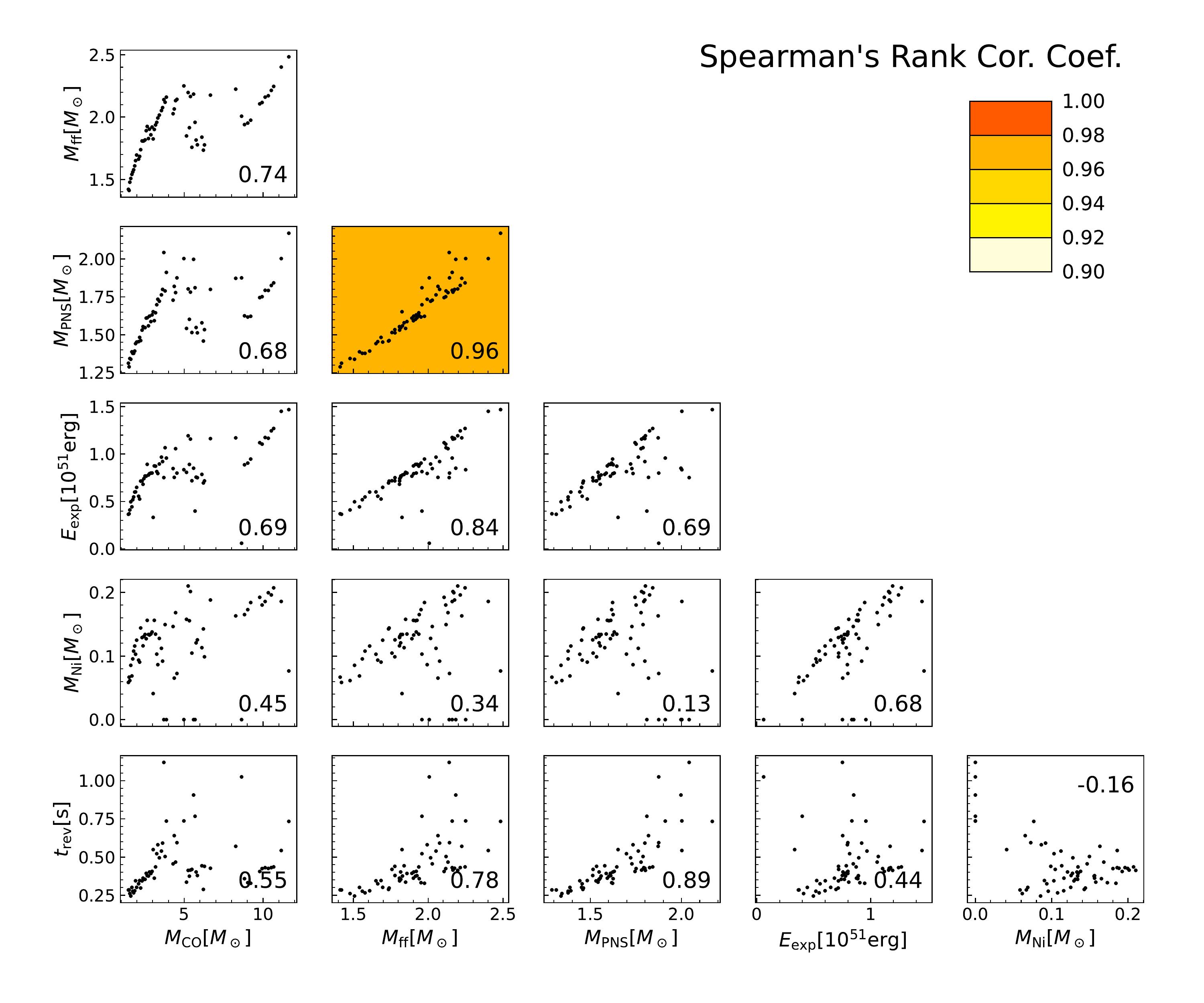}
    \caption{The same as Fig.~\ref{plot-matrix0}, but for CCSN explosion properties including PNS mass ($M_\odot$), explosion energy ($10^{51}$ erg), nickel ejecta mass ($M_\odot$), and shock revival time (s). The correlations shown here are made for successfully exploded models only.}
    \label{plot-matrix2}
\end{figure}
Based on the semi-analytic model of \citet{Mueller+16}, we have suggested the existence of correlations between \Mff{} and explosion properties such as explosion energy, $^{56}$Ni ejecta mass, shock revival time, and especially PNS mass (Fig.~\ref{plot-matrix2}). This should be interpreted as the result of the more general monotonicity, in particular the correlation between \Mff{} and the global density structure of the core. In this sense, the monotonicity of the core provides a unified understanding of the progenitor-explosion connection that has been investigated in the past decade. Furthermore, as long as the assumed explosion mechanism is linked to the density distribution of the progenitor's core and does not have an irregular dependency, the outcome of any theoretical investigations will also be characterized by the same parameter that is connected to the monotonicity of the progenitor's core. In a real explosion, however, progenitor properties other than the density distribution, such as the stellar rotation, the convective turbulence, and the magnetic fields may have an equally important influence.

The monotonicity will be useful for some aspects. For example, in order to reduce the computational cost, population synthesis studies have used simple prescriptions to determine the fate of the star that is based on the initial, the final, and the He- and CO-core masses \citep[e.g.,][]{Belczynski10, Kinugawa14, Spera19, Rodriguez16, Banerjee17, Tagawa20}. Utilizing the core monotonicity will further reduce the complexity and may improve the accuracy of such a prescription. It will be also useful for constructing a parametric model of CCSN progenitors \citep[e.g.][]{Suwa+16}. The monotonicity will be particularly substantial as a sanity check, and it may also improve the efficiency of parametric studies by setting a constraint to the parameter space.

The monotonicity we have shown is far from perfect, and many outliers have been found. The scatter might be due to some physical effects, but equally possible is that they originate from numerical errors. Improving numerical accuracy \citep[e.g., increasing the spatial resolution,][]{Sukhbold+18} will be worthwhile to disentangle the possibilities. Besides, it will be interesting to search for higher-order correlations.

The last note about the robustness of our result is that our calculation is based on 1D stellar evolution simulations, in which significant simplifications are involved in many aspects. One critical issue will be the treatment of convection.
In our calculation, both energy and chemical transport due to the convective turbulence relies on the traditional mixing-length theory \citep{Boehm-Vitense58}. Applying a more sophisticated theory \citep[e.g.,][]{Arnett+19, Arnett+18, Yokoi+22} with a more reliable treatment for the convective boundary mixing may affect the result. Similarly, it will be interesting to investigate the effect of stellar rotation \citep{Maeder&Meynet00, Heger+00} and stellar magnetic fields \citep[e.g.,][]{Takahashi&Langer21} on the monotonicity.

\vspace{5mm}
The authors appreciate the referee for carefully reading the manuscript and providing many valuable comments.
K.T. would like to thank Masaru Shibata and Norbert Langer for stimulating discussions.
The authors are grateful to Alexander Heger for providing a set of progenitor data, M16, which is available at \url{https://2sn.org/DATA/MHLC16/presn/} and to Bernhard M\"{u}ller for providing data for detailed comparison shown in the Appendix.
This study was supported in part by Grants-in-Aid for Scientific Research of the Japan Society for the Promotion of Science (JSPS, Nos. 
JP17H06364, 
JP21H01088, 
JP22H01223, 
JP22K20377) 
and JICFuS as “Program for Promoting researches on the Supercomputer Fugaku” (Toward a unified view of 
the universe: from large scale structures to planets, JPMXP1020200109).
Numerical computations were in part carried out on a PC cluster at Center for Computational Astrophysics, National Astronomical Observatory of Japan.

The progenitor models calculated in this work are available for download at \url{https://doi.org/10.5281/zenodo.6665557}.

\vspace{5mm}
\facilities{PC cluster (NAOJ, CfCA)}

\software{HOSHI \citep{Takahashi+18, Takahashi&Langer21}}

\begin{appendix}
\section{Muller's semi-analytic model} \label{Appendix1}

Although the complete physical concept is well described in \citet{Mueller+16}, we have observed that subtle differences in the implementation can affect the result considerably. In order to improve the traceability of our work, here we describe how the semi-analytic model is implemented and provide results of the comparison between the original work.
The physical constants applied are
$
	c = 3.0\times10^{10}\,{\rm cm\,s^{-1}},
$
$
	G = 6.67408\times10^{-8}\,{\rm cm^3\,s^{-2}\,g^{-1}},
$
$
	M_\odot = 1.9884\times10^{33}\,{\rm g},
$
$
	m_u = 1.66054\times10^{-24}\,{\rm g},
$
and
$
	a_{\rm rad} = 7.5657\times10^{-15}\,{\rm erg\,cm^{-3}\,K^{-4}}
$
for the speed of light, the gravity constant, the solar mass, the unified atomic mass unit, and the radiation constant, respectively.

\subsection{Basic equations}

Throughout the post-collapse evolution, time evolutions of the PNS mass $M_{\rm PNS}$, the explosion energies $E_{\rm imm}$ and $E_{\rm diag}$, and the ejected nickel mass $M_{\rm Ni}$ are evaluated.
Before shock revival, the mass of the PNS is identical to the stellar mass,
$
	M_{\rm PNS}(i) = M(i),
$
where $i$ is the grid number and $M(i)$ is the (cell-surface) enclosed mass.
Other quantities are set to zero.

We assign a time for each grid, with which the mass shell reaches the stellar center after the core collapse, as
\begin{eqnarray}
	t(i) = \sqrt{ \frac{\pi}{4 G \rho_{\rm ave}} },
\end{eqnarray}
with the average density 
$
	\rho_{\rm ave} = M(i)/(4 \pi r^3(i)/3)
$
and the (cell-surface) radius
$
	r(i).
$
Consequently, the mass accretion rate is given by
\begin{eqnarray}
	\dot{M}(i) = \frac{2 M(i)}{t(i)} \frac{\rho(i)}{\rho_{\rm ave}(i) - \rho(i)},
\end{eqnarray}
where $\rho(i)$ is the (cell-center) density.

The gain and shock radii at time $t(i)$ are estimated by
\begin{eqnarray}
	r_{\rm g}(i) = \sqrt[3]{r_0^3 
	+ r_1^3 
		\biggl( \frac{\dot{M}(i)}{M_\odot} \biggr)
		\biggl( \frac{M_{\rm PNS}(i)}{M_\odot} \biggr)^{-3}
	}
\end{eqnarray}
with the parameters
$
	r_0 = 12\,{\rm km}
$
and
$
	r_1 = 120\,{\rm km}
$
and 
\begin{eqnarray}
	r_{\rm sh}(i) = \alpha_{\rm turb} \times {0.55\,\rm{km}}
		\times
		\biggl( \frac{L_\nu(i)}{10^{52}\,\rm{erg\,s^{-1}}} \biggr)^{4/9}
		\biggl( \frac{M_{\rm PNS}(i)}{M_\odot} \biggr)^{5/9}
		\biggl( \frac{r_{\rm g}(i)}{10\,{\rm km}} \biggr)^{16/9}
		\biggl( \frac{\dot{M}(i)}{M_\odot} \biggr)^{-2/3}
		( \alpha_{\rm redshift}(i) )^{6/9} \label{eq-rshock}
\end{eqnarray}
with
$
	\alpha_{\rm turb} = 1.18.
$
The neutrino luminosity is estimated as
$
	L_\nu(i) = L_{\rm acc}(i) + L_{\rm diff}(i),
$
which consists of the accretion component 
\begin{eqnarray}
	L_{\rm acc}(i) = \zeta \times \frac{G M_{\rm PNS}(i) \dot{M}(i) }{ r_{\rm g}(i) }
\end{eqnarray}
with
$
	\zeta = 0.8
$
and the diffusion component
\begin{eqnarray}
	L_{\rm diff}(i) = 0.3 \times \frac{E_{\rm bind}(i)}{\tau_{\rm cool}(i)}  
		\exp\left( -\frac{t(i)}{\tau_{\rm cool}(i)} \right).
\end{eqnarray}
Note that the factor $1/\tau_{\rm cool}(i)$ is missing in \citet{Mueller+16}.
For the diffusion component, the binding energy of the PNS is
\begin{eqnarray}
	E_{\rm bind}(i) = a \times 
		\biggl( \frac{M_{\rm PNS}(i)}{M_\odot} \biggr)^{2}
		M_\odot c^2
\end{eqnarray}
with
$
	a = 0.084
$
and the cooling time is
\begin{eqnarray}
	\tau_{\rm cool}(i) = {\rm max} \biggl[
		0.1\,{\rm s},
		\tau_{15} 
		\biggl( \frac{M_{\rm PNS}(i)}{M_\odot} \biggr)^{5/3}
	\biggr]
\end{eqnarray}
with
$
	\tau_{15} = 1.2\,{\rm s}.
$
The redshift correction is given by
\begin{eqnarray}
	\alpha_{\rm redshift}(i) = 1 - \frac{2GM_{\rm PNS}(i)}{r_{\rm PNS}(i) c^2},
\end{eqnarray}
where the PNS radius is estimated to be
$
	r_{\rm PNS}(i) = \frac{5}{7} r_{\rm g}(i).
$
Note that the power of the redshift correction in eq.~(\ref{eq-rshock}) is increased from the value written in the original work of 2/9 to obtain results consistent with them.

These radii are used to estimate the advection timescale as
\begin{eqnarray}
	t_{\rm adv}(i) = 18 \, {\rm ms} \times
		\biggl( \frac{ r_{\rm sh}(i) }{100\,{\rm km}} \biggr)^{3/2}
		\biggl( \frac{M_{\rm PNS}(i)}{M_\odot} \biggr)^{-1/2}
		\ln \biggl( \frac{r_{\rm sh}(i)}{r_{\rm g}(i)} \biggr),
\end{eqnarray}
and it is compared with the heating timescale given by
\begin{eqnarray}
	t_{\rm heat}(i) = 150 \ {\rm ms} \times
		\biggl( \frac{e_{\rm g}(i)}{10^{19} \ {\rm erg}} \biggr)
		\biggl( \frac{r_{\rm g}(i)}{100 \ {\rm km}} \biggr)^{2}
		\biggl( \frac{L_\nu(i)}{10^{52} \ \rm{erg s^{-1}}} \biggr)^{-1}
		\biggl( \frac{M_{\rm PNS}(i)}{M_\odot} \biggr)^{-2}
		( \alpha_{\rm redshift}(i) )^{-3/2}, \label{eq-theat}
\end{eqnarray}
where 
\begin{eqnarray}
	e_{\rm g}(i) = \frac{3}{4} e_{\rm diss}
		+ \frac{1}{4} \frac{GM_{\rm PNS}(i)}{ {\rm max} [r_{\rm g}(i), r_{\rm sh}(i)] } 
\end{eqnarray}
with
$
	e_{\rm diss} = 8.8 \ {\rm MeV/}m_u
$
being the post-shock binding energy without rest-mass contributions, to yield the condition of shock-revival: the bounce shock revives if $t_{\rm heat}(i) < t_{\rm adv}(i)$. Also, note that the power of the redshift correction in eq.~(\ref{eq-theat}) is changed from the original value of $-1/2$. 
At shock revival, the PNS mass one time-step before is recorded as the `initial' mass of the PNS, $M_{\rm ini} = M_{\rm PNS}(i-1)$.

In the earlier phase after shock revival, equations solved are
\begin{eqnarray}
	M_{\rm PNS}(i+1) &=& M_{\rm PNS}(i) +
		(1-\alpha_{\rm out})
		\biggl( 1-\frac{\eta_{\rm acc}(i)}{e_{\rm g}(i)} \biggr)
		\times \Delta M(i) \\
	E_{\rm imm}(i+1) &=& E_{\rm imm}(i) +
		e_{\rm rec}
		\biggl( \frac{\eta_{\rm acc}(i)}{e_{\rm g}(i)} \biggr)
		{\rm min} \biggl[ 1.0, \frac{\dot{M}(i)}{4 \pi r^2(i) \rho(i) v_{\rm sh}(i) } \biggr]		
		\times \Delta M(i) \nonumber \\
	&&	+ \alpha_{\rm out} (\epsilon_{\rm bind}(i) + \epsilon_{\rm burn}(i))
		\times \Delta M(i) \\
	E_{\rm diag}(i+1) &=& E_{\rm diag}(i) +
		e_{\rm rec}
		\biggl( \frac{\eta_{\rm acc}(i)}{e_{\rm g}(i)} \biggr)
		(1-\alpha_{\rm out})
		\times \Delta M(i) \nonumber \\
	&&	+ \alpha_{\rm out} (\epsilon_{\rm bind}(i) + \epsilon_{\rm burn}(i))
		\times \Delta M(i) \\
	M_{\rm Ni}(i+1) &=& M_{\rm Ni}(i) +
		X_{\rm Ni}(i)
		\times \Delta M(i)
\end{eqnarray}
with $\Delta M(i) = M(i+1)-M(i)$ and the parameters are
$
	\alpha_{\rm out} = 0.5
$
and
$
	e_{\rm rec} = 5 \ {\rm MeV/}m_u.
$
$\eta_{\rm acc}(i)$ is an efficiency parameter relating the mass accretion rate and the neutrino heating rate and is evaluated as
\begin{eqnarray}
	\eta_{\rm acc}(i) = e_{\rm g}(i) \biggl( \frac{t_{\rm adv}(i)}{t_{\rm heat}(i)} \biggr),
\end{eqnarray}
and $v_{\rm sh}(i)$ is the shock velocity evaluated as
\begin{eqnarray}
	v_{\rm sh}(i) = 0.794 \times
		\biggl( \frac{E_{\rm imm}(i)}{M(i)-M_{\rm ini}} \biggr)^{1/2}
		\biggl( \frac{M(i)-M_{\rm ini}}{\rho(i) r^3(i)} \biggr)^{0.19}.
\end{eqnarray}
$\epsilon_{\rm bind}(i)$ and $\epsilon_{\rm burn}(i)$ are the binding energy per unit mass of the unshocked material and the added energy due to nuclear burnings. They are estimated as
\begin{eqnarray}
	\epsilon_{\rm bind}(i) = e_{\rm therm}(i) - \frac{GM(i)}{r(i)}
\end{eqnarray}
with the thermal energy, $e_{\rm therm}(i)$, and as
\begin{eqnarray}
	\epsilon_{\rm burn}(i) = \Sigma_k (X_k(i) - X'_k(i)) \epsilon_{{\rm rm}, k},
\end{eqnarray}
where 
$X_k(i)$ is the chemical composition after the explosive nucleosynthesis, $X'_k(i)$ is the initial composition, and $\epsilon_{{\rm rm}, k}$ is the rest mass contributions per unit mass for nucleus $k$.
Note that the definition of $\epsilon_{\rm bind}(i)$ is not explicitly provided in \citet{Mueller+16}.

$X_k(i)$ is determined using the post-shock temperature $T_{\rm sh}(i)$, which is given by
\begin{eqnarray}
	T_{\rm sh}(i) = \biggl[ \frac{ 3\beta-1 }{ a_{\rm rad}\beta }\rho(i) v_{\rm sh}^2(i) \biggl]^{1/4}
\end{eqnarray}
with 
$
	\beta = 4.
$
Using the temperature, $X_k(i)$ is given as
\begin{enumerate}
	\item changing elements lighter than O into $^{16}$O
		if $T_{\rm sh}(i) \in [2.5\times10^9, 3.5\times10^9)$ K.
	\item changing elements lighter than Si into $^{28}$Si
		if $T_{\rm sh}(i) \in [3.5\times10^9, 5\times10^9)$ K.
	\item changing all elements into $^{56}$Ni
		if $T_{\rm sh}(i) \ge 5\times10^9$ K.
\end{enumerate}
Note that this post-shock temperature is the same as eq.~(46) in \citet{Mueller+16} and is different from 
$
	T_{\rm sh}(i) = [ ( 3(\beta-1) / (a_{\rm rad}\beta) )\rho(i) v_{\rm sh}^2(i) ]^{1/4}
$
that is implied from eq.~(45) in \citet{Mueller+16}.

The first explosion phase ends when the post-shock velocity,
\begin{eqnarray}
	v_{\rm post} = \frac{\beta-1}{\beta} v_{\rm sh},
\end{eqnarray}
exceeds the local escape velocity,
\begin{eqnarray}
	v_{\rm esc} = \sqrt{ \frac{2GM(i)}{r(i)} },
\end{eqnarray}
thus 
$
	v_{\rm post} > v_{\rm esc}.
$
Thereafter, the second explosion phase begins, and the evolution equations 
\begin{eqnarray}
	M_{\rm PNS}(i+1) &=& M_{\rm PNS}(i) \\
	E_{\rm imm}(i+1) &=& E_{\rm imm}(i) +
		(\epsilon_{\rm bind}(i) + \epsilon_{\rm burn}(i))
		\times \Delta M(i) \\
	E_{\rm diag}(i+1) &=& E_{\rm diag}(i) +
		(\epsilon_{\rm bind}(i) + \epsilon_{\rm burn}(i))
		\times \Delta M(i) \\
	M_{\rm Ni}(i+1) &=& M_{\rm Ni}(i) +
		X_{\rm Ni}(i)
		\times \Delta M(i)
\end{eqnarray}
are solved.

We set four possibilities for judging BH formation.
Firstly, a BH forms if the model never meets the shock revival condition.
Secondly, a BH forms if the (baryonic) mass of the PNS exceeds $2.40301 \ M_\odot$, which corresponds to the maximum gravitational mass of $M_{\rm grav} = 2.05 M_\odot$ under a relation
\begin{eqnarray}
	M_{\rm PNS} = M_{\rm grav}
	 + 0.084 \biggl( \frac{ M_{\rm grav} }{ M_\odot } \biggl)^2 M_\odot.
\end{eqnarray}
Thirdly, a BH forms if the diagnostic explosion energy, $E_{\rm diag}(i)$, becomes negative.
Lastly, a BH forms if the redshift correction, $\alpha_{\rm redshift}(i)$, becomes negative.

\subsection{Comparison with the original work} \label{Appendix2}

\begin{figure}[t]
 \centering
  \begin{minipage}[b]{0.45\linewidth}
	\includegraphics[width=\hsize]{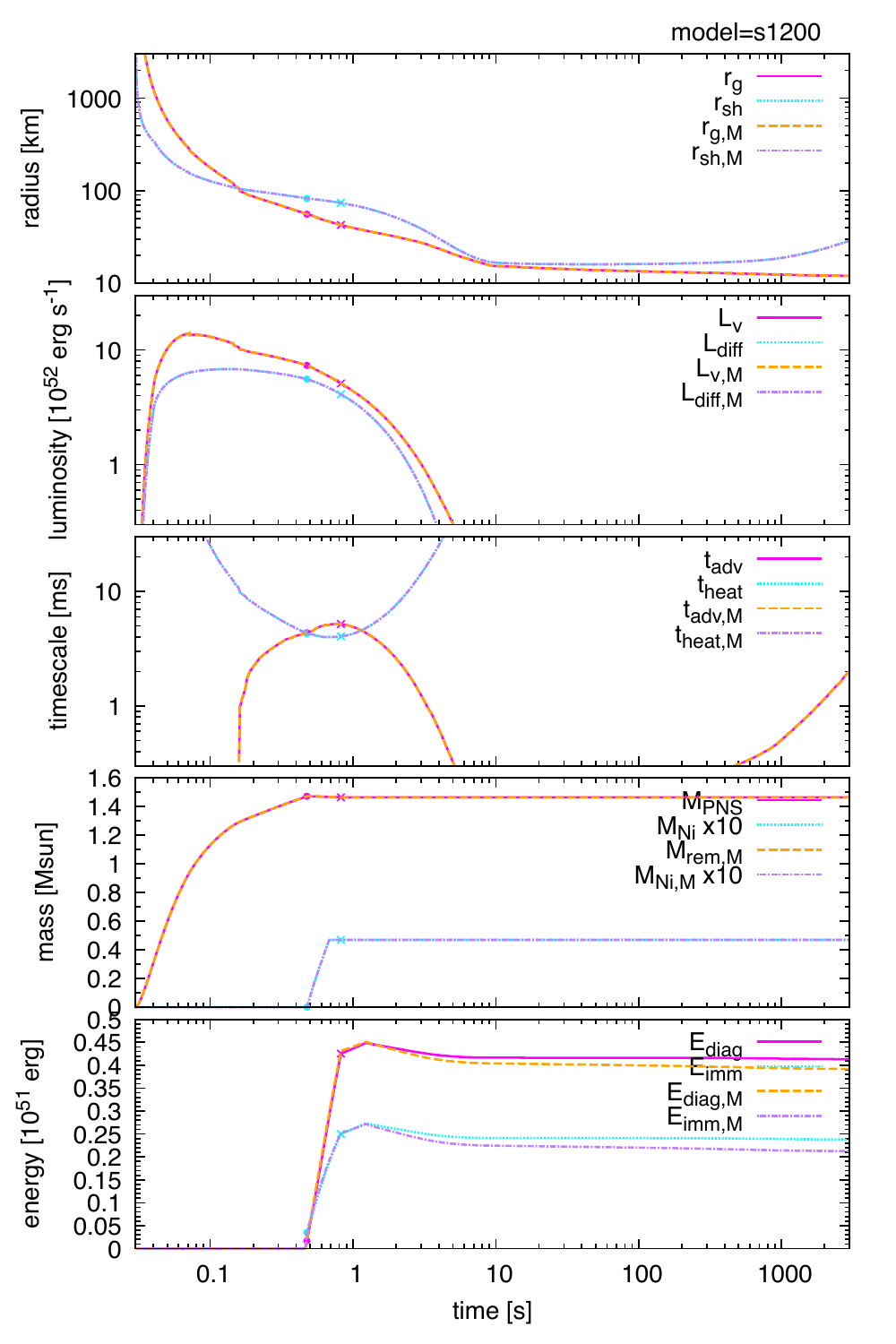}
  \end{minipage}
  \begin{minipage}[b]{0.45\linewidth}
	\includegraphics[width=\hsize]{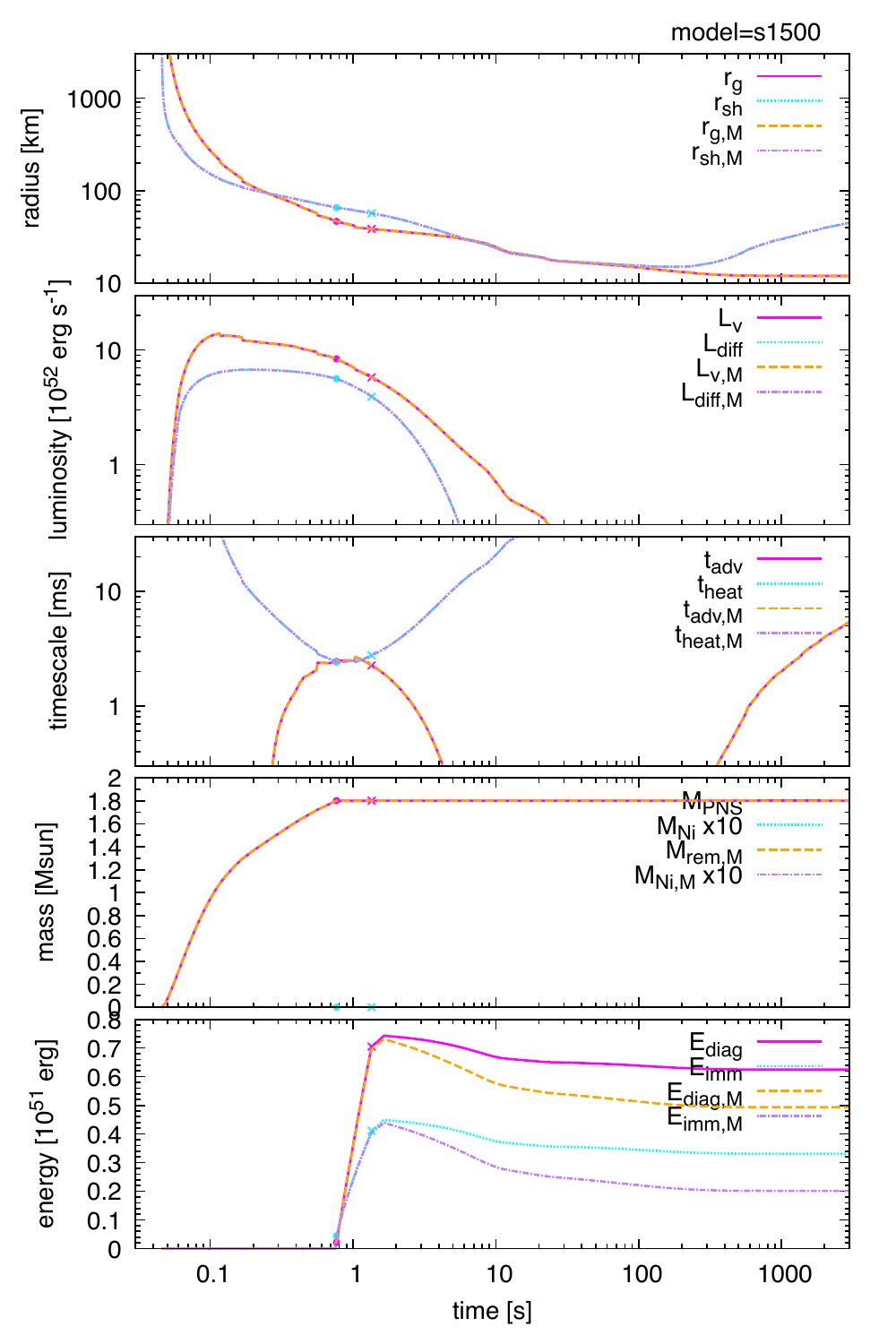}
  \end{minipage}
 \caption{Detailed comparisons for models with the initial masses of 12 (left) and 15 $M_\odot$ (right) that are taken from \citet{Mueller+16}. From the top to bottom panels, time evolutions of 
 	$r_{\rm g}$ and $r_{\rm sh}$ (top),
	$L_\nu$ and $L_{\rm diff}$ (2nd),
	$t_{\rm adv}$ and $t_{\rm heat}$ (3rd),
	$M_{\rm PNS}$ and $M_{\rm Ni}$ multiplied by a factor of 10 (4th), and
	$E_{\rm diag}$ and $E_{\rm imm}$ (bottom)
are shown.
As for the comparison, results obtained with our implementations are shown by magenta solid and cyan dotted curves, and that of the original work are by orange dashed and purple dash-dotted curves. In the legends, results from the original work are also indicated by the subscript `$_{\rm M}$'. The timing of shock revival is indicated by filled points, while the timing of $
	v_{\rm post} = v_{\rm esc}
$ is indicated by crosses.}
 \label{plot-comparison}
\end{figure}

Detailed comparisons between our and the original implementations for models with the initial masses of 12 and 15 $M_\odot$ are shown in Fig.~\ref{plot-comparison}. We have confirmed that, throughout the evolution, except for at the very beginning, more than 5-digit consistency is achieved for the quantities shown in the top four panels. On the other hand, the two energies shown in the bottom panels involve $\sim 1$\% inconsistencies for the first explosion phase, which increase to $\sim 10$\% order differences for the second explosion phase.

\begin{figure}[t]
 \centering
  \begin{minipage}[b]{0.45\linewidth}
    \includegraphics[width=\hsize]{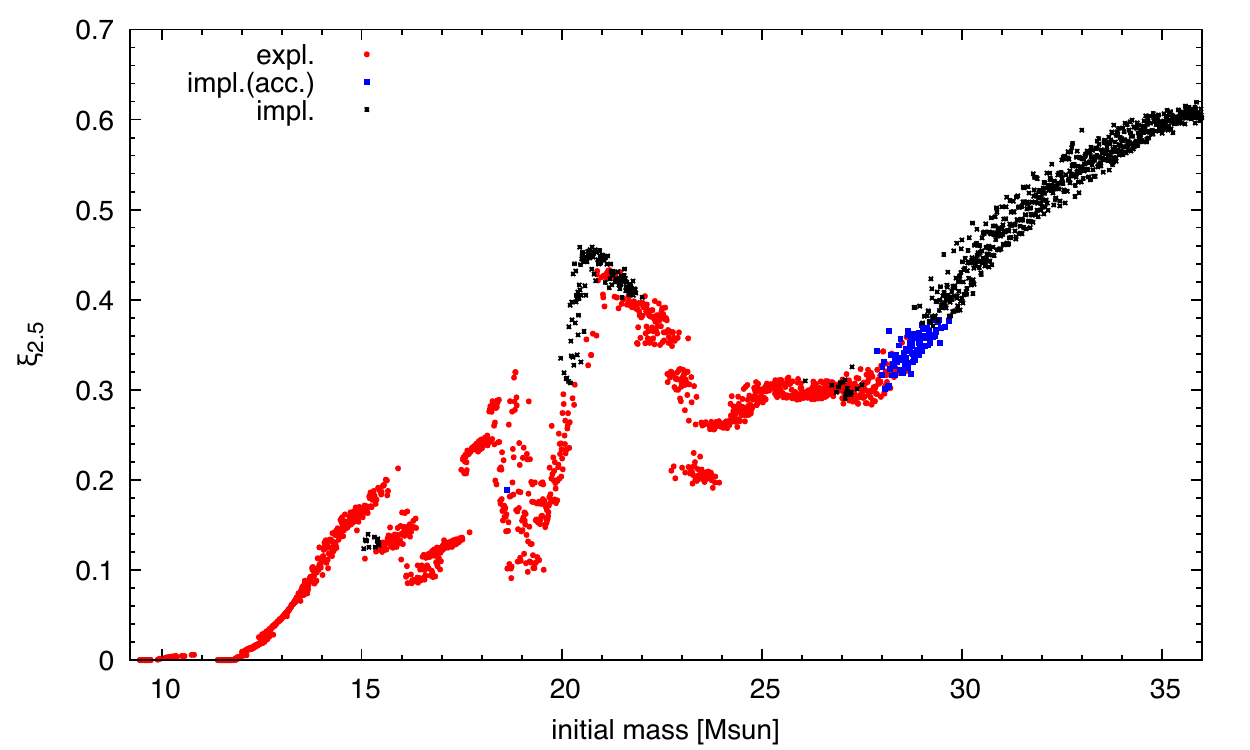}
  \end{minipage}
  \begin{minipage}[b]{0.45\linewidth}
	\includegraphics[width=\hsize]{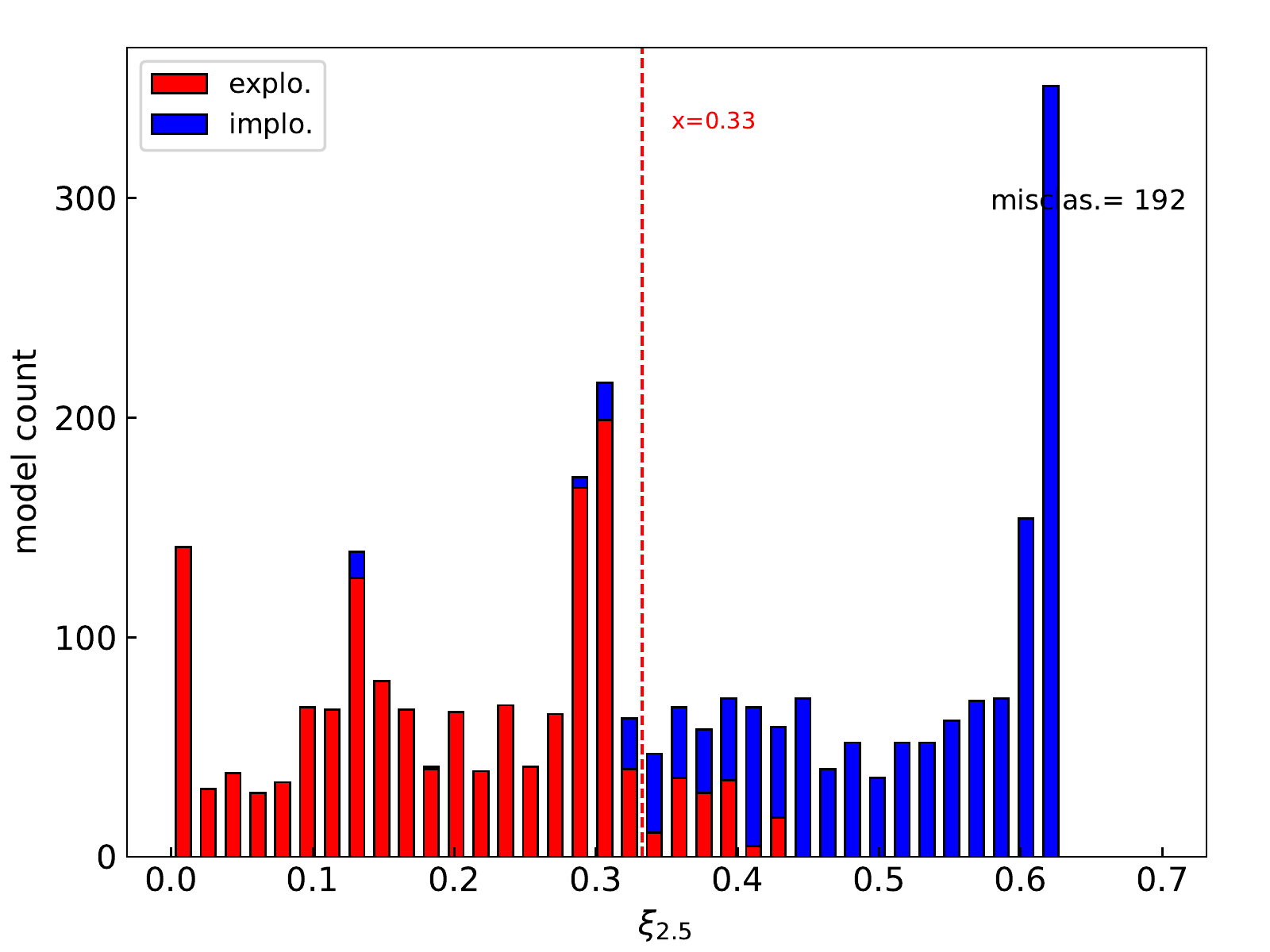}
  \end{minipage}
 \caption{(left) $\xi_{2.5}$ distribution of M16 model set. The colors indicate the fates expected from the M\"{u}ller's semi-analytic model of successful explosion (red), BH formation due to accretion after shock revival (blue), and BH formation before shock revival (black). This figure is comparable to Fig.~6 of \citet{Mueller+16}. (right) Same as Fig.~\ref{plot-explodability2} but for the \xx{2.5} histogram of progenitor models of \citet{Mueller+16}.}
 \label{plot-explodability2-M16}
\end{figure}

\begin{figure}
    \centering
	\includegraphics[width=0.5\hsize]{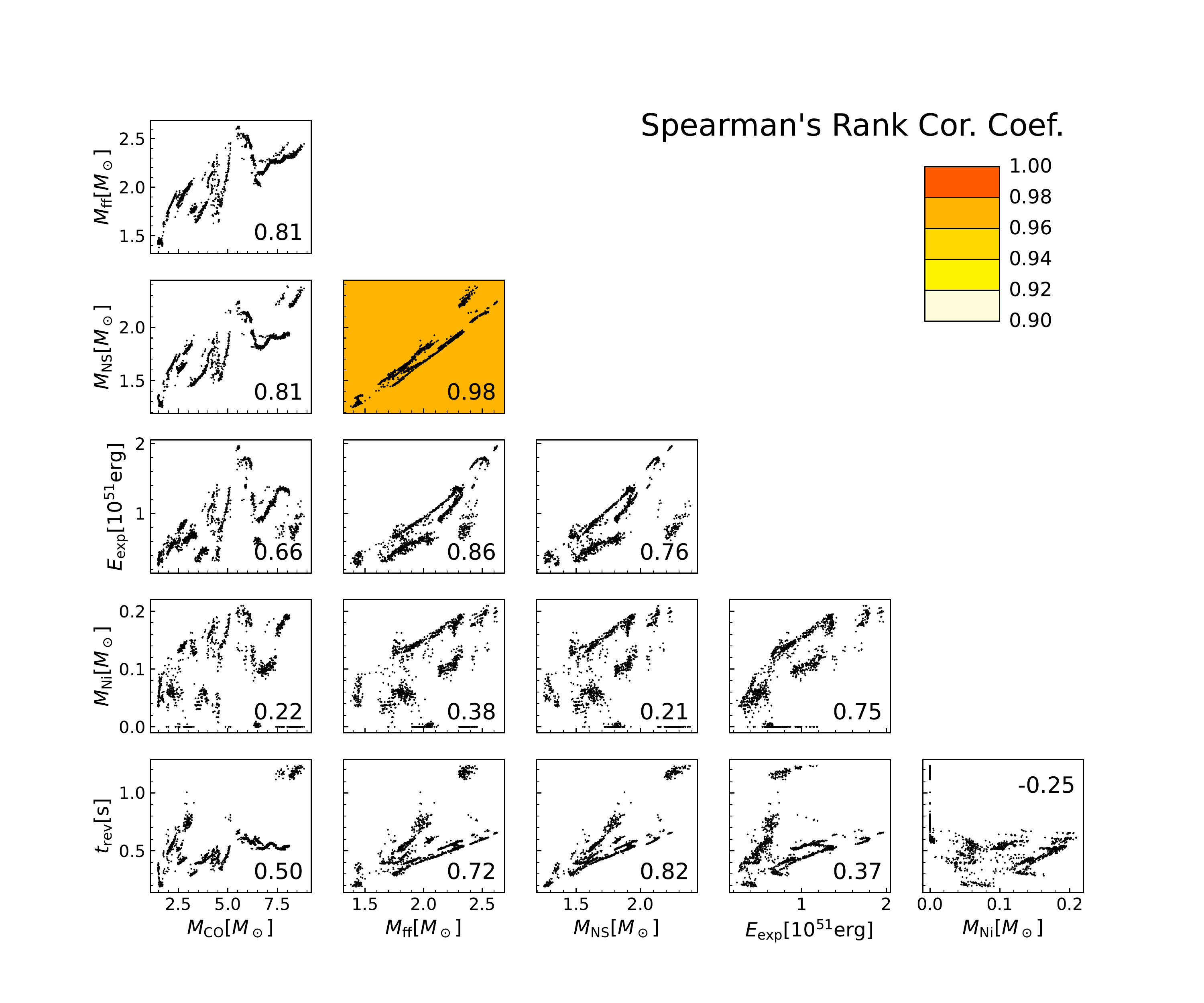}
    \caption{The same as Fig.~\ref{plot-matrix2}, but for CCSN explosion properties calculated for the M16 set using our implementation of the M\"{u}ller's semi-analytic model.}
    \label{plot-matrix3}
\end{figure}

An estimate of the explodability is shown in the left panel of Fig.~\ref{plot-explodability2-M16}, which is comparable to Fig.~6 of \citet{Mueller+16}. This distribution includes a region of implosion around the peak at $M_{\rm ini} \sim 20.5$ $M_\odot$ as well as a region of BH formation due to late time accretion at $M_{\rm ini} \sim 29$ $M_\odot$, and these are qualitatively consistent with the original. The region of explosion between them ($M_{\rm ini} \sim$ 22--28 $M_\odot$) is wider in this work, and this may be due to the disagreement of $E_{\rm imm}$ and $E_{\rm diag}$ in the latter explosion phase. Qualitatively speaking, this disagreement enlarges the window of exploding models in our implementation. This is illustrated in the right panel of Fig.~\ref{plot-explodability2-M16}, which shows the histogram of \xx{2.5} of progenitor models in M16. The threshold value, \xx{2.5} $\sim$ 0.33, is larger than the original estimate of \xx{2.5} = 0.278. As in Fig.~\ref{plot-matrix2}, a summary for the M16 set is shown in Fig.~\ref{plot-matrix3}. A strong correlation between \Mff{} and the explosion properties, in particular, the NS mass is also found in the result, hence we obtain almost the same trends as with our own model set.

\section{Comparison of the explosion properties}

The M\"{u}ller's semi-analytic model we have used to predict the property of CCSN explosion, like any other theoretical calculation, involves a certain degree of uncertainty. Therefore, it is important to know how robust the obtained results are. Accordingly, although predicting the properties of CCSNe is not the main topic of this study, we compare the results obtained here with previous studies and summarize their similarities and differences. In particular, we investigate whether the explosion can be judged by \xx{2.5} and similar parameters and whether there are correlations between the various quantities that characterize the explosion.

We first compare the results of three studies using the M\"{u}ller's model \citep{Mueller+16, Schneider21, Aguilera-Dena22}. Regarding the estimate of the explodability, it was stated that the exploding models can be roughly judged with \xx{2.5} $\lesssim 0.278$ \citep{Mueller+16} or with \xx{2.5} $\lesssim 0.35$ \citep{Aguilera-Dena22}. Results in \citet{Schneider21} also seem to indicate that the explosion is more successful for models with small compactness, for example, looking at their Fig. 7. In other words, all of these studies show that the explodability can be judged in a semi-empirical (compactness-based) manner to first order, although the discrimination is not perfect and the critical value is not definitive.

In all of these works, correlations between explosion properties (explosion energy, nickel ejecta mass, and PNS mass) were found. Furthermore, \citet{Schneider21} and \citet{Aguilera-Dena22} have shown that the nature of the explosion, in particular the PNS mass, correlates with \xx{2.5}.
Although \citet{Mueller+16} stated that there is no strong correlation between compactness and the explosion properties, their results (e.g., Fig. 12) also show that the strongest explosions come from the most compact stars, so it seems likely that a loose correlation could be found. Besides, when progenitor models of \citet{Mueller+16} are analyzed using the M\"{u}ller's model with our implementation, a correlation is found, especially between PNS mass and \Mff{} (see Appendix).
From these facts, we consider that the properties of the CCSN explosion estimated by the M\"{u}ller's model are correlated with each other and that they are also correlated with the compactness, especially the PNS mass to some extent. These properties are in good agreement with our results.

By combining simulations and analytical relations, \citet{Pejcha&Thompson15} investigated the explodability and the nature of the CCSN explosion.
They first simulated the gravitational collapse of a massive star including the neutrino emission processes using the 1D general relativistic neutrino radiation hydrodynamical code GR1D \citep{OConnor10} and determined the time evolution of $L_\nu/L_{\rm crit}$. Here, $L_{\rm crit}(t)$ is the analytically estimated critical neutrino luminosity required for the explosion, and $L_\nu(t)$ is the neutrino luminosity obtained from the simulation. Then, the explosion was assumed to occur when $L_\nu/L_{\rm crit}(t)$ exceeds a certain threshold value, and the explosion energy and nickel mass were further estimated by a semi-analytical method for exploding models. This threshold, which was given as a simple function of the mass accretion rate $\dot{M}$, includes parameters, and various explosion conditions were considered by changing the parameters.
They argued that explodability is not determined by compactness alone. However, their result includes a region where it is difficult to find exploding parameters at $M_{\rm ZAMS} = 22$--$26 M_\odot$ (their Fig. 13), which corresponds to the peak of the compactness distribution. Besides, for a specific model set (the parameter (a)) that mimics the explosion fraction in \citet{Ugliano+12}, the compactness-based judgment was able to separate models that explode from those that do not with 88\% accuracy. So there seems to be a loose correlation between compactness and explodability also in their result.
For explosion properties, they found a correlation between the explosion energy and the nickel ejecta mass. Their estimate of the nickel ejecta mass was based on the assumption that radiation energy is dominant inside the shock, and the correlation seems to be a direct consequence of this robust but ad-hoc assumption. The explosion energy was also correlated with the NS mass, but it should be noted that this is an inverse correlation (see their Fig. 19). The inverse correlation is probably due to their method for energy estimation, where the less compact models explode earlier, having stronger neutrino winds, and therefore have larger explosion energies estimated by integrating the power of the neutrino winds. For compactness, it was stated that compactness roughly correlates with NS mass.

\citet{Ugliano+12}, \citet{Ertl+16}, and \citet{Sukhbold+16} performed parametric 1D simulations calibrated with observations. A NS model was incorporated as an energy source, and the neutrino luminosity was controlled by parameters to yield explosions even in the 1D hydrodynamical simulations.

\citet{Ugliano+12} calibrated the parameters for SN 1987A, the most closely observed supernova. They found that the NS mass correlates with the mass at the bottom of the oxygen-burning shell (equivalent to our \Mob{}), but there is no clear correlation between the various quantities characterizing the explosion. 
Later, it was recognized that calibration with 1987A alone would cause the less compact models to explode very strongly, which is contrary to the sophisticated simulations of Crab-like supernovae. Accordingly, \citet{Ertl+16} and \citet{Sukhbold+16} treated the parameters as variables that vary in proportion to compactness (or a similar measure), rather than constants, for stars with small compactness, so that less massive stars have weak neutrino luminosities correlated with compactness.
This modification appears to affect the correlation between the properties of the explosion; \citet{Sukhbold+16} reported that, for the heavier mass models using constant parameters, nickel ejecta mass correlates with the compactness while the explosion energy is roughly constant meaning no correlation. On the other hand, a correlation between nickel ejecta mass and explosion energy exists for the less massive stars using a linear function of compactness for the engine parameters.

\citet{Perego+15} and \citet{Ebinger18, Ebinger20} performed 1D explosion simulations using the PUSH method, which incorporates the effect that the efficiency of neutrino heating is increased by multidimensional convective motion. In Push, heavy-flavor neutrinos are used as the effective additional source of energy, and the region where convection is likely to occur is heated for the time that convection is likely to occur. Parameters are included for the heating efficiency and the convection generation time. Three types of outcomes were considered for calibration: crab-like supernovae with less compactness, SN 1987A-like supernovae with intermediate compactness, and compact stars, which are thought to create BHs. Then, by interpolating these three points with a quadratic function of compactness, they determined a parameter function for the neutrino heating efficiency.

\citet{Ebinger18, Ebinger20} found that there is a correlation between the properties of the explosion (explosion energy, nickel emission mass, NS mass) and also between compactness and, in particular, NS mass. They also studied the chemical composition of the supernova ejecta in detail and found that the amount of $^{56}$Ni and $^{44}$Ti correlate with compactness \citep{Ebinger20}.

Finally, we compare the trends obtained from multi-D explosion simulations conducted by \citet{Nakamura15} and \citet{Burrows20}. Among many works that conduct multi-D simulations, these are particularly relevant to our work, as they discuss how the characteristics of the explosion depend on the structure of the progenitor star. 
While these simulations still rely on approximate treatments of neutrino transport and general relativity, there is no artificial engine, and the explosion is driven by neutrino heating in accordance with the delayed explosion mechanism. For this reason, estimates of explodability and explosion features may be more realistic than results from parametric models.
On the other hand, due to the computational costs, especially for \citet{Burrows20}, where 3D simulations were performed, the number of calculated models is small. For the same reason,  long simulations were not performed in these works, and the estimation of explosion energy is not yet certain \citep[However, see][]{Murphy19}. This is why the correlation between the explosion energy and the nickel ejecta mass cannot be confirmed from these works.

\citet{Nakamura15} calculated 2D axisymmetric simulations for 378 progenitor models with three metallicities: solar metalicity, ultra metal-poor, and zero metallicity. In their calculations, most of the models exploded, so there is no discussion of what determines the explodability. On the other hand, a number of explosion indices, such as accretion luminosity and nickel ejecta mass, were shown to correlate with compactness. In particular, PNS mass had the strongest correlation. The shock revival time, defined as the time when the shock front passes 400 km, was shown to have a weak positive correlation with compactness, although it has a large scatter.

\citet{Burrows20} performed 3D simulations for 19 progenitor models. An important conclusion is that their results show that models with small or large compactness explode, while models with intermediate compactness (their $M_{\rm ZAMS}=13, 14, 15 M_\odot$ models) do not, i.e., the explodability cannot be separated by compactness. On the other hand, if we restrict ourselves to exploded models, many of their properties appear to be correlated with compactness. For example, neutrino luminosity and neutrino energy deposition rate are smaller for less compact models and larger for more compact models (their Figs. 4 and 5). PNS mass is also highly correlated with compactness (Table 3). One exception is that shock revival time does not appear to correlate with compactness (Fig. 2). This property may be related to explodability.

In summary, we have found similar trends to our results in many previous studies. In particular, the correlation between explosion energy and nickel ejecta mass and the correlation between PNS mass and compactness are common features. 
The former correlation indicates that the equation (9) in \citet{Pejcha&Thompson15} is robust. 
As for the correlation between PNS mass and compactness, compactness correlates with the density structure of the entire core as shown by this work, and thus compactness is a highly predictive indicator of the time evolution of the mass accretion rate. Hence, it suggests that the PNS mass determined as a result of the explosion can be predicted solely from the evolution of the mass accretion rate and does not sensitively depend on the details of the explosion mechanism.
Conclusions about the relation between explodability and compactness depend on the modeling method.
In other words, parametric 1D simulations have shown that explodability can be determined by compactness, while more self-consistent 3D simulations by \citet{Burrows18, Burrows20} have concluded that compactness does not predict the explodability, but is determined by differences in the entropy jump at the base of the O layer and the non-uniformity of density due to convection.
It should be noted that many parametric calculations implicitly assume that the engine property depends on the compactness of the progenitor, which may be why the explosion properties are correlated with compactness. To accurately determine explodability, a more realistic engine model should be used in parametric calculations.
On the other hand, regardless of the modeling method, explosion properties, such as explosion energy, tend to correlate with compactness when limited to models that experience a successful explosion.

\end{appendix}

\bibliography{biblio}

\end{document}